# The microstructural evolution of water ice in the solar system through sintering

J. L. Molaro[1,2*], M. Choukroun[2], C. B. Phillips[2], E. S. Phelps[2], R. Hodyss[2], K. L. Mitchell[2], J. M. Lora[3], and G. Meirion-Griffith[2]

[1]Planetary Science Institute, 1700 East Fort Lowell, Suite 106, Tucson, AZ 85719, USA.

[2]Jet Propulsion Laboratory, California Institute of Technology, 4800 Oak Grove Drive, Pasadena, CA 91109, USA.

[3]University of California, Los Angeles, 595 Charles Young Drive East, Los Angeles, CA 90095, USA.

* Corresponding author: jmolaro@psi.edu

**Abstract**

Ice sintering is a form of metamorphism that drives the microstructural evolution of an aggregate of grains through surface and volume diffusion. This leads to an increase in the grain-to-grain contact area ("neck") and density of the aggregate over time, resulting in the evolution of its strength, porosity, thermal conductivity, and other properties. This process plays an important role in the evolution of icy planetary surfaces, though its rate and nature are not well constrained. In this study, we explore the model of Swinkels and Ashby (1981), and assess the extent to which it can be used to quantify sintering timescales for water ice. We compare predicted neck growth rates to new and historical observations of ice sintering, and find agreement to some studies at the order of magnitude level. First-order estimates of neck growth timescales on planetary surfaces show that ice may undergo significant modification over geologic timescales, even in the outer solar system. Densification occurs over much longer timescales, suggesting some surfaces may develop cohesive, but porous, crusts. Sintering rates are extremely sensitive to temperature and grain size, occurring faster in warmer aggregates of smaller grains. This suggests that the microstructural evolution of ices may vary not only throughout the solar system, but also spatially across the surface and in the near-surface of a given body. Our experimental observations of complex grain growth and mass redistribution in ice aggregates point to components of the model that may benefit from improvement, and areas where additional laboratory studies are needed.

**Key Points**

- Sintering drives the microstructural evolution of ice through growth of grain-to-grain contacts (or necks) and aggregate density.
- We use a model to predict ice sintering timescales on planetary surfaces and compare the results to experimental observations.
- We find that densification occurs over longer timescales than neck growth, suggesting surfaces may develop cohesive but porous crusts.

**Plain Language Summary**

Ice sintering is a process that occurs to fresh ice grains deposited onto a planetary surface which causes them to stick to each other and diffuse together. The contact regions (or "necks") between individual grains and the density of the aggregate increases, leading the ice to become stronger and more cohesive over time. This process plays an important role in how ice surfaces evolve, which has implications for predicting their surface characteristics, interpreting spacecraft and telescopic observations, and developing technology to land on and sample these bodies. In this study, we use a numerical model to calculate the rate that sintering occurs in ice grains of varying size and temperature. We compare the predicted sintering rates to experimental observations and calculate estimates of sintering timescales on planetary surfaces. Our results suggest that ice on planetary surfaces can undergo significant modification over geologic timescales, even in the outer solar system where the cold temperatures result in slow sintering rates. We find that many bodies may develop a cohesive, but porous, surface crusts. Due to the temperature dependence of the process, the evolution of ices is likely to vary significantly throughout the solar system, as well as spatially across a given surface.

# 1. Introduction

Ice sintering is a form of frost metamorphism that drives the microstructural evolution of an aggregate of grains. Surface and volume diffusion driven by the surface curvature of the grains causes them to diffuse together, increasing the grain-to-grain contact area (or "neck") and aggregate density. Different diffusion mechanisms (Fig. 1) may dominate the process during different stages, under different temperature and atmospheric conditions, at depth or on the surface, or in different materials. The process is driven thermodynamically, as the rounding and diffusion of grains minimizes the surface area of the system and lowers its overall energy state. This results in the evolution of the strength, porosity, thermal conductivity, and other properties of the ice over time. Evidence suggests that sintering plays an important role in the evolution of icy surfaces throughout the solar system, though the rate and nature of the process on different worlds is not well constrained. In this study, we will use an existing model to estimate the sintering rate of water ice on planetary surfaces and discuss the implications for their microstructural evolution and near-surface properties. We will explore the model in detail to assess its limitations, identify how improvements can be made to its accuracy, and highlight areas where further research is needed.

While sintering of water ice has been studied extensively in terrestrial environments (e.g., Blackford, 2007; Colbeck, 1998; 1997; Cuffey and Paterson, 2010; Kingery, 1960), limited research has been done for planetary environments. Eluszkiewicz (1991) modeled nitrogen ice sintering on Triton and found that the formation of non-porous ice slabs over seasonal timescales (~100 yr) is consistent with observations of its absorption features. Similar features have also been explained by sintering of ices on Mars and Pluto (Eluszkiewicz, 1993; Eluszkiewicz and Moncet, 2003). Other studies have found that sintering can drive the development of hard, sub-surface layers on comets, influencing their strength and thermal conductivity (Kossacki, 2015; Kossacki et al., 2015). Schaible et al. (2016) find that neck growth from radiation-driven diffusion can explain the thermal anomalies observed on Mimas and Tethys. Overall, these studies suggest that sintering plays a key role in the microstructural evolution of planetary ices, and that understanding its nature is critical to understanding the landscape evolution of icy worlds. Sintering rates also vary with temperature, grain size, and composition (e.g., Blackford, 2007; and references therein), suggesting that the process varies widely throughout the Solar System. Characterizing this process also has critical implications for both the interpretation of remote sensing data and the successful *in-situ* exploration of these worlds, as it affects numerous properties of ice, such as its density, pore size and shape, cohesion, angle of repose, strength, roughness, thermal conductivity, effective dielectric constant, and coefficients of friction.

Water ice sintering on Earth (also called snow metamorphism) is typically studied using field samples of snow, firn (densified snow), and glacial ice (e.g., Keeler, 1969; Maeno and Ebinuma, 1983; Marshall and Johnson, 2009; Schneebeli et al., 1999). (Note that all references in this paper to ice will refer only to water ice unless explicitly stated.) Such studies characterize the structure of ice and snow *in-situ*, providing insight into landscape geomorphology and seasonal and climatic processes on Earth. Much may be learned from the terrestrial literature generally, but the nature of the sintering process at low (mechanical and atmospheric) pressures and temperatures is not well understood. Many such works are difficult to relate to other planetary bodies since only limited remote sensing data are able to probe their surfaces at microstructural scales. The review by

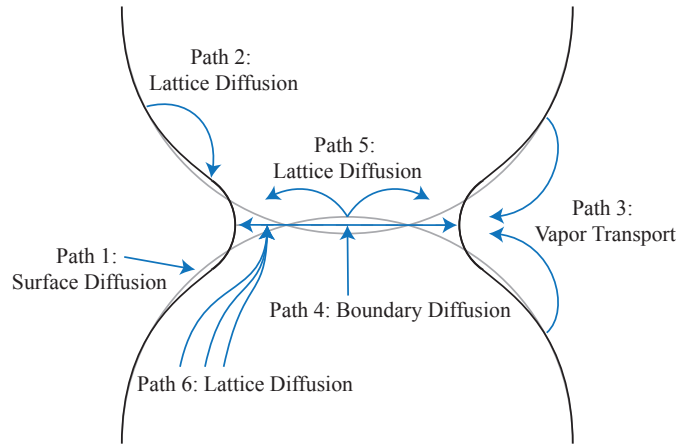

*Figure 1. A 2D cross section of two grains sintering and their saddle-shaped neck. The diffusion mechanisms described by the rate equations in Table 3 are labeled and arrows indicate the associated paths of mass flow.*

Blackford (2007) also notes that the majority of both field and laboratory studies on ice sintering focus on its end state, rather than on its evolution, exacerbating the challenge to understanding how the nature and rate of sintering may differ across solar system bodies. While models of pressure-sintering of subsurface ice have been adapted to study glaciology on other planets (e.g., Ebinuma and Maeno, 1985; Goldsby and Kohlstedt, 2001; Ligtenberg et al., 2011; Wilkinson, 2017; Schleef and Löwe, 2013), there are fewer options for modeling sintering of surface snow and frost (in the absence of overburden pressure) that can be applied to planetary surfaces.

Early models of ice sintering were power-law expressions describing neck growth as a function of empirically measured parameters (Hobbs and Mason, 1964; Hobbs and Radke, 1967; Kingery, 1960; Kuroiwa, 1961). There is a debate in the literature over which diffusion mechanisms dominate sintering in ice, and under what conditions. Most of these early studies focused on the effects of vapor transport and observed only the earliest stages of the process. Ultimately, the empirical nature of their models made it difficult to capture the fundamental physics of sintering, and the influence of different environmental factors and diffusion mechanisms. Later studies found both vapor transport and surface diffusion to be important, the latter being primarily dominant at temperatures below -15 C when neck sizes are small (e.g., Maeno and Ebinuma, 1983; Löwe et al., 2005; Vetter et al., 2010). Many modern models of snow metamorphism use x-ray tomography and other measurements of snow and firn to provide initial parameters and validate results (e.g., Legagneux, and Domine, 2005; Flin et al., 2004; Taillandier et al., 2007; Kaempfer and Plapp, 2009; Pinzer et al., 2012; Blackford, 2007; and references therein). These have similar issues to their empirical predecessors, as while they can be extremely sophisticated, their basis in field observations means they necessarily, inherently incorporate effects related to melting and pre-melting, local and macroscopic humidity, bulk vapor transport, wind, temperature gradients, and other factors from their terrestrial environment. Without deconvolving how each of these factors influences the sintering process, they are difficult to apply in a planetary context where the environment and materials may be very different. We instead take the opposite approach, beginning with a model based as closely on physical first principles as possible, with the intent to (in future work) incorporate more complexity as needed to suit different planetary environments.

One such sintering model came out of the field of metallurgy. Swinkels and Ashby (1981) developed a widely-accepted model to study the sintering of metal powders, which calculates the increase in neck size (Fig. 2, $x$), decrease in inter-particle distance or densification (Fig. 2, $a - y$), and change in grain and pore morphology (Fig. 3) driven by several diffusive mass fluxes (Fig. 1) in a 3D aggregate of spherical particles. This model was validated with laboratory measurements of eight different metal alloys, and since then has been used extensively in the metallurgy literature (e.g., Parhami et al., 1999; Svoboda and Riedel, 1995; Wadley et al., 1991). Many modern sintering models are based, in some part, on this work. In this study, we use the model of Swinkels and Ashby (1981), henceforth referred to as SA81, to quantify the microstructural evolution of surface and near-surface ice on planetary bodies.

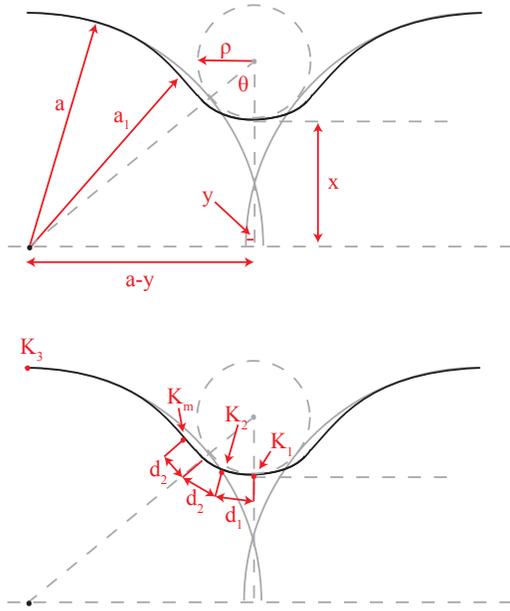

*Figure 2. Diagram of geometrical parameters in a 2D cross section of two sintering grains of radius a, with the neck at center of size x. Definitions and units for all parameters are given in Table 1.*

Swinkels and Ashby (1981) characterize the sintering process by four stages. In Stage 0, instantaneous adhesion of grains occurs via inter-atomic forces as the grains are brought into contact, which forms an initial neck between them. Stage 1 is characterized by growth of the necks via several surface and volume diffusion mechanisms (Table 1), until they reach ~0.5-0.8 times the grain radius. In this stage, the grains are still distinguishable from each other as the aggregate becomes more cohesive, and neck growth slows as mass is redistributed to form round pore spaces between the grains. Some densification may occur during this stage, but it is not the dominant effect. The geometry of the grains becomes difficult to describe at the end of Stage 1, and so it is assumed that a transitional Stage 2 occurs to transform the aggregate into the starting configuration of Stage 3 (densification). This final stage begins with a solid slab of ice containing isolated spherical pores, which undergoes complete densification and pore shrinkage over a long time period (relative to the neck growth timescale). The geometry and rate equations that define each stage are described in detail in the following section. As we will show in section 3.2, some problems arise in the way that Stage 2 is handled by the model, and so we will focus primarily on Stages 0 and 1 in this study.

Grain to grain sintering models based on local mass flux and grain geometry have also been developed specifically for ice. For example, Gubler (1985) presented a compelling model of grain growth due to vapor transport in snow subject to a temperature gradient, and Colbek (1998) put forward a model characterized by a grain-boundary groove rather than the saddle shape of SA81. However, we favor SA81 over other options for several reasons. First, it is thermodynamically derived and relies on only a few empirical parameters (the diffusion coefficients). It is well validated for other materials, suggesting that, while imperfect, its physical basis is robust. Second, given the geologic timescales of interest and the uniqueness of planetary environments and materials, a model that includes many diffusion mechanisms is most versatile while other ice sintering models consider only the subset of diffusion mechanisms that are most important in terrestrial environments. SA81 also offers the most flexibility in terms of developing future modifications and improvements. For example, new mechanisms such as radiation driven diffusion (Schaible et al., 2016) can be accounted for with the addition of a mass flux term to those in Table 1.

Finally, we are motivated to explore SA81 because several studies have already used it to predict the effects of sintering on Triton, Mars, Pluto, comets and other bodies (Eluszkiewicz, 1993; 1991; Eluszkiewicz and Moncet, 2003; Kossacki, 2015; Kossacki et al., 2015; 1994; Schaible et al., 2016). None of these studies implemented the full model, or performed experimental validations for ice of any composition. As a result, its ability to accurately predict the behavior of ice has not been explored in detail until now. As we will see in section 3.2, some aspects of the model do not behave well for ice and may result in an underestimation of sintering timescales. Nevertheless, it has great potential as a tool to investigate planetary surfaces. In this study, we explore how well the model reproduces the behavior of ice by analyzing its individual components and comparing predicted sintering rates to both new and historical experimental measurements. We discuss the implications that sintering has for the evolution of icy planetary surfaces, as well as the implications of our analysis for previous works.

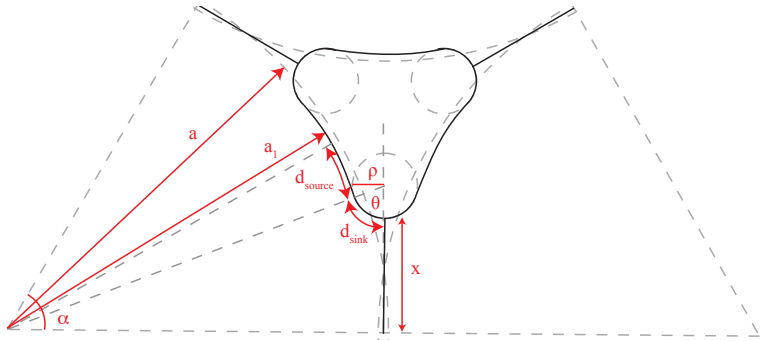

*Figure 3. Diagram of geometrical parameters in a 2D cross section of three grains sintering, and enclosed pore space. Definitions and units for parameters are given in Table 1 and discussed in section 2.3.2.*

## 2. Model

The following sections describe the geometry and components of SA81 during each stage of the sintering process. In their original study, Swinkels and Ashby (1981) included derivations of the model for multiple geometries and many sections of the publication are abstruse. For the benefit of the reader, we outline here in detail the model for an aggregate of uniform spheres and take the opportunity to clarify certain aspects of the original description. Thus, the following is a (paraphrased) repetition of what can be found in Swinkels and Ashby (1981), and interested readers can find expanded discussion on certain points in their study. No modifications have been made to SA81 in this paper. For clarity, a list of variable names, definitions, and values is provided in Tables 1 and 2, and the symbols used are consistent with the original model derivation. The rate equations for neck growth and densification in each stage are provided in Table 3.

2.1 Model Geometry

The model assumes an aggregate of uniform, spherical particles with some initial density. Mass flow between grains is calculated using the geometry shown in Figure 1. As the grains diffuse together, the amount and distribution of mass that is transported during the process is controlled by differences in the surface curvature at various points along the neck. The system moves towards a more energetically favorable state, and the difference in curvature along the neck goes to zero when the pore becomes spherical. The various surface curvatures ($K_1, K_2, K_3, K_m$) at play are shown in Figure 2.

The surface curvature is the inverse of the radius of curvature. It has a positive value if the radius of curvature is inside the object (e.g., a circle) or a negative value if the radius of curvature is outside the object (e.g., a convex arc). For a 3-dimensional object, the curvature at a given point is a sum of curvatures along the two relevant axes. Thus, the surface curvature of the original spherical grains ($K_3$) is positive and equal to:

$$K_3 = \frac{2}{a} \tag{1}$$

where a is the grain radius. The mean curvature of the neck ($K_m$) is the sum of the curvatures on both saddle axes:

$$K_m = -\frac{1}{\rho} + \frac{1}{x} \tag{2}$$

where $x$ is the neck radius and $\rho$ is radius of the imaginary circle between grains. The surface curvatures $K_1$ and $K_2$ have a negative value and must be calculated dynamically as material is redistributed along their diffusion distances $d_1$ and $d_2$ (see Section 2.3.1). The value of $\rho$ can be calculated from its geometrical relationship to $x$:

$$\rho = \frac{x^2 + y^2 - 2ay}{2(a-x)} \tag{3}$$

where $y$ is the distance of interpenetration between grains. The angle ($\theta$), made by rotating $\rho$ from the direction of $x$ towards the grain center, is given by:

$$\theta = \tan^{-1}\left[\frac{a-y}{x+\rho}\right] \tag{4}$$

2.2 Stage 0: Adhesion

When ice grains are brought into contact, inter-atomic forces act between their surfaces drawing them together. They deform elastically, causing a small, circular contact region (the "neck") to develop between the grains. This adhesion occurs instantaneously, and the neck radius ($x$) is approximately given by:

$$x \cong \left(\frac{\gamma_{eff} a^2}{\mu}\right)^{1/3} \tag{5}$$

where $a$ is the grain radius and $\mu$ is shear modulus. In an ideal case, the effective surface energy ($\gamma_{eff}$) is given by the clean surface approximation, $\gamma_{eff} \cong 2\gamma_s - \gamma_B$, which is the change in free energy when two surfaces (with free energy $\gamma_s$) are brought together to form a single grain boundary (with free energy $\gamma_B$). However, most particles are covered by an organic film or oxide that causes their surface to be, in some sense, dirty. In this case, the adhesive forces involved are weaker. A lower limit is obtained by assuming they are Van der Waals forces and using the dirty surface approximation the effective surface energy is assumed to be $\gamma_{eff} = \gamma_s/10$ (Ashby, 1974). The dirty surface approximation may be less accurate for non-terrestrial environments, but it eliminates a free energy variable and ultimately the change in value of $\gamma_{eff}$ will have a negligible effect on sintering timescales. The value of $y$ is assumed to be approximately zero at this stage.

*Table 1. Model Variables*

| | | |
|---|---|---|
| $a$ | m | initial grain radius |
| $a_1$ | m | modified grain radius |
| $x$ | m | radius of contact region |
| $y$ | m | one half of the interpenetration between particles |
| $\rho$ | m | radius of curvature of the neck |
| $K_m, K_1, K_2, K_3$ | m$^{-1}$ | surface curvatures (Fig. 2) |
| $d_1, d_2$ | m | diffusion distances for surface diffusion and redistribution |
| $T$ | K | absolute temperature |
| $\Delta_o$ | kg/m$^3$ | theoretical density of solid ice |
| $\Delta_i$ | kg/m$^3$ | initial density of aggregate |
| $\Delta/\Delta_o$ | | relative density |
| $\dot{V}$ | m$^3$/s | diffusive currents |
| $\dot{x}$ | m/s | rate of neck growth |
| $\dot{y}$ | m/s | rate of approach of particle centers |
| $\dot{\Delta}/\Delta_o$ | | densification rate |
| $\dot{n}$ | m/s | rate of normal displacement to the surface |
| S | | smoothing function for non-densifying mechanisms |
| $\alpha$ | | minimum angle between grains |
| $\alpha_i$ | | initial minimum angle of the starting packing density |

*Table 2. Material Parameters*

| | | | |
|---|---|---|---|
| $T_m$ | | K | melting temperature |
| $x_o$ | $(\gamma_s a^2/10\mu)^{1/3}$ | m | initial neck size |
| $y_o$ | 0 | m | initial interpenetration distance |
| $\delta_s$ | | m | effective surface thickness |
| $D_{os}$ | | m$^2$/s | surface diffusion coefficient prefactor |
| $Q_s$ | | J/mol | activation energy for surface diffusion |
| $\delta_s D_s$ | $\delta_s D_{os}\exp(-Q_s/RT)$ | m$^3$/s | surface diffusion coefficient: |
| $\delta_b$ | | m | effective grain boundary thickness |
| $D_{ob}$ | | m$^2$/s | grain boundary diffusion coefficient prefactor |
| $Q_b$ | | J/mol | activation energy for surface diffusion |
| $\delta_b D_b$ | $\delta_b D_{ob}\exp(-Q_b/RT)$ | m$^3$/s | grain boundary diffusion coefficient |
| $D_{ov}$ | | m$^2$/s | lattice (volume) diffusion coefficient prefactor |
| $Q_v$ | | J/mol | activation energy for lattice diffusion |
| $D_v$ | $D_{ov}\exp(-Q_v/RT)$ | m$^2$/s | lattice diffusion coefficient |
| $P_o$ | | Pa | vapor pressure prefactor |
| $Q_{vap}$ | | J/mol | activation energy for sublimation |
| $P_v$ | $P_o\exp(-Q_{vap}/RT)$ | Pa | vapor pressure |
| $\gamma_s$ | | J/m$^2$ | surface free energy |
| $\gamma_b$ | | J/m$^2$ | grain boundary free energy |
| $\Omega$ | | m$^3$ | molecular volume |
| k | 1.38x10$^{-23}$ | J/K | Boltzman's constant |
| R | 8.31 | J/mol | gas constant |
| μ | | N/m$^2$ | shear modulus |
| N | | m$^{-2}$ | dislocation density |

*Table 3. Rate Equations*

| Stage 1 Rate Equations | |
|---|---|
| ***Non-Densifying Mechanisms*** | |
| 1. Surface diffusion from a surface source | $\dot{V}_1 = S \dfrac{3\pi x \delta_s D_s \gamma_s \Omega (K_3 - K_2)}{d_2 kT}$ |
| 2. Lattice (or volume) diffusion from a surface source | $\dot{V}_2 = S \dfrac{3\pi x D_v \gamma_s \Omega (K_3 - K_m)}{kT}$ |
| 3. Vapor transport from a surface source | $\dot{V}_3 = S 2\pi x \rho \theta P_v \dfrac{\gamma_s \Omega}{kT} \left[ \dfrac{\Omega}{2\pi \Delta_o kT} \right]^{1/2} (K_3 - K_m)$ |
| ***Densifying Mechanisms*** | |
| 4. Grain boundary diffusion from a boundary source | $\dot{V}_4 = \dfrac{16\pi x \delta_b D_b \gamma_s \Omega}{xkT}\left(1 - \dfrac{K_1 x}{2}\right)$ |
| 5. Lattice diffusion from a boundary source | $\dot{V}_5 = \dfrac{32\pi \rho \theta D_v \gamma_s \Omega}{xkT}\left(1 - \dfrac{K_m x}{2}\right)$ |
| 6. Lattice diffusion from dislocation sources | $\dot{V}_6 = \dfrac{8\pi x^2 \rho \theta N D_v \gamma_s \Omega}{9kT}\left(-K_m - \dfrac{3\mu x}{2\gamma_s a}\right)$ |
| ***Redistribution Mechanisms*** | |
| Surface diffusion | $\dot{V}_{st} = \dfrac{12\pi x \delta_s D_s \gamma_s \Omega (K_1 - K_2)(d_1 + 2d_2)}{kT d_1 (d_1 + 3d_2)}$ |
| Neck growth rate | $\dot{x} = \dfrac{1}{2\pi x \theta \rho} \sum_{i=1}^{6} \dot{V}_i$ |
| Linear shrinkage rate | $\dot{y} = \dfrac{1}{\pi x^2} \sum_{i=4}^{6} \dot{V}_i$ |
| Densification rate | $\dfrac{\dot{\Delta}}{\Delta_o} = \dfrac{3\dot{y}/a}{(1 - \dot{y}/a)^4} \dfrac{\dot{\Delta}(\alpha)}{\Delta_o}$ |

2.3 Stage 1: Neck-growth dominated sintering

The rate equations for Stage 1 sintering are given in Table 3 along with the diffusion mechanisms ($\dot{V}_{1-6}$), which are listed as either densifying or non-densifying. All mechanisms cause mass to flow to the neck, but those that have a surface source do not cause the density to increase. Thus, all six mechanisms contribute to the neck growth rate ($\dot{x}$), but only mechanisms 4, 5, and 6 contribute to the densification rate ($\dot{y}$). Each term is characterized by a temperature dependent diffusion coefficient, some dependence on surface curvature, and geometrical variables to specify the change in volume of the grain and neck region. Even though Figure 2 shows only a 2D cross-section of grain diffusion, the terms are derived under the assumption that the grains are spherical in shape. Since neck growth is dominated by non-densifying mechanisms, the aggregate as a whole need not be considered when calculating the change in neck morphology between two grains. That is, the neck growth between the two grains in Figure 2 mirrors the neck growth between the grains and each of their other neighbors. However, the densification rate does depend on the packing density of the aggregate, and thus is not sufficiently described by Figure 2. This is discussed in Section 2.3.2.

For brevity, derivations for the terms and equations in Table 3 are not included in this paper, with the exception of the vapor transport term ($\dot{V}_3$) (Appendix A). This is one of the most important diffusion mechanism for ice during Stage 1, but is the term with the most uncertainty. We have included a description of the issues with the assumptions made in its derivation in Appendix A, with additional discussion of implications for our results in Section 3.2. The derivations of the remaining terms can be found in the original version of this model (Ashby, 1974), and references therein. Readers interested in this level of detail are encouraged to read both Ashby (1974) and Swinkels and Ashby (1981), as some changes are made to the model in the latter.

To calculate neck growth and densification, the rate equations (Table 3) must be solved iteratively over time using numerical methods. Each iteration, mass flux ($\sum \dot{V}$) into the neck is quantified based on the current surface curvatures ($K_m, K_1, K_2, K_3$). Then, new values for the geometrical parameters ($x, y, a_1, \rho, \theta$) are calculated based on the change in morphology of the grains, which then lead to new values of the surface curvatures and their diffusion distances ($K_1, K_2, d_1, d_2$). Finally, the densification rate is calculated and the new relative density determines the value of the smoothing function (S) that is applied to non-densifying mechanisms. Swinkels and Ashby (1980) chose to iterate over the grain radius rather than time, but a time-based calculation is generally more applicable in this context. Model results are discussed and presented in terms of the relative neck size (neck radius divided by grain radius), relative density (density of the aggregate divided by the density of the material), and homologous temperature (absolute temperature over the melting temperature). Sections 2.3.1–2 described these calculations in more detail.

2.3.1 *Rate equations for mechanisms 1 and 4*

Mechanisms 2, 3, 5, and 6 depend on the mean surface curvature of the neck (Eq. 1) and are straightforward to calculate using traditional methods. However, calculating mechanisms 1 and 4 is more complex. Boundary diffusion (mechanism 4) can only deliver matter into the pore space if surface diffusion (mechanism 1) can redistribute it over the pore surface. There must then exist a curvature difference ($K_2 - K_1$) to drive this surface redistribution, where the boundary diffusion depends on the lesser curvature ($K_1$). In turn, the boundary diffusion affects the rate of surface diffusion, which is now dependent on the curvature difference $K_3 - K_2$. These surface curvatures ($K_1$ and $K_2$) and the distances along which they occur ($d_1$ and $d_2$) must be calculated together and are related by four equations.

The first equation requires that the (geometric) mean curvature be equal to $-\frac{1}{\rho} + \frac{1}{x}$ (Eq. 2):

$$K_1 K_2 = K_m^2 \qquad (6)$$

The second equation defines the sum of the diffusion distances ($d_1$ and $d_2$) as the arc length given by:

$$d_1 + d_2 = \rho\theta \tag{7}$$

In this case, $\rho$ takes the more complex form:

$$\rho = \frac{x^2 + y^2 + a^2 - a_1^2 - 2ay}{2(a_1 - x)} \tag{8}$$

where $a_1$ is the modified grain radius. Eq. (8) collapses to (3) in the case where the curvature depends only on the unmodified radius, or when no modification has occurred ($a_1 = a$). The modified radius $a_1$ is calculated by applying conservation of volume to the non-densifying mechanisms:

$$\frac{4}{3}\pi(\Delta a)^3 = \sum_{i=1}^{3} \dot{V}_i \tag{9}$$

The remaining two relations that determine $K_1$, $K_2$, $d_1$, and $d_2$ are continuity equations. All of the material driven by boundary diffusion ($\dot{V}_4$) must be redistributed over the neck surface area out to $d_1$ on either side of the boundary:

$$\dot{V}_4 = \dot{V}_{st} \tag{10}$$

where $\dot{V}_{st}$ is the flow of mass away from the center of the neck (at $K_1$) by surface diffusion (see Table 3). Finally, the growth rate of the surface at the location of $K_2$ caused by the redistribution of one half of $\dot{V}_4$ along $d_1$ must be equal to the growth rate caused by surface diffusion from $\dot{V}_1$ along $d_2$:

$$\frac{\dot{V}_{st}}{2d_1} = \frac{\dot{V}_1}{d_2} \tag{11}$$

Thus, the four equations (6, 7, 10, 11) may be solved to obtain the precise values of the four unknowns ($K_1, K_2, d_1, d_2$).

Since the value of these variables changes throughout the sintering process, their calculation can be computationally burdensome (see Appendix B). Thus, it is helpful to derive approximate analytical solutions as well. Swinkels and Ashby (1981) simplify the expressions for $V_1$ and $\dot{V}_4$ by neglecting terms on the order of $1/a$ and $1/x$, and for $\dot{V}_{st}$ by assuming $d_1 + d_2 \approx d_2$, to obtain:

$$\dot{V}_1 = \frac{-3\pi x \delta_s D_s \gamma_s \Omega}{d_2 kT} K_2 \tag{12}$$

$$\dot{V}_4 = \frac{-8\pi \delta_b D_b \gamma_s \Omega}{kT} K_1 \tag{13}$$

$$\dot{V}_{st} = \frac{-8\pi x \delta_s D_s \gamma_s \Omega}{kT d_1}(K_2 - K_1) \tag{14}$$

Substituting Eqs. 11–13 into Eqs. 10 and 11 yields:

$$\frac{K_2-K_1}{K_1} = \frac{Ad_1}{x} \tag{15}$$

$$\frac{K_2-K_1}{K_2} = \frac{3d_1^2}{4d_2^2} \tag{16}$$

where

$$A = \frac{\delta_b D_b}{\delta_s D_s} \tag{17}$$

A further approximation can be made that $d_1 + d_2 \cong \rho$, and by writing $\Delta K = K_2 - K_1$ Swinkels and Ashby (1981) obtain:

$$d_1 = \rho \left[1 - \frac{1}{1+\left(\frac{4\Delta K}{3K_2}\right)^{1/2}}\right] \tag{18}$$

$$d_2 = \frac{\rho}{1+\left(\frac{4\Delta K}{3K_2}\right)^{1/2}} \tag{19}$$

$$\frac{\Delta K}{K_1} = \frac{3A\rho}{x} + \frac{3}{2} - \frac{3}{2}\left[3\left(\frac{A\rho}{x}\right)^2 + 4\left(\frac{A\rho}{x}\right) + 1\right]^{1/2} \tag{20}$$

which, together with Eq. 6, completely determine $K_1$, $K_2$, $d_1$, and $d_2$. We note that Eq. 15 differs from SA81 by a numerical factor (on the right-hand side) of 3/4, as the original derivation was performed for a geometry of wires rather than spheres. However, we were unable to reproduce the algebra leading to Eq. 20 in order to carry forward this correction, and thus as a result (20) is the approximation for a geometry of wires, rather than spheres. Swinkels and Ashby (1981) state that the equivalent relation for a geometry of spheres only differs by a numerical factor of order 1, and thus Eq. 20 may be used as an approximation for either geometry. We have followed this recommendation and left Eq. 20 unchanged in our implementation.

2.3.2 *Calculation of the densification rate and smoothing function*

The non-densifying diffusion during Stage 1 sintering is driven by the differences in surface curvature along the neck between ice grains. In a 2D view of this process (Fig. 3), this diffusion would slow as the pores become more rounded, and it must go to zero once they become circular. The geometry and size of the pore can be described using the relations above, as presented by Swinkels and Ashby (1981) for the sintering of closely packed wires. They apply a smoothing function $S$ to the non-densifying mechanisms ($\dot{V}_{1-3}$) to force them to reach zero when the appropriate geometrical state has been reached:

$$S = \frac{d_{source}}{d_{source}+d_{sink}} \tag{21}$$

*Figure 4. Diagram of (left to right) simple cubic, body-centered cubic, and face-centered cubic sphere packing structures.*

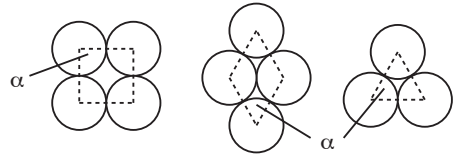

The source and sink regions make up the entire length of the neck. When sintering has just begun and the neck radius is small, $d_{source}$ is large and the value of S is close to 1. As the neck grows, $d_{source}$ (and thus S) diminishes until it becomes zero when the pore becomes circular. In the case of the sintering of wires, the 3D aggregate reflects this 2D geometry. Thus, the density of the aggregate can be determined geometrically throughout the process, and Stage 1 ends when the pores between wires become cylinders.

However, the density of an aggregate of spheres is more difficult to describe geometrically because the packing density varies continuously throughout the sintering process. To account for this, Swinkels and Ashby (1981) formulate a smoothing function based on the relative density ($\Delta/\Delta_o$) of possible packing arrangements of spheres (Fig. 4). The least dense packing arrangement is simple cubic ($\Delta/\Delta_o = 0.52$), the second is body-centered cubic ($\Delta/\Delta_o = 0.68$), and the densest is face-centered cubic ($\Delta/\Delta_o = 0.74$). The corresponding inter-particle angle (α) varies roughly linearly with density, decreasing from 90° (simple cubic) to 60° (face-centered cubic). The equation of a line between these two states is used to track the increase in $\Delta/\Delta_o$ as a function of the α:

$$\frac{\Delta(\alpha)}{\Delta_o} = -2.38\alpha + 2.808 \tag{22}$$

The initial value of $\Delta(\alpha_i)/\Delta_o$ is the initial relative density of the aggregate ($\Delta_i/\Delta_o$), and the initial angle $\alpha_i$ is calculated by rearranging Eq. 22. In order to incorporate the increase in density both from particle rearrangement (characterized by Eq. 22) and from neck growth during each time step, the densification rate must also be modified by the change in interparticle distance (y). First, the value of $\Delta(\alpha)/\Delta_o$ is determined from Eq. 22. Then, the current value of the density ($\Delta/\Delta_o$) as also modified by neck growth is calculated from:

$$\frac{\Delta}{\Delta_o} = \frac{\Delta(\alpha)/\Delta_o}{(1-y/a)^3} \tag{23}$$

Swinkels and Ashby (1981) explain that they have then "*kept track of the density throughout the computation…and caused α to decrease from this initial value towards 60° in a way that varies linearly with density, and would cause α to become 60° when $\Delta/\Delta_o$ reaches 1.*" Our interpretation of this is that at the beginning of the calculation, the equation of a straight line is calculated between the initial ($\alpha = \alpha_i$, $\Delta/\Delta_o = \Delta_i/\Delta_o$) and ending states ($\alpha = \pi/3$, $\Delta/\Delta_o = 1$). The initial point of this line is located on that described by Eq. 22, but departs from it for an ending state at a relative density of 1. During a given timestep, this is used with the new value of $\Delta/\Delta_o$ from Eq. 23 to determine the new value of α, which is then fed back into Eq. 22 during the subsequent timestep. In this way, the increase in density of the aggregate is tracked along the line between its initial value and an ending state of 1, and along the way it is modified by both particle rearrangement (Eq. 22) and the contribution by neck growth (Eq. 23). This methodology was not clearly described by Swinkels and Ashby (1981), however our implementation of the model produces results consistent with the examples in their study (see Section 3.1).

The values for $d_{source}$ and $d_{sink}$ can then be approximated using the 2D geometry used for wires (Fig. 3) and the angle α, yielding:

$$d_{sink} = \rho\theta \tag{24}$$

$$d_{source} = a_1 \left(\frac{\alpha}{2} + \theta - \frac{\pi}{2}\right) \tag{25}$$

If we assume that the grains are at maximum packing density ($\alpha = \pi/3$) when the pore becomes circular, then $\theta$ becomes half of $\alpha$ when this occurs. Thus, the criterion for the end of Stage 1 is:

$$\theta + \frac{\alpha}{2} = \frac{\pi}{2} \tag{26}$$

Using Eqs. 23–25, the model uses geometrical constraints to force the non-densifying mechanisms to ramp down in a way that is physically meaningful. As we will see in Section 3.2, however, problems arise with this approach in our application.

2.4 *Stages 2 & 3: Transitional and late stage sintering*

Since the packing density of the aggregate is assumed to be changing, the 2D geometry (Fig. 3) used to derive the smoothing function can only approximately describe what is really happening. For this reason, Swinkels and Ashby (1981) name Stage 2 as a transitional stage following Stage 1, during which the aggregate goes through a transformation from being a lattice of heavily sintered grains with interconnecting pore spaces to a densely packed aggregate of indistinguishable grains with isolated, spherical pores. Even though the criterion for the end of Stage 1 assumes that the pore is approximately spherical, in reality the final "closing off" of these pores occurs during Stage 2. Stage 2 is assumed to take a negligible amount of time, and the change in density that occurs is linearly interpolated between the value at the end of Stage 1 and the beginning of Stage 3.

At the beginning of Stage 3, SA81 assumes that all grains in the aggregate have 14 contact neighbors and have the shape of a tetradecahedron (14-faced solid). There is a spherical pore at each of the 24 corners (Fig. 5) of the tetradecahedron, which is shared with three other neighboring particles. No further changes in the packing density occur. The new pore radius ($\rho$) at the beginning of Stage 3 is calculated by equating the volume ($V_g$) of a grain with an edge of length ($l$):

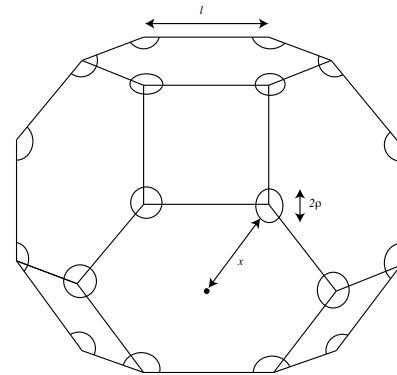

*Figure 5. A tetradocedahedral (14-sided) particle representing the state of a particle at the beginning of Stage 3, with pore spaces shown at each vertex. Each particle contains 24 pores that are each shared with 3 neighboring particles.*

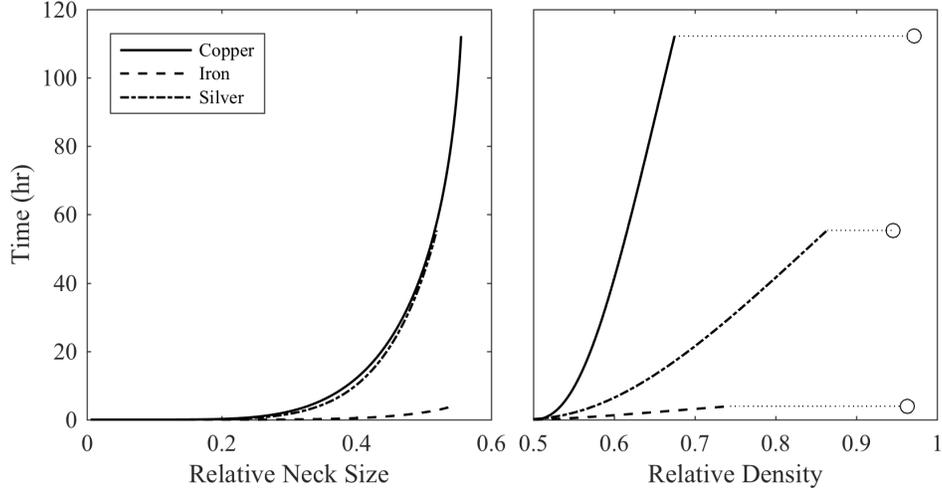

*Figure 6. Relative neck size (left) and density (right) over time for an aggregate of 25 μm radius grains composed of copper (solid), iron (dashed), and silver (dot-dashed) with an initial relative density of 0.5 at a homologous temperature of 0.8.*

$$V_g = 8\sqrt{2}l^3 \tag{27}$$

to the volume of the matter within each particle:

$$V_m = \frac{4}{3}\pi a^3 \tag{28}$$

plus the volume of 6 pores:

$$V_p = 8\pi\rho^3 \tag{29}$$

and noting that $l = x + \rho$. The relative density at the beginning of Stage 3 is then:

$$\frac{\Delta}{\Delta_o} = \frac{V_g - V_p}{V_g} \tag{30}$$

Further progress of Stage 3 sintering is calculated using new rate equations, however these are not included in Table 3. As we will see in Section 3.2, problems arise in quantifying Stage 2 for ice, requiring that modifications to be made to the model in order to quantify late stage sintering on planetary surfaces. As a consequence, exploration of Stage 3 will be left to future work.

## 3. Model validation

3.1 *Behavior of metals*

Swinkels and Ashby (1981) constructed sintering diagrams (their Fig. 16-34) of neck size and density as a function of temperature and time for a number of metals in order to validate the model's behavior against experimentally measured sintering rates in the literature. In turn, we used the material properties of these metals included in their study in order to validate our own implementation of the model against their sintering diagrams. Here, we use these to explore the variation in sintering rate and behavior between different materials. For example, we calculated the neck growth and densification rates for aggregates of copper, iron, and silver grains (Table 4) with a radius of 25 $\mu$m, at a homologous temperature of 0.8 (Fig. 6). Each material exhibits similar overall behavior, though their sintering rates vary. Iron, for example, reaches the end of Stage 1 after only a few hours, whereas copper and silver take several days to reach the same state. Additionally, while their neck sizes at the end of Stage 1 are similar, their final relative densities vary more significantly. This occurs because the dominant diffusion mechanisms and diffusion coefficients vary in each material. Figure 7 shows the strength of each diffusion mechanism in copper and iron over time. While surface ($\dot{V}_1$) and grain-boundary ($\dot{V}_4$) diffusion dominate throughout much of Stage 1 for both materials, the efficacy of the other mechanisms varies. Additionally, Figure 7 shows copper at two homologous temperatures, demonstrating that the relative efficacy of diffusion mechanisms also varies with temperature regime.

The relative density of each aggregate at the beginning of Stage 3 is shown in the right panel of Figure 6. The approximation made by the model to skip Stage 2 in the calculation (Section 2.4) predicts overall sintering timescales to varying levels of accuracy for each material. Iron sinters very quickly and the amount of time it would take for the aggregate to change from the ending density of Stage 1 to the beginning density of Stage 3 appears negligible. Silver sinters more slowly, but the two densities are closer in value and thus the approximation still works at an order of magnitude level. Copper, on the other hand, both sinters slowly and ends Stage 1 at a relative density substantially lower than it is assumed to start with at Stage 3. In this case, it is clear that skipping Stage 2 of the calculation would provide a much less accurate estimate of total sintering time. This being said, sintering rates vary exponentially with grain size and temperature (see Section 4.1). Thus, for manufacturing processes and laboratory studies where these parameters can be chosen to maximize efficiency, this model can work well to estimate sintering times and study diffusion regimes.

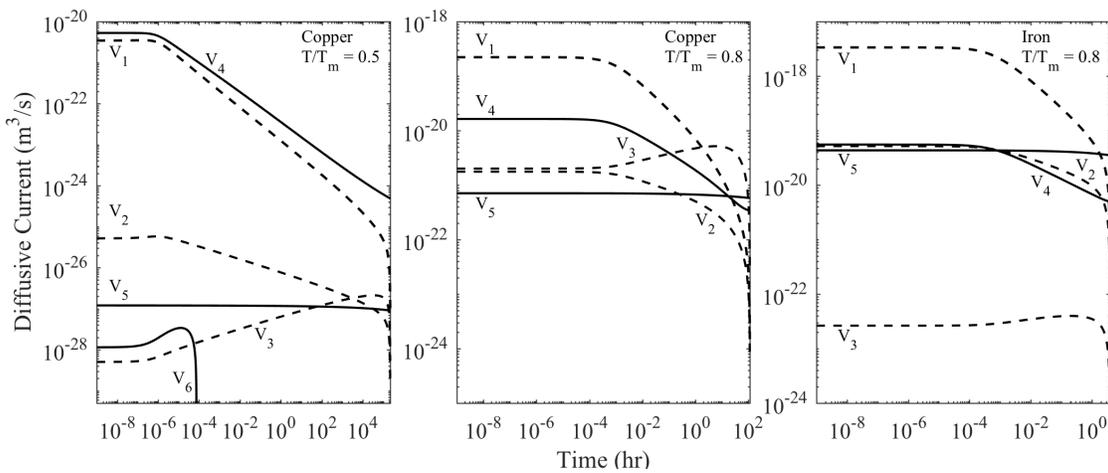

Figure 7. Diffusive currents over time for an aggregate of 25 μm copper grains at a homologous temperature of 0.5 (left) and 0.8 (center), and iron grains at homologous temperature of 0.8 (right). Aggregates have an initial relative density of 0.5. Non-densifying mechanisms are denoted by dashed lines, and densifying mechanisms by solid.

## 3.2 *Behavior of water ice*

While the validation performed above provides some confidence in the model's ability to quantify sintering behavior, it has not been validated for water ice to date. Laboratory studies measuring ice sintering rates in the literature are limited, which presents a challenge to determining how accurately the model may reflect its behavior. Moreover, the sintering rates reported by existing studies seem to be inconsistent with each other. Figure 8 shows the neck growth rate of ice from three separate studies (Hobbs and Mason, 1964; Kingery and Berg, 1955; H. Thomas et al., 1994) compared to the model's prediction using the material properties given in Table 5. Sensitivity to material properties is discussed in Appendix B. In each panel, the grain radius and temperature of the simulation matches that of the corresponding study. Neck growth measurements from these studies were not provided in tables and had to be manually extracted from images, introducing some uncertainty due to human error in the data. Nonetheless, these estimates allow us to demonstrate how the model compares to historical measurements.

Our model most closely matches the data from Hobbs and Mason (1964) (Fig. 8, a), though sintering rates are still somewhat faster than the authors measured. They performed their study in a small chamber that was partially evacuated and filled with atmosphere (no pressure is stated), whose walls were coated with ice to ensure an ice-saturated environment. Swinkels and Ashby (1981) note that theory underestimates neck growth by surface diffusion at temperatures very close to the melting temperature because, unlike other mechanisms, surface diffusion is often characterized by more than one activation energy. If corrected, this may decrease the discrepancy between the model prediction and data in (a). However, the model differs dramatically from the other two studies (Fig. 8, b, c). It predicts neck growth to be significantly slower than (b) Kingery (1960) and significantly faster than (c) Thomas et al. (1994), in spite of the fact that both studies used comparable grain sizes and temperatures. Both studies state that the experiments were performed in a cold room at normal atmospheric pressure, offering no

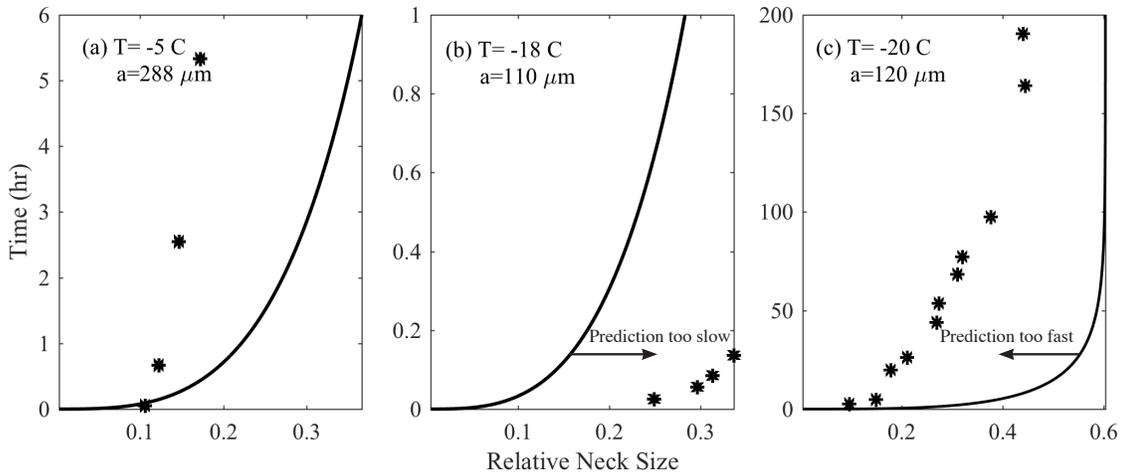

*Figure 8. Estimated neck growth (lines) over time compared to measurements (*) from (a) Hobbs and Mason (1963), (b) Kingery (1960), and (c) Thomas et al. (1994). The temperature and grain radius for each experiment is indicated in each panel. Panel a represents the best match to the model, while panels b and c show predicted growth to be slower and faster, respectively, than measured values.*

*Table 4. Material properties for metals (Swinkels and Ashby, 1981; and references therein), as well as nitrogen and methane ices (Eluszkiewicz, 1991; and references therein).*

|  |  | Silver | Copper | Iron | Nitrogen | Methane |
|---|---|---|---|---|---|---|
| $\Omega$ | m$^3$ | 1.71 x 10$^{-29}$ | 1.18 x 10$^{-29}$ | 1.18 x 10$^{-29}$ | 4.7 x 10$^{-29}$ | 5.1 x 10$^{-29}$ |
| $T_m$ | K | 1234 | 1356 | 1810 | 63.148 | 90.67 |
| $\Delta_o$ | kg/m$^3$ | 1.05 x 10$^4$ | 8.96 x 10$^3$ | 7.65 x 10$^3$ | 995 | 515 |
| $\mu$ | N/m$^2$ | 2.64 x 10$^4$ | 4.21 x 10$^4$ | 8.1 x 10$^4$ | 10$^9$ | 10$^9$ |
| N | m$^{-2}$ | 10$^{14}$ | 10$^{14}$ | 10$^{14}$ | 10$^{16}$ | 10$^{16}$ |
| $\gamma_s$ | J/m$^2$ | 1.12 | 1.72 | 2 | 0.02 | 0.02 |
| $D_{ov}$ | m$^2$/s | 4.4 x 10$^{-5}$ | 6.2 x 10$^{-5}$ | 1.80 x 10$^{-4}$ | *1.6 x 10$^{-7}$ | *10$^{-3}$ |
| $Q_v$ | kJ/mol | 185 | 207 | 270 | 8.6 | 15.9 |
| $\delta_b D_{ob}$ | m$^3$/s | 6.94 x 10$^{-15}$ | 5.12 x 10$^{-15}$ | 7.50 x 10$^{-14}$ | ** | ** |
| $Q_b$ | kJ/mol | 89.8 | 105 | 159 | 5.7 | 10.6 |
| $\delta_s D_{os}$ | m$^3$/s | 1.05 x 10$^{-6}$ | 6.0 x 10$^{-10}$ | 1.10 x 10$^{-10}$ | ** | ** |
| $Q_s$ | kJ/mol | 266 | 205 | 220 | 8.6 | 15.9 |
| $P_o$ | Pa | 9.53 x 10$^4$ | 1.23 x 10$^5$ | 3.67 x 10$^5$ | 5.5 x 10$^9$ | *** |
| $Q_{vap}$ | kJ/mol | 272 | 324 | 382 | 324 | *** |

*Esteve and Sullivan (1981), Chezeau and Strange (1979)

**Assumed by Eluszkiewicz (1993; 1991) and Eluszkiewicz et al. (2007) to be $2\Omega^{1/2}D_{ov}$.

***Vapor pressure from Ziegler (Ziegler, 1959): $log_{10}P_v[mm\,Hg] = 7.69540 - 532.20/(T[K] + 1.842)$

*Table 5. Material properties for water ice.*

|  |  | Water Ice | Reference |
|---|---|---|---|
| $\Omega$ | m$^3$ | 1.181 x 10$^{-29}$ | 4/3 * π * (van der Waals radius)$^3$ |
| $T_m$ | K | 273.15 |  |
| $\Delta_o$ | kg/m$^3$ | 917 | e.g., Mellor (1974) |
| $\mu$ | N/m$^2$ | 10$^9$ | e.g., Mellor (1974) |
| N | m$^{-2}$ | 10$^{14}$ | Swinkels and Ashby (1981) |
| $\gamma_s$ | J/m$^2$ | 0.06 | Kossacki et al. (1994), Schaible et al., (2016) |
| $D_{ov}$ | m$^2$/s | 9.1 x 10$^{-4}$ (1.5 x 10$^{-2}$ – 7.5 x 10$^{-23}$) | Goldsby and Kohlstedt (2001), Livingston et al. (1997), *Ramseier (1967), Mizuno and Hanafusa (1987), Nasello et al. (2007), Nie et al. (2009) |
| $Q_v$ | kJ/mol | 5.94 | *Ramseier (1967) |
| $\delta_b, \delta_s$ | m | 9.04*10$^{-10}$ | Frost and Ashby (1982), Goldsby and Kohlstedt (2001) |
| $D_{ob}$ | m$^2$/s | 8.4 x 10$^{-4}$ (0.03 – 10$^{-12}$) | Goldsby and Kohlstedt (2001), Maeno and Ebinuma (1983), Nasello et al. (2005) |
| $Q_b$ | kJ/mol | 4.9 | *Goldsby and Kohlstedt (2001), Nasello et al. (2005) |
| $D_{os}$ | m$^2$/s | 1.4 x 10$^{-8}$ (1.4 x 10$^{-8}$ – 1.35 x 10$^{-13}$) | Livingston et al. (1997), Mizuno and Hanafusa (1987), *Nasello et al. (2007) |
| $Q_s$ | kJ/mol | 23.156 | Nasello et al. (2007) |
| $P_o$ | Pa | 5.5 x 10$^9$ | Murphy and Koop (2005) |
| $Q_{vap}$ | J/mol | 324 | **Murphy and Koop (2005) |

*Primary reference for material property used in this calculation.

**Heat of sublimation from Murphy and Koop (2005): $Q_{vap}[J/mol] = 46782.5 + 35.8925 * T - 0.07414 * T^2 + 541.5 * \exp(-(T/123.75)^2)$

other details that might indicate why their measurements of neck growth were so vastly different. This demonstration highlights our lack of understanding of ice sintering at the individual grain scale and emphasizes the need for more laboratory studies on this topic. With modern techniques and equipment, significant progress may be made in measuring sintering rates in water ice and other materials relevant in planetary science.

A reasonable hypothesis for the discrepancy in measurements shown in Figure 8 (b) and (c) is differences in humidity or local vapor pressure conditions during each experiment. The lack of a match to model predictions, then, would also be consistent with inaccuracy or oversimplification of the vapor transport diffusion mechanism ($\dot{V}_3$) in the calculation. There are a number of issues with the way this term is derived, not the least of which is that it assumes that sintering occurs in a vacuum. Hobbs and Mason (1964) found that a decrease in sintering rate occurs for ice grains submersed in kerosene or silicone oil, which they attribute to suppressed vapor diffusion. It is reasonable to assume that vapor transport in an inert gas atmosphere may also be diffusion limited. This would be reflected in an overestimation of sintering rates predicted by the model, consistent with panel (c) in Figure 8. The model also assumes that mass is conserved, neglecting both ambient sources of water molecules from surrounding atmosphere and potential mass loss from the system. These effects are likely to influence calculations differently in ambient, low pressure, and vacuum environments, and may present challenges in fully validating the model. A full derivation of the term for $\dot{V}_3$ is included in Appendix A, along with additional discussion of its limitations.

We can further explore the sintering of ice by comparing its behavior to that of the metals in Section 3.1. Figure 9 (left) shows the entire neck growth stage of water ice grains 25 $\mu$m in radius at a homologous temperature of 0.8 (-55 C), matching the metal particles (Fig. 6). The neck growth in the ice occurs over a period of several days, which is similar to copper and silver except that it ramps down to the end of Stage 1 over a much longer period. However, Figure 9 (right) shows that almost no densification occurs in the ice over this period, which is in stark contrast to the behavior of the metals. The reason for this difference lays in the dominant diffusion mechanisms at play. In metals (Fig. 7), Stage 1 is largely dominated by surface diffusion ($\dot{V}_1$), a non-densifying mechanism, and grain-boundary diffusion ($\dot{V}_4$), a densifying mechanism. Thus neck growth and densification occur over similar timescales. However, Stage 1 in ice is strongly dominated by surface ($\dot{V}_1$) and vapor diffusion ($\dot{V}_3$), both non-densifying mechanisms (Fig. 10). This drives neck growth to occur over a much shorter period than densification. So while the neck growth timescale predicted by the model for ice appears accurate within an order of magnitude, clearly the approximation made by eliminating Stage 2 in the densification calculation would not produce a meaningful estimate of the total sintering timescale. In addition, since the smoothing function (section 2.3.2) that slows neck growth at the end of Stage 1 depends on the density, its value will influence the maximum neck size that grains can reach, with lower density aggregates reaching slightly larger necks. It is unclear whether this trend is physically realistic, or an artifact of how the model handles packing arrangement.

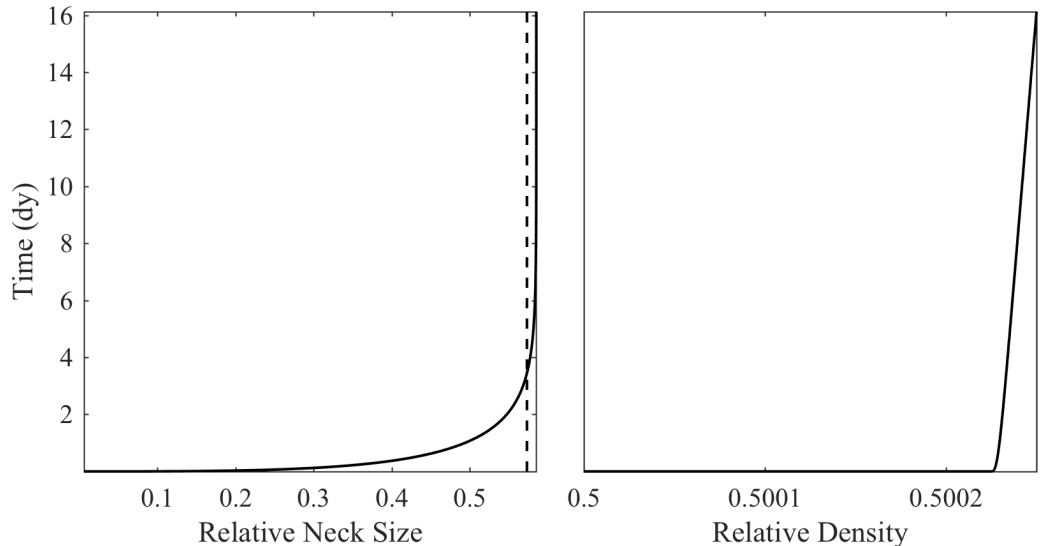

*Figure 9. Relative neck size (left) and density (right) over time for an aggregate of ice grains 25 μm in radius with an initial relative density of 0.5 at a homologous temperature of 0.8. The dotted line shows when 98% completion of Stage 1 has been achieved.*

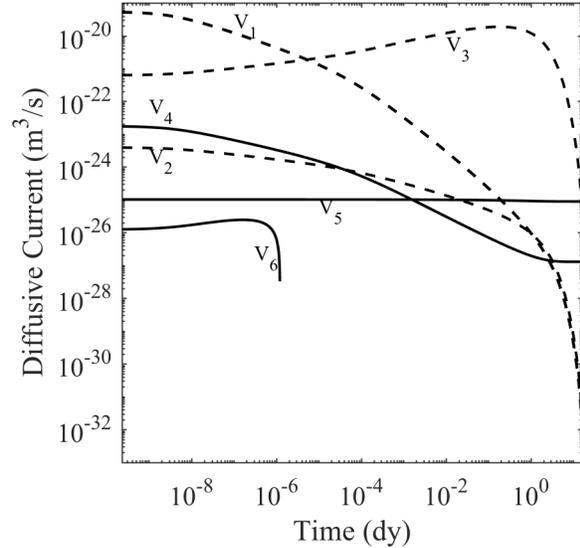

*Figure 10. Diffusive currents over time for an aggregate of ice grains 25 μm in radius with an initial relative density of 0.5 at a homologous temperature of 0.8. Non-densifying mechanisms are denoted by dashed lines, and densifying mechanisms by solid.*

To ignore the issues described above related to how SA81 handles the transitional Stage 2 of the process would result in an underestimation of total sintering timescales for ice by an unknown (but likely large) amount, and thus any quantification of the late stage densification (Stage 3) timescales would have an extremely large uncertainty. Further, since the geometrical state of the ice at the end of Stage 1 is incompatible with the assumptions made by Eq.'s 27-30, calculation of Stage 3 timescales would have to be performed using arbitrary starting conditions. For this reason, we have elected not to quantify total sintering timescales or explore Stage 3 of the sintering process in this study. A better method will be needed to quantify the densification that occurs during Stage 2 and the starting conditions of Stage 3, capturing more realistic changes in morphology and packing arrangement between the early period of neck growth and late-stage densification in ice aggregates.

## 4. New Experimental Observations

Two different sets of experiments were performed to compare to model predictions. One set monitored neck growth between isolated grain pairs, and the other changes in the structure and grain size distribution in bulk ice aggregates. These experiments used different ice synthesis methods and cryostage systems, described in each subsection below. For both experiments, the cryostages were placed under the objective turret of an Olympus BX51 microscope, and sample images using reflected illumination and an Olympus DS-30 camera. The Olympus Stream software was used to acquire images in Extended Focal Imaging mode (each image is a mosaic of images built and stitched during acquisition while varying focus, to compensate for the narrow depth-of-field of the microscope), as well as to process images and measure particle and neck sizes after scale calibration using a micrometer-graduated slide.

4.1 Grain pair experiments

In the first set of experiments, we monitored the rate of neck growth between two individual grains of ice, similar to the historical experiments discussed in the section 3.2. We observed grain pairs in a Peltier-cooled Instec TS102V/G cryostage at -5 and -20 C. The vacuum ports of the stage were capped in order to seal the chamber during the experiment. Within the cryostage, a glass slide is secured to a cooling plate by a cover equipped with a glass window. The clearance between the slide and cover is ~1 mm. Though this area does not seal off from the rest of the chamber, both the cap and the plate are thermoelectrically cooled, allowing the air temperature and glass slide to be brought into thermal equilibrium. The surface of the slide was coated with ice to help saturate the atmosphere above it, except for its center, which was left empty. Spherical ice grains were created by spraying water directly into a bath of liquid nitrogen (LN2) using a conventional spray bottle. The LN2 bath contained a sieve with a 212 $\mu$m mesh, on which the ice grains collected. The sieve was then tapped to deposit ice grains onto the center of the slide, after which the chamber was sealed and placed under the microscope. At each temperature, we located an isolated pair or cluster of grains (Fig. 11) that had come into contact when transferred to the slide, and monitored the growth of their necks over time.

Figure 11 shows the grain pair at -20 C at the time of our first observation (left), and again after 57 minutes (right). The larger and smaller grains began with radii of 101 and 73 $\mu$m, shrinking by 3% and 4%, respectively, due to sublimation during the experiment. During this time, the neck size grew from 33 to 65 $\mu$m, or a relative neck size of 0.17–0.33 and 0.24–0.45 for the large and small grain, respectively. The grain pair monitored at -5 C began with radii of 99 and 78 μm, respectively, shrinking by 9% due to sublimation. Their neck size grew from 32 to 44 μm, or a relative neck size of 0.16–0.22 and 0.2–0.27 for the large and small grain, respectively. Fewer measurements were obtained of the latter experiment, as the warmer temperature caused the grains to sublimate faster. We ceased taking measurements when enough sublimation occurred to cause the neck to decrease in size. Figure 12 compares these measurements to predicted values from the model. Measurement uncertainties are not shown in Figure 12 for image clarity, but are described in Appendix B. It is difficult to know exactly when the grains we observed came into contact during the few minutes it took to create and transfer them to the cryostage, so we anchored the first measurement to the time at which the model shows the same neck size. In this way, we can assess how subsequent neck growth compares to the prediction. The grains at -20 C show a good match to the model, particularly when the neck size is calculated relative to the larger grain (Fig. 12, dashed, o). The grains at -5 C show a slower sintering rate than predicted by the model, though these measurements also have higher uncertainty. The observed sintering rates appear to be the same order of magnitude as that in Figure 8 (a, b), though it is difficult to compare directly given the difference in grain size in each experiment.

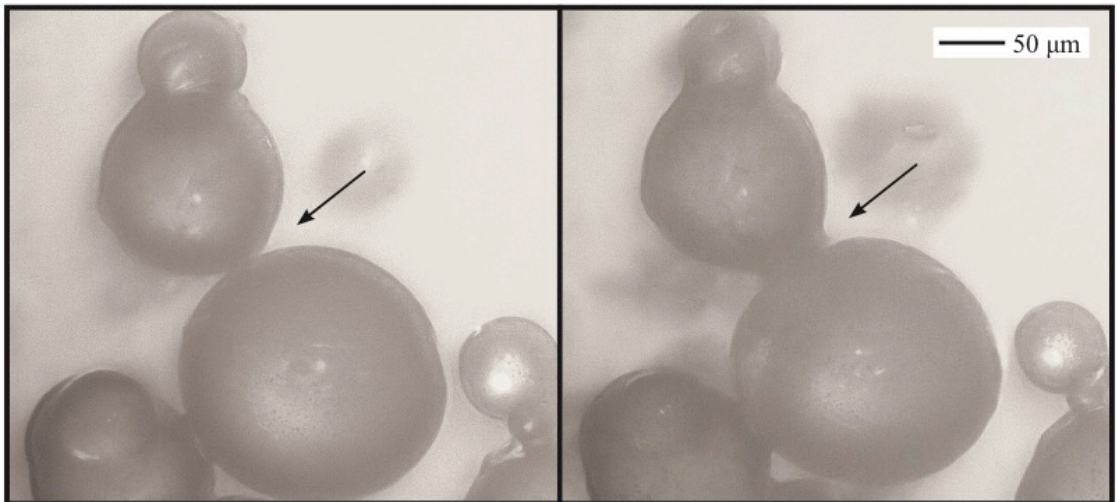

*Figure 11. Water ice grains sintering at -20 C at the first observation (left) and after 57 minutes (right), showing growth of the neck (arrow) from 33 to 65 µm. The grains have starting radii of 101 and 73 µm, shrinking by 3% and 4%, respectively, after 57 minutes.*

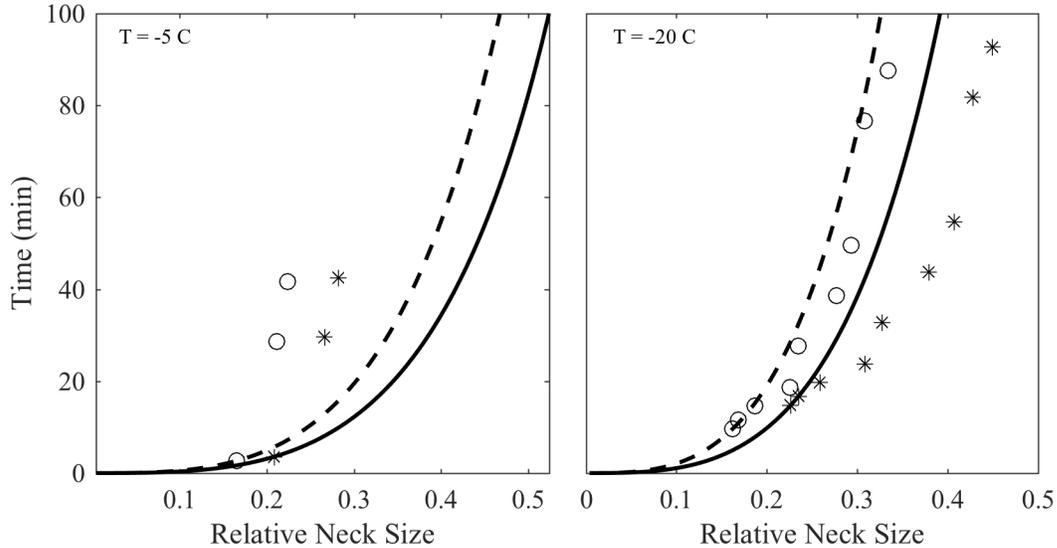

*Figure 12. Estimated neck growth (lines) compared to measurements (symbols) for ice grains with (left) radii of 99 μm (dashed, o) and 78 μm (solid, *) at -5 C over a period of 50 minutes, and (right) radii of 101 μm (dashed, o) and 73 μm (solid, *) at -20 C over a period of 100 minutes. The measured relative neck sizes (symbols) are equal to the shared neck size between grains divided by each of their diameters.*

Unfortunately, it is difficult to capture the entire neck growth process because, at warm temperatures, the grains sublimate even as sintering occurs and the molecules will tend to redistribute throughout the cryostage chamber. Using larger grains or lower temperatures slows the sublimation, but also increases the amount of time the process takes to occur, which introduces other challenges to performing the experiment. As a result, while we can observe that the model qualitatively reflects the neck growth process, the uncertainty in the predicted time scale is not yet well constrained. Interestingly, we also noted that some necks behaved predictably while others showed no growth at all, suggesting some inter-grain effect at play that controls the relative efficacy of neck growth between individual grains. Such an effect would have important implications for the way that the structure of the aggregate evolves as a whole (see discussion of mass redistribution in the following section). No densification (decrease in interparticle distance) was observed between grains during these experiments, which is consistent with the prediction that the timescale for densification is much longer than for neck growth.

Chen and Baker (2010), Chen et al. (2013), and Wang and Baker (2017) performed experiments similar to these by creating mm scale ice spheres and observing their neck growth under isothermal conditions (-10 C) and thermal gradients. While the morphological changes observed under an optical microscope look very similar to ours, their observations under SEM and x-ray tomography revealed that protrusions developed as a result of vapor transport between grains formed the porous necks between grains. This behavior is clearly a departure from what is predicted by SA81, but the size of the protrusions (10s of μm) was on the order of the grain sizes in our experiment, making it unclear whether or not this is a macroscale, vapor transport process or whether the same may be occurring in our own experiments. Such experimental results will provide a valuable point of comparison to efforts focused on smaller grains and lower temperatures, providing important insight into scaling effects.

4.1 Bulk ice experiment

In a second experiment, two bulk samples of ice grains were monitored for changes in aggregate structure. Spherical ice grains were created by spraying water into a bath of liquid nitrogen (LN2) using a custom-made gas and liquid handling system. The system uses gaseous nitrogen (GN2) to transport liquid water to an AutoJet1550+ controller, with a separate line in for pure GN2. These are then fed through a pneumatically-actuated drip-free atomizing nozzle to achieve a spray of fine, spherical particles. Both the AutoJet and atomizing nozzle were made by Spraying Systems, Inc. The nozzle used produces a mean grain radius of ~10 $\mu$m, and range of ~5-50 $\mu$m (Fig 13). Each sample had a mass of ~700 g and density of ~50% and was created in a plastic bucket and placed in a cold room precooled to -20 C for approximately one hour to allow the LN2 to boil off. The lids for the buckets had rubber gaskets, allowing us to then enclose the samples, though the seals were not airtight. One bucket was stored at -20 C (the "Warm Sample") and the other at -80 C (the "Cold Sample"). The samples were created two days apart, and then observed approximately 2, 4, 6, and 10 weeks later. The Warm Sample was observed after 14, 26, and 40 days, and the Cold Sample after 12, 24, 38, and 68 days. During each observation, ice grains from each sample were collected and placed onto a glass slide in an LN2-cooled Linkam LTS350 cryostage at -100 C and imaged with the Olympus BX51microscope (Fig. 13). Each sample was probed with a metal screwdriver to qualitatively assess its strength, and the mass and approximate volume of each was measured. The Warm Sample was observed to lose ~30% of its mass due to sublimation after two days. The mass was lost largely from the surface and edges of the sample and resulted in only a ~10% decrease in density. While this may have had some effect on sintering rates, we do not believe it changed our experimental observations qualitatively. Each closed bucket was then also placed inside a plastic bag to minimize further losses due to sublimation. No further noticeable changes in density of either sample were observed.

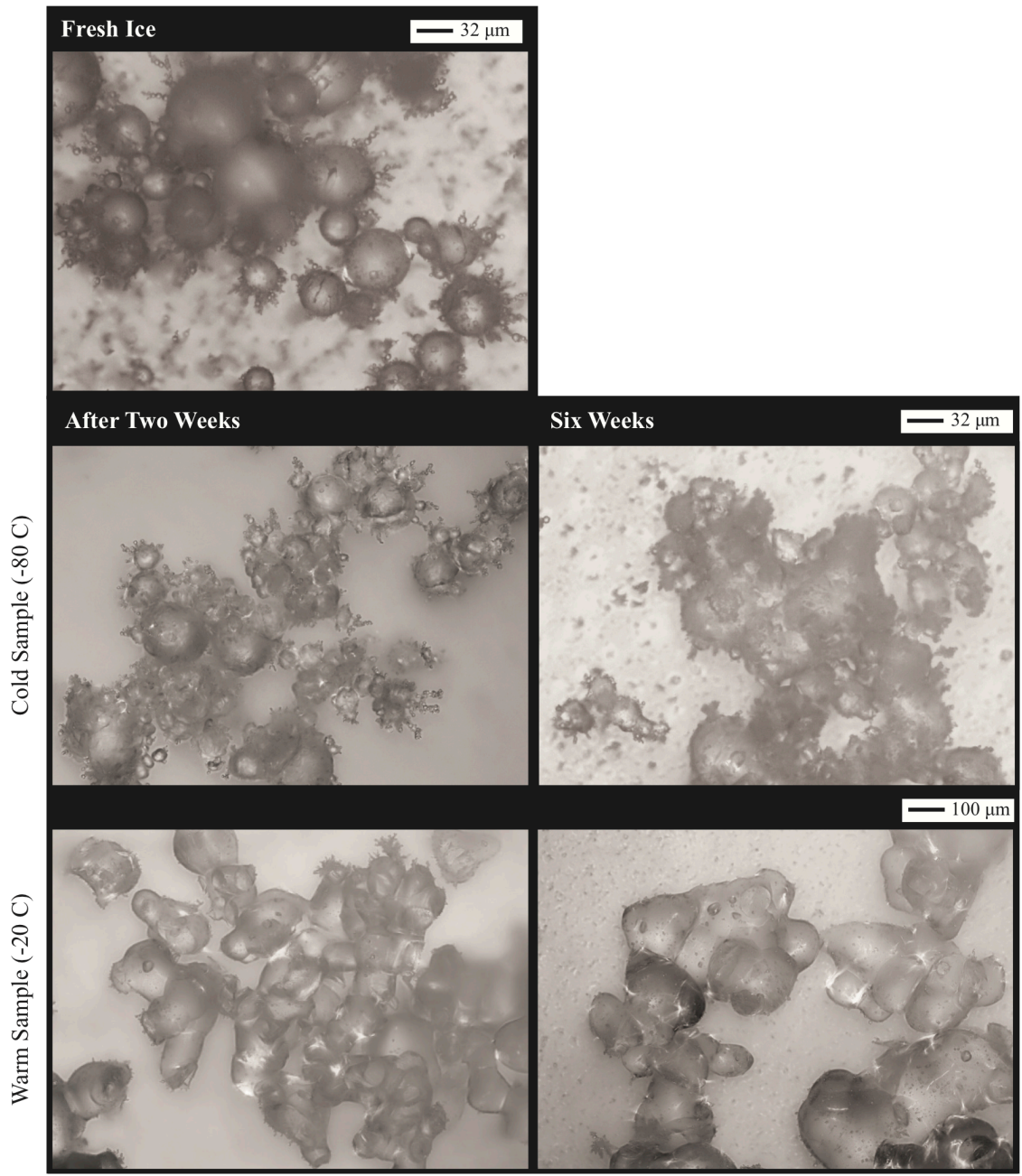

*Figure 13. Microscope images of ice grains from the Warm (top) and Cold (bottom) samples after two (left) and six (right) weeks. The images were taken under 10x magnification and processed to enhance their color contrast. Additional images may be found in the Supplemental Information.*

As one might expect due to the larger number of grains and longer experiment duration, these samples showed much more complex structure and modification than the isolated grain pairs in the previous experiment. Figure 13 shows microscope images of ice grains from both samples after two (left) and six (right) weeks (see also the Supporting Information). After two weeks, grains in the Cold Sample (Fig. 13, middle left) sintered together into clumps but remained distinguishable from each other. They evolved from nearly spherical to slightly subrounded in shape. The overall texture and bulk structure of the ice showed minimal modification even though some neck growth occurred. Sintering rates are temperature and grain size dependent, with the model predicting neck sizes of 0.59 and 0.43 for grains of 5 and 50 $\mu$m, respectively. This includes both the one hour the sample spent at -20 C as the LN2 boiled off (during which the smallest grains are predicted to complete neck growth), and the subsequent two weeks at -80 C. This is roughly consistent with the neck sizes observed in the sample, though some larger grains have necks that reach ~0.5 suggesting that the prediction may be an underestimate of actual sintering times in aggregates. After six weeks, some growth of grains is observed, fewer small particles are visible, and necks between larger grains have continued to grow. An assessment (see Appendix B) of the grain size distribution (Fig. 14) showed that the small grains (<10 $\mu$m) disappeared over the course of our ten-week observation, resulting in an increase in mean grain size (Fig. 14, diamonds) compared to the fresh ice.

In contrast, the Warm Sample (Fig. 13, bottom left) showed that significant modification of the aggregate had occurred after two weeks, clearly demonstrating that sintering rates are strongly temperature dependent. Few small grains remained in the sample, and the mean grain size had increased from 7 to 38 $\mu$m (Fig. 14). No change in mass was measured in the sample, suggesting that absorption of the smallest grains in the aggregate contributed to growth of the larger grains. There was also no change in density, suggesting that the grain and neck growth that occurred served to redistribute mass and pore space, rather than consolidate it. All of these observations are consistent with findings from previous studies (e.g., Blackford, 2007; and references therein), but provide a useful demonstration to which we can compare our model. The model predicts that neck growth should be complete in two weeks for 5 and 50 $\mu$m grains, with a final neck size of 0.59. This is somewhat consistent with the necks still visible in the sample, but many of the grains have formed closely packed agglomerates with larger and/or less distinct grain boundaries, making a comparison to the model difficult. In theory, the state of this sample should reflect the transitional (Stage 2) part of the sintering process after neck growth has completed. Pore spaces can be seen preserved inside some of the agglomerates, highlighting that the observed modification was primarily driven by non-densifying mechanisms. Examination of the bulk structure of the aggregate and the nature of the agglomerates indicates that the starting conditions of Stage 3 (section 2.4) assumed by SA81 are not necessarily representative, as the number of neighboring particles that an individual grain has varies throughout. Overall, given the model's idealized geometry and uniform grain size, the complex dynamics of grain growth and mass redistribution apparent in Figure 14 (top) are not well reflected by its predictions. After six weeks, the agglomerates had grown in size and the individual grain boundaries within them were less distinguishable, consistent with some recrystallization taking place.

The strength of both Warm and Cold samples was tested qualitatively by probing the ice with a flathead screwdriver during each observation and found to increase in both samples over time. During the first observation, the Cold Sample was noticeably stronger than the Warm Sample, which went against our expectation that the sample that underwent more neck growth would be stronger. We interpret this to be the result of increased strength of the Cold Sample due to its temperature, and/or decreased strength of the Warm Sample due to a growth in pore size from mass redistribution. During the second observation, the Cold Sample's strength had increased further, perhaps due to an increase in neck size but without significant mass redistribution. When probed, a linear fracture formed between two holes made by the screwdriver, a behavior never observed in the Warm Sample. The fact that fractures can form in only lightly sintered material at low temperatures has significant implications for the macroscopic-scale properties of icy surfaces and the interaction between sintering and other active processes. However, we emphasize that more experiments are needed to quantify and explore such effects.

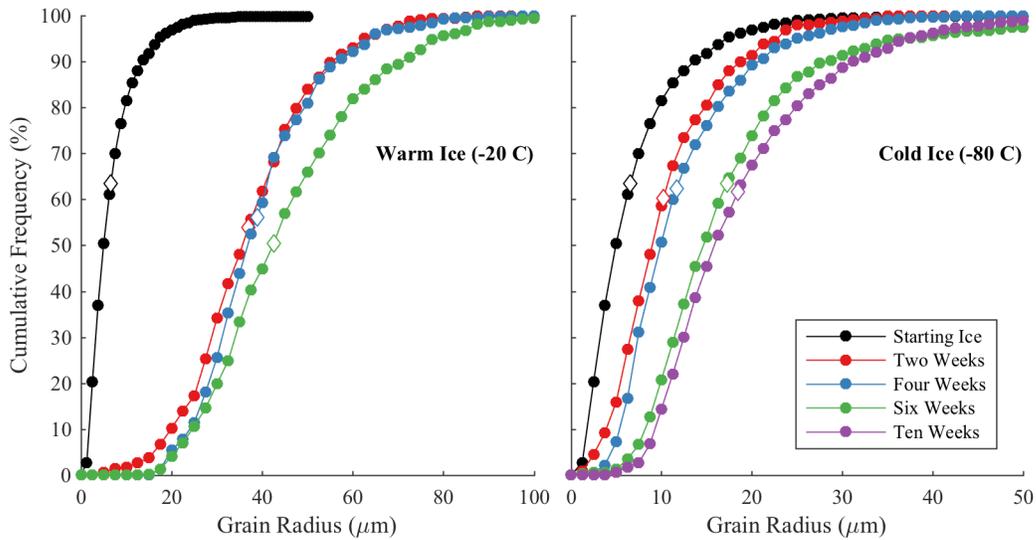

*Figure 14. The cumulative frequency of ice grains of a given radius in the starting ice (black), and after two (red), four (blue), six (green), and ten (purple) weeks, for the Warm (left) and Cold (right) samples. The value of the cumulative frequency indicates the percentage of grains measured that are smaller than a given radius. The mean grain size of each sample is denoted with a diamond. See Appendix B for measurement uncertainties.*

Overall, substantial modification of bulk ice aggregates was observed over a period of weeks in our samples, including grain growth, small grain absorption, and mass redistribution, with the effects being more pronounced in the Warm ice. The standard deviation of the grain size population (see Appendix B) increased along with mean grain size, showing that sintering occurred at all scales and increased the spread of the grain size distribution. As they sintered, the morphology of individual grains changed from nearly spherical to sub-rounded/sub-angular, and the development of agglomerates composed of multiple individual grains was apparent in the warmer sample. Further discussion on both the temperature dependence and mass redistribution are included in sections 5.1 and 5.3, respectively.

## 5. Implications for other worlds

Using SA81, we can begin to apply what is known about sintering from terrestrial studies to the surfaces of other worlds. Given the uncertainties in quantifying Stages 2 and 3 of the process (section 2.4, 3.2), we will focus only on Stage 1 (neck growth) for the remainder of the manuscript. For brevity, we will refer to Stage 1 sintering timescales simply as "sintering timescales," or $\tau$. Further, the exponential tail at the end of Stage 1 means it can take a considerable amount of time to quantify a negligible change in neck size, making an estimate of the sintering timescale less meaningful for constraining planetary surface properties. As such, Figures 15–17 present timescales to achieve 98% completion of neck growth. For comparison, this would lower the sintering timescale in Figure 9 (left, dotted line) from 15 to 3.3 dy and the final neck size from 0.58 to 0.57. The surface porosity is estimated to range from 40 to 95% on various icy satellite surfaces (Carvano et al., 2007; Domingue et al., 1995) based on phase curve analysis from telescopic observations. The following calculations were completed using an initial porosity of 70% (or relative density of 0.3) as a midrange value, though we note that remote sensing observations can only sample the upper surface of these objects and porosities may vary with depth on different bodies. The sensitivity of these results to starting density and material properties is discussed in Appendix B and is typically within one order of magnitude.

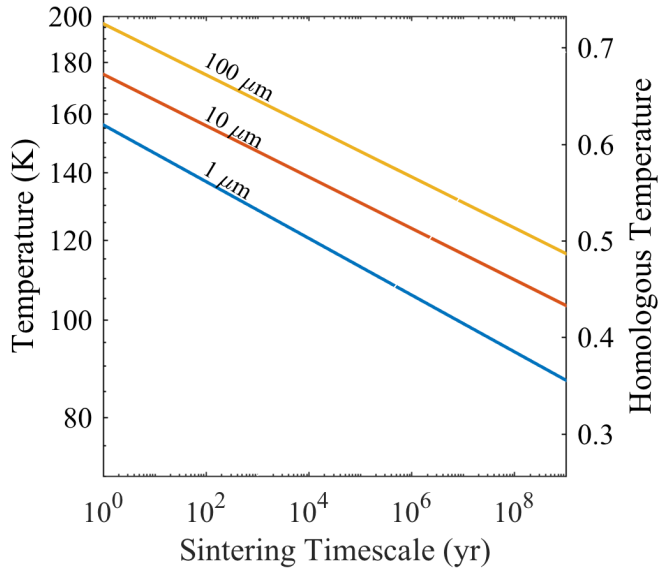

*Figure 15. Stage 1 (neck growth) sintering timescales with temperature and grain radius.*

5.1 *Sensitivity to temperature and grain radius*

Figure 15 shows the sintering timescale for water ice with varying temperature for grain radii of 1, 10, and 100 μm. Sintering rates have a power law relationship with both of these parameters, yielding shorter timescales for smaller grains and warmer temperatures. This suggests that sintering rates will vary significantly across different surfaces in the solar system, both with distance from the Sun and dominant grain size. This temperature sensitivity also suggests timescales on a given surface should vary globally with latitude and season and has important implications for the effect of diurnal thermal cycling. Even in the outer solar system, surface temperatures can vary by 10s of degrees throughout the day, leading cyclic changes in sintering rates and local spatial variations due to topographic shadowing effects. While the model assumes that the ice grains are at a constant temperature, it can still be used to constrain sintering rates on these surfaces. Since the majority of sintering will occur during a daily burst of activity at the warmest time of day, we can use the maximum surface temperature to obtain a lower limit on the sintering timescale. For example, the warmest, daytime equatorial temperature at Europa is 132 K (Pappalardo et al., 2009), which yields a timescale of $10^4$ yr for grains 10 μm in radius. On the other hand, the surface spends the most amount of time close to its mean temperature of 102 K, yielding an upper limit timescale of $10^9$ yr. In reality, the net sintering timescale will fall somewhere between these two bounds. Further, sintering timescales increase strongly with grain size (Fig. 16). Overall, this suggests that smaller grains in equatorial regions will have completed neck growth over Europa's 30 My surface age (Pappalardo et al., 2009), but larger grains may only be partially sintered.

Calculations at very low temperatures (see Appendix B) are computationally inefficient, so we must extrapolate sintering timescales at temperatures below 130 K. The relationship between sintering timescale ($\tau$) and temperature for grains of a given radius is best fit by:

$$\tau = f \mathrm{T}^g \qquad (31)$$

where the values of $f$ and $g$ are listed in Table 6. Our results are not perfectly fit by a power law (see Appendix B), which underpredicts sintering timescales at warm temperatures for grains $\geq 1$ μm. We do not recommend using (31) for temperatures >205 K, where the power law provides a poor fit to the results. At warmer temperatures, running the full calculation will provide a more accurate prediction. The power law for 0.1 μm grains overpredicts sintering timescales, and is only recommended at temperatures $\leq 180$ K. One consideration when applying such predictions is that the relevant threshold state of the ice aggregate may vary with application. For example, the full sintering timescale as given by (31) may be most relevant for evaluating surface strength, though perhaps only a modest increase in neck size is needed to influence remote sensing observations. Further, no adjustments to these calculations have been made to account for the presence or lack of atmosphere on any of these surfaces (see appendix A).

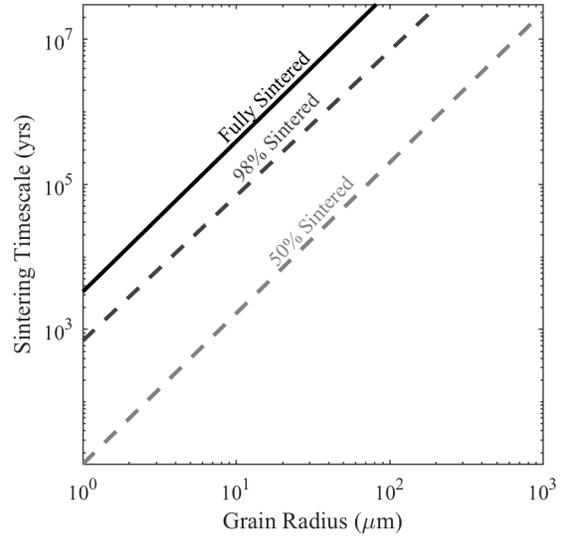

*Figure 16. Stage 1 (neck growth) sintering timescale with a 10 μm grain radius at a temperature of 132 K.*

| Radius (μm) | $f$ (yr) | $g$ |
|---|---|---|
| 0.1 | 3.189 x $10^{39}$ | -19.18 |
| 1 | 8.945 x $10^{77}$ | -35.54 |
| 10 | 1.204 x $10^{88}$ | -39.26 |
| 30 | 3.743 x $10^{89}$ | -39.51 |
| 50 | 1.043 x $10^{90}$ | -39.51 |
| 100 | 4.159 x $10^{90}$ | -39.51 |

*Table 6. Power law parameters for the Stage 1 (neck growth) sintering timescale for grains of varying radii in units of Earth years. The timescale is defined as achieving 98% of the neck growth process.*

5.2 *Implications for surface properties*

Our results predict that substantial changes in grain morphology and aggregate microstructure due to sintering will occur on icy surfaces throughout the inner and outer solar system. However, while neck growth is predicted to occur relatively quickly on many bodies, densification occurs at a much slower rate. This was demonstrated in section 3.2 (Fig. 9), and the effect is even more extreme at outer solar system temperatures. Using the example from the previous section, negligible densification is predicted to occur over Europa's young surface age, suggesting that ice regolith that develops on its surface may form a cohesive, but porous, sintered crust. Below we discuss implications for the strength, subsurface structure, thermal conductivity, and spatial variation of such a crust on different bodies.

The strength of a sintered crust will depend on the extent of neck growth, porosity, and packing structure, and may vary in different thermal and pressure environments. This is not well constrained due to the considerable parameter space occupied by these variables. Further, there are many different metrics for material strength, each of which is relevant in different contexts. Measurements of the compressive strength or penetration resistance of snow do exist in the literature (e.g., Jellinek, 1959; Schneebli et al. 1999), but it is difficult to interpret them broadly due to the uniqueness of the conditions under which each sample evolved. It is also unclear how representative these may be of other planetary surfaces, as the sintering process on Earth is strongly influenced by melting effects, wind, gravity driven creep and compaction, and the shape of snowflakes. In general, sintered ice is expected to be stronger with age. We noted this qualitatively during our bulk ice experiments, though it is unclear whether the strength increase was caused by a change in neck size or particle rearrangement. The most relevant laboratory measurements to date are those of Grün et al. (1993) and Kömle et al. (2001), which report penetration resistance up to 10 MPa in unconsolidated ice irradiated to temperatures >273 K to simulate cometary conditions. This is consistent with Spohn et al. (2015), who report that the penetration resistance of material at the Philae landing site on comet 67P is >4 MPa. Thomas et al. (1994) measured the penetration resistance of ice aggregates under isothermal conditions at -20 C, and reported 40 and 160 kPa after 3 minutes and 25 hours, respectively. Sintered ice is also expected to be stronger with decreasing porosity, as has been documented for certain rock types (e.g., Chang et al., 2006; Palchik and Hatzor, 2004). The unconfined compressive strength of solid, polycrystalline ice has a roughly linear dependence on temperature, ranging from 3–50 MPa at temperatures from 263 K down to 77 K, respectively (Arakawa and Maeno, 1997; Durham et al., 1983; Parameswaran and Jones, 1975), which represents the end member case for ice that has sintered into a cohesive, high density mass.

Changes in strength due to sintering may have different implications for the evolution of different planetary surfaces. For example, sintering of ice and ice-dust mixtures may contribute to the hardening of the near-surface of comets as they enter the solar system or approach perihelion. The strong seasonal increase in sintering rate as the surface warms will cause a sintering front to move downward into the subsurface, increasing material cohesion and strength under the loose regolith layer at the surface. Kossacki et al. (2015) estimated that a sintered layer 1 meter thick could develop in the near subsurface of comet 67P over 10s of years if the ice has a small grain size, increasing its strength with each successive perihelion pass. Such effects may also contribute to cometary activity by helping drive the development of strong, near-surface ice cap that competes with the buildup of volatile pressure in the subsurface, eventually leading to an outburst. Of course, sintering effects will interact with larger-scale vapor transport processes within the subsurface in ways that are not well understood and are not considered by this model. However, such interactions may have important implications for cometary evolution and deserve careful consideration.

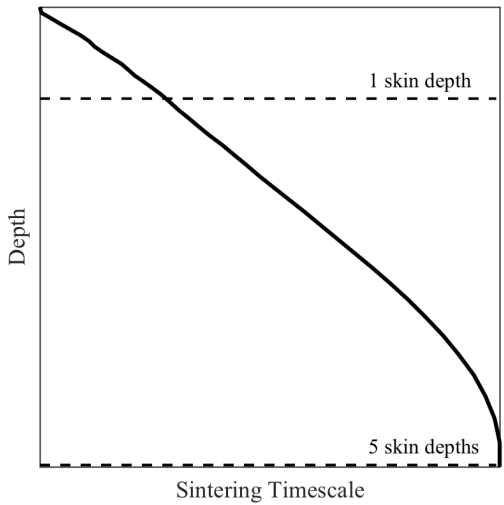

*Figure 17. Stage 1 (neck growth) sintering timescale with subsurface depth for an aggregate of ice undergoing an arbitrary thermal cycle. The dashed and dotted lines indicate depths of one and five thermal skin depths, respectively.*

Due to its extreme sensitivity to temperature, sintering may also drive subsurface evolution of different planetary bodies depending on their diurnal and seasonal thermal cycles. Ice at different depths reaches the same mean temperature for a given cycle, but experiences different temperature amplitudes. This will drive the development of subsurface gradients in neck size and density at the scales of the diurnal and seasonal thermal skin depths of the ice. In its current form, the sintering model assumes a constant temperature. However, by incorporating spatially and temporally varying temperatures into it, we can assess how sintering rates vary in the near surface. We generated surface and subsurface temperature profiles using a thermal model (Molaro et al., 2015) for a material with arbitrary constant density and thermal conductivity, on a body with an arbitrary solar distance and rotation rate. By using these time-varying temperatures in the sintering model instead of a constant value, we calculated the resulting profile of sintering timescale with depth (Fig. 17). The maximum sintering rate (and thus minimum sintering timescale) occurs at the surface, where the ice spends the most time at the highest temperature. The interaction between diurnal and seasonal thermal cycles, as well as the effects of continuous or intermittent deposition of fresh ice grains may lead to complex gradients in microstructure on certain bodies. Such subsurface structure could be a driving mechanism for a variety of phenomena observed on icy surfaces.

As neck growth occurs, the surface area of solid-to-solid contact between ice grains will increase, causing an increase in the thermal conductivity of the aggregate. This change will be reflected in the temperature and thermal inertia observed at the ice surface. This indicates that the thermal behavior of icy surfaces is not only a function of density and grain size, as it is for rocky surfaces, but also of age. This has important implications for how we interpret observations of tectonically (or otherwise) active surfaces such as Europa, though such interpretations are complex in that the age of surface regolith may or may not correspond to the age of an underlying surface unit. We note that the demonstration shown in Figure 17 does not account for this effect, suggesting that the profile may increase even more steeply with depth. There are several models for variation in thermal conductivity with neck size (Ferrari and Lucas, 2016; Sirono and Yamamoto, 1997; Steiner and Kömle, 1991), but incorporation of this into SA81 is beyond the scope of the present study.

Additionally, the effects of sintering on thermal conductivity suggest spatial variations in thermal inertia may be observed on and between some surfaces, with less sintered ice at higher latitudes and larger solar distances, reflected in lower thermal inertia values. This is consistent with the observation that saturnian satellites have a thermal inertia 2–6 times lower than Galilean satellites (Howett et al., 2010). This difference is typically attributed to the deposition of fine E-ring particles on their surfaces, but lower sintering rates in the saturnian system may serve to contribute to or reinforce the effect. However, this is opposite of the trend observed on Europa, which shows increased thermal inertia at higher latitudes (Rathbun et al., 2010; Spencer et al., 1999). Ultimately, whether or not such trends are observed is a matter of the relative rates of sintering and regolith production. They may not be observed on surfaces where sintering occurs very slowly and is not a dominant metamorphism process, or where it occurs fast enough such that the entire surface has completed the neck growth process. In the latter case, variations may still be observed due to subsequent densification stages of sintering, but such timescales are not calculated in this study. Such trends will also be influenced by endogenic heat fluxes, the effects of microstructure on absorption and scattering of light, and a variety of other factors. This highlights how complex these icy surfaces are, and how little we understand about their active processes and characteristics at small scales.

5.3 *Grain size distribution and mobilization*

Given that sintering is strongly size-dependent, the size-frequency distribution (SFD) of ice grains that make up an aggregate will significantly influence its evolution. For aggregates that have a strongly bimodal or wide range of grain sizes, the sintering state of the largest may be slow or quasi-static compared to the changes experienced by the smallest. As such, very small grains can become absorbed into the mass of larger grains, as observed in early laboratory studies (Kuroiwa, 1961) and shown in Figure 14, which will cause an increase in the mean grain size of the aggregate over time. The location where a small grain contacts a larger grain will determine where it is absorbed and to some extent how its mass is redistributed, so absorption of small grains at a contact between larger grains may also serve to effectively enhance their sintering rate than drive grain growth. It is unclear what diffusion mechanisms are at play in each of the large and small grains during absorption, and whether it reflects the (relatively) rapid progress of small grains all the way through Stage 3 of the sintering process or whether the grain size difference contributes in another way. Grain boundaries do not appear to be retained in this process. A grain will also sinter to each other grain with which it is in contact, so in addition to the SFD, an aggregate's initial density and packing arrangement also have a controlling effect on its evolution through the formation of agglomerate structures. Agglomerates do retain grain boundaries, which affect the sintering rates of individual grains and the behavior of the aggregate as a whole. In its current form, the model predicts sintering rates for an aggregate of uniformly sized spheres in an ideal packing structure (Fig. 4). Its assumptions about how the pores in such an aggregate should evolve to become spherical during Stage 1 is reasonable, and physically should lead to a stable lattice of sintered grains with large necks that have ceased growing. However, such an ideal system does not allow for formation of agglomerates or grain absorption. In this context, clearly the model is inadequate to describe a realistic surface.

Grain absorption and agglomerate formation have separate but overlapping effects on the characteristics of the aggregate. On one hand, since grains of a similar size sinter at similar rates, any absorption that occurs would increase the mean grain size of the aggregate until the difference in sintering timescales between small and large grains is no longer sufficient for the latter to be considered slow or quasi-static. In a scenario in which the grains were neatly packed into an arrangement that prevented agglomerate formation, one might expect the aggregate to evolve towards a larger, but narrower, range of grain sizes and eventually stabilize such that it resembles the idealized case in the model. On the other hand, a perfect packing structure is hardly expected on a natural surface, and as we observed in our experiments, agglomerate formation can facilitate grain mobilization and mass redistribution within the aggregate that drives grain growth beyond the idealized scenario. The control that grain SFD and packing arrangement have on if, where, and how agglomerates form, how many grains can be involved, their resulting size and characteristics, and the extent to which they can continue to grow is not well understood. Depending on the context, one may also need to distinguish between the aggregate's grain and agglomerate SFD, as grain boundaries within agglomerates may influence, e.g., its scattering properties but not necessarily its strength. Ultimately, the bulk evolution of a non-uniform aggregate will be driven by the convolution of sintering stages and timescales experienced by each of its constituent grains, and the dependence of the rate and morphological end state of this process on initial grain SFD has not been quantitatively constrained. This contributes to the difficulty in characterizing changes in packing density and particle arrangement in the sintering model, and its effect on densification timescales. In this area, laboratory studies and models of snow metamorphism and bulk vapor transport under warmer isothermal and terrestrial conditions (e.g., Gubler, 1985; Kaempfer and Schneebeli, 2007; Kaempfer and Plapp, 2009; Pinzer et al., 2012) may offer valuable insight into how complex structures evolve in ice aggregates. Incorporating effects such as Ostwald ripening and grain coarsening described in the terrestrial literature may lead to a more sophisticated approach in quantifying grain growth on planetary surfaces. Further laboratory experiments will also be needed at colder temperatures and in low pressure environments.

The grain SFD, density, and packing arrangement on any given surface will be determined by the processes that create the ice, and the evolution of a relatively static grain population will vary from one with a high deposition rate of fresh, unsintered material. In the latter case, the relative rate of modification and deposition must be considered in combination with the grain SFD. If the fresh material has a different grain SFD than the initial or underlying aggregate, and/or the surface has already undergone significant modification prior to its deposition, grain absorption effects may drive evolution beyond what can occur in a static grain population. This may be important on Enceladus, as fine (0.25–2.5 μm radius) grained plume fallout and E-ring material is being deposited on the surface (Kempf et al., 2010) where particles 10s to 100s of μm are observed (Jaumann et al., 2008). At low deposition rates, absorption of fine material onto the necks of grains in an aggregate of larger particles may contribute to an enhancement to their sintering rate. Such an enhancement would have significant implications for its surface evolution, as sintering timescales are long at such temperatures. At higher deposition rates, absorption of fine particles may cause enough growth in grain volume to both increase neck sizes and fill in pore space between adjacent grains, driving densification of the surface layer of ice. However, no estimates are available (to the authors knowledge) of the deposition rate of the large grain population, making it difficult to quantify how the grain SFD exposed at the surface may evolve in this context. Further, it is unknown how long the plumes have been active, or how their mass flux may vary. The way that an aggregate evolves will also be influenced by other active surface processes, such as sputtering, micrometeorite bombardment, and thermal segregation, which may serve to inhibit or enhance the sintering process on different bodies or at different locations.

*5.4 Ices of other composition*

While pure water ice does exist in the solar system, many of the surfaces on which sintering is relevant are also thought to contain salts such as sodium chloride and magnesium sulfate. If such impurities are present in small volumes, their presence will not change the diffusion coefficients of the ice. However, they will lower the ice's melting temperature ($T_m$) and thus increase the rate at which sintering occurs. The right-hand axis of Figure 15 shows the homologous temperature ($T/T_m$), which can be used to estimate the change in sintering timescales for salty ices. For example, Enceladus has a maximum equatorial surface temperature of ~80 K (Howett et al., 2010) or a homologous temperature of 0.293. For a 0.1 μm grain deposited by plume fallout, the sintering timescale at this temperature is ~$10^3$ yr. However, the ice is thought to contain up to 2% NaCl by mass (Postberg et al., 2009), which would lower its melting temperature to ~270 K and raise its homologous temperature to 0.296. This corresponds to a temperature of 81 K for pure water ice, reducing the timescale to ~$10^2$ yr. This effect is small for larger grains due to Enceladus' cold surface temperature, though a eutectic mixture of water and NaCl has a melting temperature of 253 K and would decrease the sintering rate of 1 μm grains from above to below the age of the solar system. On Europa, 10 μm grains at the mean temperature of 132 K with 2% NaCl are reduced by a factor of 1.5 relative to pure water ice. While the timescales are of the same order of magnitude, surface units that are still undergoing neck growth may exhibit differences in surface properties due to variation in salt content. Overall, the presence of salts can have a significant impact on sintering timescales, which suggests that local and global variations in composition across a surface or between bodies could lead to substantial variations in microstructural properties.

Water ice dominates the surfaces of Europa and Enceladus, but other bodies are dominated by nitrogen and methane ices. Qualitatively, nitrogen and methane in vacuum environments behave similarly to water ice because vapor diffusion dominates neck growth in all cases. However, their diffusion coefficients and melting temperatures vary from that of water (Table 5), leading to significant differences in sintering timescales. Triton, for example, is dominated by nitrogen ice and has a mean surface temperature of 38 K (Cruikshank et al., 1993). While this is substantially colder than the Galilean and saturnian satellites, the melting temperature of nitrogen ice is only 63 K, yielding sintering timescales of only hours for 100 μm grains. Sintering is likely to be an important process on surfaces like Triton, Titan, and Pluto, though the diffusion coefficients for non-water ices (Table 4) are less well constrained in the literature. Sintering is also likely to play an important role in martian polar processes, but the authors were unable to find any published measurements of $CO_2$ ice diffusion coefficients. Eluszkiewicz and Moncet (2003) and Eluszkiewicz et al. (2005) estimated these parameters based on measurements of $N_2$ (Estève and Sullivan, 1981) and other crystalline materials (Chezeau and Strange, 1979). The effects of atmosphere on vapor transport are also not well understood (section appendix A), which prevents us from quantifying sintering rates on their surfaces at this time. Additional laboratory studies are needed to address these issues.

*5.5 Implications for previous works*

To the authors' knowledge, the earliest application of Swinkels and Ashby (1981) to a non-terrestrial surface is Eluszkiewicz (1991). They performed calculations for both the neck growth and densification stages of sintering of nitrogen ice on the surface of Triton. Unfortunately, they neglected vapor transport during neck growth, which is the most efficient diffusion mechanism during this stage. They pointed out this error in a later paper (Eluszkiewicz and Moncet, 2003), but stated their conclusions still held because neck growth makes a negligible contribution to the total sintering timescale. Their analysis found that densification can occur over seasonal timescales (~100 Earth years) on Triton if the starting grain size is <0.5 µm. However, our findings suggest that since the timescale to complete Stage 2 of the process is not quantified by the model, they may underestimate densification timescales by multiple orders of magnitude. Similarly, the densification timescales calculated by Eluszkiewicz (1993), Eluszkiewicz and Moncet (2003), and Eluszkiewicz et al. (2005; 2007) on Mars, Pluto, and Io are also likely underestimated. Little can be said about their densification calculations themselves, since we do not quantify Stage 3 in this study. Any future efforts wishing to provide a comparison to their results should be aware that the densification expression they used does not follow SA81, but is a modified form from "an unpublished internal report by Ashby (1988)," (Eluszkiewicz and Moncet, 2003). (We have not included the quoted reference in our citations list because we could neither find nor verify the existence of the report.) We also note that all six of these studies appear to mistake particle radius for diameter, and their expression for neck growth due to vapor transport (Eluszkiewicz, 1993; Eluszkiewicz et al., 2005; Eluszkiewicz and Moncet, 2003) matches SA81 except for a missing geometrical factor of order $\sim 10^{16}$. It is unclear if these are mistakes in their calculation or typos.

Kossacki et al. (2015) and Kossacki (2015) used a partial implementation of this model to calculate sintering rates and material strengths on comets. They considered neck growth via vapor transport and, in the latter case, densification via lattice diffusion from a boundary source, both terms of which appear consistent with SA81. Since they only calculate sintering rates for small neck sizes (up to 0.3), their results are unlikely to be affected by the problems arising from the Stage 2 calculation. However, by neglecting other diffusion mechanisms (surface diffusion in particular, see Appendix B) they may have somewhat underestimated sintering rates. Beyond this, it is unclear what effect our findings have on their results because their calculation of porosity (on which their strength calculation is based) diverges from this model. Kossacki et al. (1994; 1997) also used a partial implementation of this model to investigate the thermal conductivity of sintered ice. The full model of Kossacki et al. (1997) incorporated sintering due to vapor transport from SA81 along with other components related to bulk vapor transport processes. They found good agreement when comparing their model to bulk measurements of temperature with depth in laboratory samples of ice undergoing irradiation. Unfortunately for us, the microstructural properties of the ice (e.g., neck size) in their model were fitted parameters rather than measured from their samples, and thus cannot help to validate this sintering model.

Schaible et al. (2016) suggested that sintering on saturnian satellites can explain their observed thermal inertia anomalies. While they included most of the diffusion mechanisms in Table 3, they neglected vapor transport in their calculation. They also extended SA81 to incorporate the effects of sputtering- and radiation-induced diffusion that are likely to be present on some icy satellites in the outer solar system. A direct comparison between our results is challenging since they only calculated sintering timescales for small neck sizes. Their reported timescales (their Fig. 6) are approximately the same for 5 µm grains, but two orders of magnitude lower than what we find for 25 µm grains. On one hand, this suggests that sputtering- and radiation-induced diffusion are of comparable efficiency to vapor transport for smaller grains, and more important for larger grains. Their timescales should still reflect overestimates since they did not include vapor transport, but they are underestimates relative to our values since we quantified a longer part of the sintering process. Overall, it is unclear what quantitative conclusion may be drawn from this. Qualitatively, however, this highlights the strong role that sputtering and radiation play in the sintering process on these surfaces and the need to incorporate their effects into future implementations of SA81.

## 6. Conclusions

In this study, we have provided a detailed exploration of the sintering model of Swinkels and Ashby (1981), which can be used to study the microstructural evolution of ice throughout the solar system. We compared the predicted Stage 1 (neck growth) rates to historical measurements of sintering ice and found that it agrees within approximately one order of magnitude, which is consistent with the high uncertainty in sintering rates reported in the literature. Our own experimental observations of an isolated pair of sintering ice grains are also consistent with the model at the order of magnitude level, but more measurements are needed to fully validate the model and provide better quantitative constraints. Experimental observation of the evolution of bulk aggregates show that grain growth and mass redistribution due to grain absorption and formation of agglomerate structures occurs in aggregates with a non-uniform grain distribution. This complex behavior is not described by the idealized packing structure in the model, emphasizing the need for more laboratory studies on this topic to improve the model. The strength of sintered samples was observed to increase even over the short duration of our experiment, and fractures were observed to develop in lightly sintered material at cold temperatures. This has significant implications for the importance of this process for planetary surfaces.

    Our results demonstrate that sintering timescales for water ice are extremely sensitive to temperature and grain size, consistent with previous findings on the topic (e.g., Blackford, 2007; and references therein), increasing in aggregates with colder and larger grains. This suggests that the microstructural evolution of ices may vary both throughout the solar system and globally on a given body. Even in the outer solar system, ice is expected to undergo significant modification through neck growth (Stage 1), leading to changes in thermal conductivity, strength, and other surface properties. Further, sintering rates will vary with depth over diurnal and seasonal timescales, which may drive the development of subsurface gradients in structure on some bodies. Densification (Stage 3) occurs over much longer timescales than neck growth, suggesting that some icy surfaces may develop cohesive, but porous, sintered crusts. The potential ability for these objects to retain porosity over long geologic timescales has important implications for their surface evolution and characteristics.

    Several aspects of the model may benefit from improvement, including the accuracy of the vapor transport term and the way that the model transitions between the neck growth (Stage 1) and densification (Stage 3) dominated stages of the sintering process. In its current form, the model provides estimates of neck growth timescales that appear accurate at the order of magnitude level, but our analysis suggests it may underestimate total sintering timescales including late stage densification by multiple orders of magnitude. The model would also benefit from a more sophisticated geometry that can include grains of multiple sizes. Future work will address these topics, including additional laboratory

measurements of ice sintering rates under various environments to help better validate and develop improvements to the model. Laboratory measurements of the diffusion coefficients of non-water ices are limited in the literature and will be needed for application of the model to planetary bodies where water is not the dominant form of ice. There is much to learn about the process of sintering, and the implications it has for the evolution of ice throughout the solar system.

**Appendix A: Derivation of the vapor diffusion mechanism**

Vapor transport of molecules ($\dot{V}_3$) is one of the two dominant diffusion mechanisms driving neck growth in ice (Fig. 10). However, close examination of this term reveals that certain assumptions made in its derivation are problematic, which we anticipate are partially responsible for the discrepancy between model predictions and observed sintering rates (Fig. 8). Here we derive $\dot{V}_3$, as ascertained from the sparse descriptions in Ashby (1973) and Kingery and Berg (1955) (SA81's predecessors), in an effort to discover why such a discrepancy may exist.

The expression for $\dot{V}_3$ (which we will simply call $\dot{V}$ in this appendix) is based fundamentally on the Knudsen-Langmuir (or Hertz-Knudsen) equation, which describes the mass flow rate of molecules due to disequilibrium between the saturated vapor pressure ($P_s$) and the actual vapor pressure ($P$) above a surface. The difference in pressure ($P_s - P$) drives the flow of particles either to or from the surface until equilibrium is achieved. The mass flow rate ($\dot{m}$, units of kg/m²s) is given by:

$$\dot{m} = \alpha(P_s - P)\left(\frac{M}{2\pi RT}\right)^{\frac{1}{2}} \tag{a1}$$

where $M$ is the molecular weight, $R$ is the gas constant, and $T$ is the temperature. The term $\alpha$ is the sublimation (or evaporation) coefficient, a correction factor applied to account for a lowered sublimation rate relative to the maximum sublimation in vacuum. Modern forms (Persad and Ward, 2016) of (a1) have both a sublimation (or evaporation) and deposition (or condensation) coefficient, and must be empirically determined in a given experiment. In this application, Kingery and Berg (1955) assumed that sublimation occurs with perfect efficiency and thus set ($\alpha = 1$).

They then related the mass flow rate to the resulting rate of change in volume of solid material ($\dot{V}$) by multiplying $\dot{m}$ by the area of the surface to which material is being transported ($A$), and dividing by the density of solid ice ($\Delta_o$):

$$\dot{V} = \frac{\dot{m}}{\Delta_o} A \tag{a2}$$

The value of $A$ is given by one half of the arc length of the neck cross section, multiplied by the three-dimensional neck circumference ($A = 2\pi x \rho \theta$). We use only one half of the total neck area because the model assumes symmetry across the grain-grain boundary. Combining (a1) and (a2) gives an expression for the volume flux in terms of the pressure difference ($\Delta P = P_s - P$):

$$\dot{V} = \Delta P \frac{2\pi x \rho \theta}{\Delta_o}\left(\frac{M}{2\pi RT}\right)^{\frac{1}{2}} \tag{a3}$$

Expression (a3) describes mass flux due to a disequilibrium with the saturated vapor pressure over a flat area. However, mass flow into the neck region during sintering is ultimately driven by differences in vapor pressure due to changes in surface curvature along the arc of the neck. The vapor pressure above a curved surface ($P$) is given by the Kelvin equation:

$$\ln\frac{P}{P_v} = \frac{2\gamma V_m}{rRT} \tag{a4}$$

where $\gamma$ is the surface energy, $V_m$ is the molar volume, $r$ is the radius of curvature of the surface, and $P_v$ is the vapor pressure on a flat surface. A convex surface has a positive radius of curvature, and thus pressure above a convex surface is increased relative to a flat surface. Pressure over a concave surface, which has a negative radius of curvature, is decreased relative to a flat surface. For our purposes, a more useful form of (a4) is in terms of the surface curvature ($K = 1/r$) and the molecular weight ($M = V_m/\Delta_o$):

$$\ln\frac{P}{P_v} = \frac{2\gamma M}{\Delta_o RT} K \tag{a5}$$

Since $P_v - P = \Delta P$ is small, we can approximate $\ln\frac{P}{P_v} \approx \frac{\Delta P}{P_v}$ which leads to:

$$\Delta P = \frac{2\gamma M P_v}{\Delta_o RT} K \tag{a6}$$

Since our system has more than one radius of curvature (Fig. 2), the net mass flow from the grain (with curvature $K_3$) into the neck (with mean curvature $K_m$) will result from the net difference in vapor pressure between each point. Using equation (a6) we can calculate the change in vapor pressure ($\Delta P = \Delta P_3 - \Delta P_m$):

$$\Delta P = \frac{2\gamma M P_v}{\Delta_o RT}(K_3 - K_m) \tag{a7}$$

We can then use an expression analogous to equation (a3) to determine the vapor transport mass flow rate for sintering grains by substituting in equation (a7):

$$\dot{V} = 2\pi x \rho \theta \frac{2\gamma M P_v}{RT}\left(\frac{M}{2\pi\Delta_o RT}\right)^{\frac{1}{2}}(K_3 - K_m) \tag{a8}$$

Finally, we convert equation (a8) to the form presented by Swinkels and Ashby (1983) using the relations $M = V_m/\Delta_o$, $V_m = \Omega * Na$, and $RT = kT * Na$, where k is the Boltzman constant, Na is Avagadro's number, and $\Omega$ is the molecular volume:

$$\dot{V} = 2\pi x \rho \theta \frac{2\gamma \Omega P_v}{kT}\left(\frac{\Omega}{2\pi\Delta_o kT}\right)^{\frac{1}{2}}(K_3 - K_m) \tag{a9}$$

It is unclear if the $\Delta P$'s in Eq.'s (a3) and (a7) can really be considered physically equal, calling into the question the overall approach of the derivation. A number of problems with Eq. (a1) on which (a3) is based are discussed below. Nonetheless, this is the origin of Eq. (a9) as used in SA81 and presented here for discussion. We also note that the Eq. (a9) has an extra factor of 2 relative to the same term in Table 3. Kingery and Berg (1955) dropped this factor of 2 when invoking the Kelvin equation (a4) in the original derivation for reasons we were unable to determine. We have included it here for clarity and transparency.

The Clausius-Clapeyron equation is then used to calculate the vapor pressure over a flat surface:

$$P_v = P_o \exp(-Q_v/RT) \quad \quad \quad (a10)$$

where $Q_v$ is the heat of sublimation, and $P_o$ is a pre-factor calculated from a reference pressure and temperature for water ice.

There are a number of issues with this derivation, stemming largely from the fact that the Langmuir-Knudsen expression (a1) has been shown to be erroneous and problematic. As discussed by review of Persad and Ward (2016), researchers readily admit in the literature that it is difficult to match experimental results without adjusting the empirical parameters. However, its use persists in the literature due to its simplicity and historical precedent. Primarily, the equation assumes that the evaporation and condensation coefficients are equal (rolled into a single parameter $\alpha$), and that all molecules that evaporate from the grains' surfaces are immediately deposited in the neck. This ignores gas-gas diffusion rates and prescribes mass conservation in the system, neglecting both ambient sources of water molecules from the environment and potential mass loss from the system. It also assumes that the temperatures of the liquid and vapor at their interface are equal. In addition, the derivation of $\dot{V}$ assumes both that the gas around the grains is in equilibrium at distance from the neck, and that the curvature within the neck drives mass flux. However, these are mutually exclusive assumptions, as mass flow to the center of the neck would decrease pressure at the saddle's edge, setting up a pressure gradient along the grain's surface. Finally, by setting the parameter $\alpha = 1$, the model assumes that sublimation occurs with perfect efficiency. This ignores the presence of an atmosphere, even though the density of the background environment has been shown to have a significant effect on sintering rates (Hobbs and Mason, 1964). Swinkels and Ashby (1981) found this to be a reasonable approximation for metals, for which vapor transport is not a dominant mechanism. Ultimately, however, the uncertainty introduced by these factors makes it challenging to determine how accurately the model can predict the behavior of ice in different environments, especially if forced to rely on poorly constrained empirical parameters to account for the presence of atmosphere. Future work will address this issue more directly by developing improvements to the vapor transport calculation.

**Appendix B: Uncertainty and model sensitivity**

A series of tests were performed in order to assess the uncertainty in model calculations, as well as the sensitivity of the result to a number of approximations and material parameters. To ensure we captured any trends in the results, we performed these calculations for two grain radii (1 and 100 $\mu$m), each at two temperatures (200 and 250 K). We compared the resulting prediction for the sintering timescale ($\tau$), which is defined as the time to achieve 98% completion of the Stage 1 (neck growth) process. We also compared the timescales to complete 50% ($\tau_{50}$) and 25% ($\tau_{25}$) of the process, to quantify the influence of these effects throughout the process. *Except for initial density, all cases tested below were found to have a negligible effect on $\tau$,* but not necessarily on $\tau_{25}$ and $\tau_{50}$. Uncertainties in our experimental measurements are also discussed below.

*Analytical Approximations:* We have run the model using the approximate analytical solutions given by equations (6), (18), (19), and (20). One could instead run the full solution by solving equations (6), (7), (10), and (11) symbolically at each timestep, but this is extremely computationally expensive, increasing the processing time from seconds to hours. To ensure the approximation is reasonable, we compared predicted timescales to those calculated using the full solution. For 100 $\mu$m grains, use of the approximation caused a negligible difference at all timescales and temperatures. For 1 $\mu$m grains, it caused a decrease in $\tau_{25}$ and $\tau_{50}$ of 9% and 1%, respectively, at 250 K, and of 11% and 2%, respectively, at 200 K. It is unclear if these decreases arise from the order of magnitude factor in the analytical approximation (which was calculated for wires rather than spheres, see section 2.3.1), or whether they result from the approximation itself. Given the small size ($\lesssim$10%) of the effect, we feel the computational benefits gained from using the approximations outweigh the costs.

The downside to using the analytical approximations is that they introduce computational issues whenever the model is calculating very small changes in neck size or other geometrical parameters. In these cases, it may produce values of curvature with the wrong sign, introducing imaginary numbers into the calculation. For this reason, there is a lower limit on the temperature at which the approximations may be used, depending on the specific values of material properties in a given application. In our case, this limit is ~135 K, which is why sintering timescales at lower temperatures were extrapolated using (31). Given some of the larger issues with the model (e.g., the vapor diffusion term, described above), the order of magnitude accuracy of (31) is sufficient for low temperature predictions at this time. It is also possible for the approximations to cause very low values of a given diffusion mechanism (Table 1) to have a negative value. Negative values of diffusion are not physical and should be set to zero. These are the least important mechanisms, and therefore in most cases will have a negligible effect on the result.

*Diffusion Mechanisms and Prefactor Values:* Neck growth during sintering of water ice is dominated by vapor diffusion ($\dot{V}_3$), and as such most of the other diffusion mechanisms included in Table 1 have only a small effect sintering timescales. To assess their relative influence, we performed the set of tests described above five times, each of which neglected one of the diffusion mechanisms $\dot{V}_{1,2,4-6}$. Most mechanisms had a negligible effect on all sintering timescales, suggesting that they may be neglected in applications of the model where only neck growth is of interest. The one exception was the surface diffusion term ($\dot{V}_1$), which was found to strongly influence $\tau_{25}$ and $\tau_{50}$ in some cases. For 1 $\mu$m grains at 250 K, neglecting surface diffusion only increased $\tau_{25}$ and $\tau_{50}$ by 13% and 1%, respectively. This results from the fact that surface diffusion dominates over vapor diffusion during the early stages of the process (Fig. 10). At 200 K, however, they were found to have a ~200% and ~3700% increase in $\tau_{25}$ and $\tau_{50}$, suggesting that the vapor diffusion mechanism has a stronger temperature dependence than surface diffusion. Grains 100 $\mu$m in radius showed no significant effects from neglecting surface diffusion on any timescales, suggesting that the effect increases in severity with decreasing radii.

We also performed similar tests to assess the sensitivity of the calculation to the values of the diffusion coefficient prefactors ($D_{ob}$, $D_{ov}$, $D_{os}$). Measured values of these parameters in the literature are not well constrained and can vary widely even at a given temperature. We used the maximum and minimum value of each prefactor found in the literature (Table 5) and compared the resulting timescales to those from the rest of the study, which used the (*) values. Similar to the results above, we found that these parameters have a negligible effect on the total predicted sintering timescale, regardless of temperature and grain size. However, for small grains at cold temperatures, $D_{os}$ (and to a lesser extent $D_{ob}$) influences $\tau_{25}$ by up to ~16%, and $\tau_{50}$ by a few percent.

Given that the influence of the surface diffusion becomes more pronounced at cold temperatures, we recommend always including the surface diffusion term when calculating neck growth timescales in planetary applications, especially in cases that are concerned with the early stages of the process, and where small grains are being considered. In such cases, the full range of values for $D_{os}$ and $D_{ob}$ may be used to constrain uncertainty in the predicted timescales. We emphasize, however, that the influence of these mechanisms and their prefactor values may be significantly more important in different types of ices (or other materials), different pressure environments, and during later stages of the process. In the latter case, the calculation will be dominated by mechanisms $\dot{V}_{4-6}$ and likely extremely sensitive to $D_{ov}$.

*Initial Density:* The starting density does have an influence over the rate at which sintering occurs, as more dense aggregates experience faster neck growth. For example, for 100 μm grains at 200 K, the total sintering timescale $\tau$ for an aggregate with an initial relative density of 0.3 is a factor of 2 longer than for a relative density of 0.5. The effect is more significant at higher densities, with an aggregate of density 0.3 having a timescale 10x longer than one of 0.8. Given that ice particles deposited on the surfaces of planetary bodies is typically thought to be very low density, we recommend using a low initial starting density, which will provide a more conservative estimate of sintering timescale.

*Timescale Power Laws:* While we had expected an exponential to provide the best fit to our results, the dependence of sintering timescale on temperature is best fit by a power law. The relationship given by (31) and associated parameter values (Table 6) each have an $R^2$ value of ≥0.9997. However, our results are not perfectly fit, and (31) underpredicts sintering timescales at warm temperatures for grains ≥10 μm in radius. The parameters in Table 6 give an order of magnitude match for $\tau$ to the model at temperatures ≤200 K for grains ≥10 μm, and up to 220 K for 1 μm grains. We do not recommend using (31) for temperatures warmer than this, where the power law provides a poor fit to the results. Timescales are underpredicted by one order of magnitude for grains of all sizes at temperatures 200>T<240 K, and one or more orders of magnitude at T>240 K. At these warmer temperatures, we recommend running the full calculation to predict timescales. Though, we also note that the model has the most uncertainty in this regime due to issues with the vapor diffusion term (more effective at warmer temperatures), as well as physical effects like premelting that may influence the process but are not accounted for in the model.

In contrast to the larger grains, the power law for 0.1 μm grains overpredicts sintering timescales, and only gives an order of magnitude match to the model at temperatures ≤180 K.

*Isolated Grain Pair Experiments:* The neck size between grains (Fig. 11) was measured using a straight-line tool in the Stream image analysis software for our microscope. It can be challenging to measure when very small, so we took five measurements and averaged them, using the minimum and maximum values to determine the neck uncertainty (Table B1). The grain radii were only measured once, using the 3-point circle method. Since the grains are not perfectly spherical, we tried to place the points in the same locations for each measurement and assume an uncertainty of ±1 μm. Using the uncertainty in both the grain radius and neck size, we calculated the maximum and minimum value for the relative neck size for both large and small grains, and took the difference from the average as our uncertainty. In cases where the relative neck size had different positive and negative uncertainty values, the largest (absolute) value was chosen.

Uncertainty values are generally larger for the measurements taken at -5 C, as it was difficult to capture good images due to fogging on the cryostage window caused by the difference in internal and external temperature, and a very high humidity that day. If the saturation ice we put inside the cryostage did not fully saturate the atmosphere above the slide, the higher ambient humidity could have also influenced sintering rates measured in the experiment. However, such an effect would likely increase sintering rates, and Figure 12 (left) shows slower measured values relative to the model prediction. That being said, the uncertainty in the vapor diffusion calculate described in the previous section has not yet been quantified, and is not reflected in the numbers in Table B1.

*Bulk Ice Experiments:* For each observation, ice grains were collected from the samples using an LN2-cooled scoop and placed into a Linkam LTS350 cryostage at -100 C for imaging with the Olympus BX51 microscope. Several images were taken of the grains at different locations on the microscope slide (see Fig. 13 and Supporting Information). For the Cold Sample, grains in the images were measured using a 2-point circle method for quasi-spherical grains, or closed polygon method when crystalline facets were distinguishable (typically in the later datasets). Grain counts for the two, four, six, and ten week samples were 358, 788, 693, and 675, respectively. The same method was used for the fresh, starting ice, which had a grain count of 1551. For the Warm Sample, grains were measured using a closed polygon method. Grain counts for the two, four, and six week samples were 343, 207, and 610, respectively. Uncertainty in the measurement of individual grain radii is assumed to be negligible, as small errors should balance out over the many hundreds of grains that were measured. To assess the potential for any systematic bias in the measurements, several coauthors conducted grain size analysis on the same five images containing a total of ~250 grains; this exercise showed an uncertainty in mean grain size of $\pm 1.25$ $\mu$m. The mean grain sizes and standard deviations are given in Table B2.

During the first observation, grains were collected from the top of the sample in the bucket. Since the top of the sample is briefly exposed to warmer temperatures during collection, concerns arose that this may enhance sintering rates relative to grains at depth. During the second observation, a comparison between grains collected from both the top and at depth (grain counts of 207 and 383, respectively) showed that the distribution of the former had an average radius ~7 $\mu$m larger than the latter. The same test performed on a batch of freshly created ice revealed that the distribution of the top had an average radius ~4 $\mu$m larger than at depth (grain counts of 413 and 424, respectively) which was inherent as a result of the sample creation. It is likely that the larger discrepancy in the sintered versus fresh ice is the result of both ice creation and exposure to warmer temperatures during observations. During all subsequent observations, grains were collected from depth in the samples, and overall uncertainty in the grain distribution mean is assumed to be $\pm 5$ $\mu$m. Given the limited scope of these experiments, this has only a minor effect on our results and does not qualitatively influence our conclusions.

*Table B1. Measurements obtained during the isolated grain pair experiments described in section 4.*

| | Time (min) | Neck Size (μm) | Lg. Diameter (μm) | Sm. Diameter (μm) | Uncertainty in Relative Neck Size |
|---|---|---|---|---|---|
| -20 C | 0 | 32.81 (+2.3 / -2.4) | 202.00 | 145.00 | lg. ($\pm$ 0.01), sm. ($\pm$ 0.02) |
| | 2 | 33.60 (+3.9 / -2.4) | 199.11 | 142.64 | lg. ($\pm$ 0.02), sm. ($\pm$ 0.03) |
| | 5 | 37.16 (+0.7 / -0.6) | 199.22 | 143.85 | lg. ($\pm$ 0.004), sm. ($\pm$ 0.01) |
| | 9 | 44.68 (+1.4 / -1.7) | 198.30 | 144.87 | lg. ($\pm$ 0.01), sm. ($\pm$ 0.01) |
| | 18 | 46.84 (+3.1 / -2.5) | 199.75 | 142.94 | lg. ($\pm$ 0.02), sm. ($\pm$ 0.02) |
| | 29 | 54.18 (+1.6 / -3.2) | 195.84 | 142.81 | lg. ($\pm$ 0.02), sm. ($\pm$ 0.02) |
| | 40 | 58.00 (+1.1 / -2.1) | 197.78 | 142.41 | lg. ($\pm$ 0.01), sm. ($\pm$ 0.02) |
| | 57 | 60.54 (+1.2 / -1.1) | 196.53 | 141.52 | lg. ($\pm$ 0.01), sm. ($\pm$ 0.01) |
| | 78 | 64.78 (+2.0 / -1.7) | 193.90 | 144.00 | lg. ($\pm$ 0.01), sm. ($\pm$ 0.02) |
| -5 C | 0 | 32.51 (+3.7 / -2.8) | 197 | 153 | lg. ($\pm$ 0.08), sm. ($\pm$ 0.08) |
| | 26 | 41.57 (+0.7 / -0.5) | 191 | 149 | lg. ($\pm$ 0.02), sm. ($\pm$ 0.12) |
| | 48 | 44.00 (+1.1 / -1.0) | 179 | 139 | lg. ($\pm$ 0.02), sm. ($\pm$ 0.14) |

*Table B2.*

| Sample | # Days | Radius (μm) | Standard Deviation (μm) |
|---|---|---|---|
| Starting Ice | 1 | 6.57 | 5.57 |
| -80 C, 2 wk | 14 | 10.22 | 5.91 |
| -80 C, 4 wk | 26 | 11.68 | 6.54 |
| -80 C, 6 wk | 40 | 17.23 | 10.96 |
| -80 C, 10 wk | 70 | 18.44 | 9.57 |
| -20 C, 2 wk | 12 | 36.89 | 14.08 |
| -20 C, 2 wk (bottom) | 24 | 38.84 | 13.35 |
| -20 C, 4 wk (top) | 24 | 42.55 | 13.51 |
| -20 C, 6 wk | 38 | 45.11 | 17.67 |

*Acknowledgements: Part of this work was supported by an appointment to the NASA Postdoctoral Program at the Jet Propulsion Laboratory (JPL), California Institute of Technology, under contract with NASA, administered by the Universities Space Research Association through a contract with NASA. We also acknowledge support from the JPL Advanced Concepts program.*

*The data used to generate the figures are included in the Supplemental Information, as well as additional images from the bulk ice experiments.*

# The microstructural evolution of water ice in the solar system through sintering

J. L. Molaro[1,2], M. Choukroun[2], C. B. Phillips[2], E. S. Phelps[2], R. Hodyss[2], K. L. Mitchell[2], J. M. Lora[3], and G. Meirion-Griffith[2]

[1]Planetary Science Institute, 1700 East Fort Lowell, Suite 106, Tucson, AZ 85719, USA.

[2]Jet Propulsion Laboratory, California Institute of Technology, 4800 Oak Grove Drive, Pasadena, CA 91109, USA.

[3]University of California, Los Angeles, 595 Charles Young Drive East, Los Angeles, CA 90095, USA.

## Introduction

This file contains supplementary information regarding the Bulk Ice experiment described in section 4.1 of the paper in the form of images (pages 2-9) and tables (pages 10-16).

The images (pages 2-9) represent a subset of those used to count grains and grain radii, provided here for a more detailed view of our observations. In the paper text, we presented the images in black and white to eliminate any confusion readers may have regarding their odd color. These we present in color, as some aspects and details of the grain structures may be easier to observe. The lighting conditions under the microscope are difficult to control and the contrast has been enhanced, so the color may be inconsistent across images. The color itself is not physically meaningful.

The tables on pages 10-16 provide the statistical data from our grain counts during each observation in the experiment, and supporting Figures S1-S3 accompany them to display their contents. These data feed into Figure 14 of the main text.

The Tables on pages 17-27 provide the numerical and model generated data plotted in Figures 6-10, 12, 15-17 of the main text.

Fresh Ice 50 μm

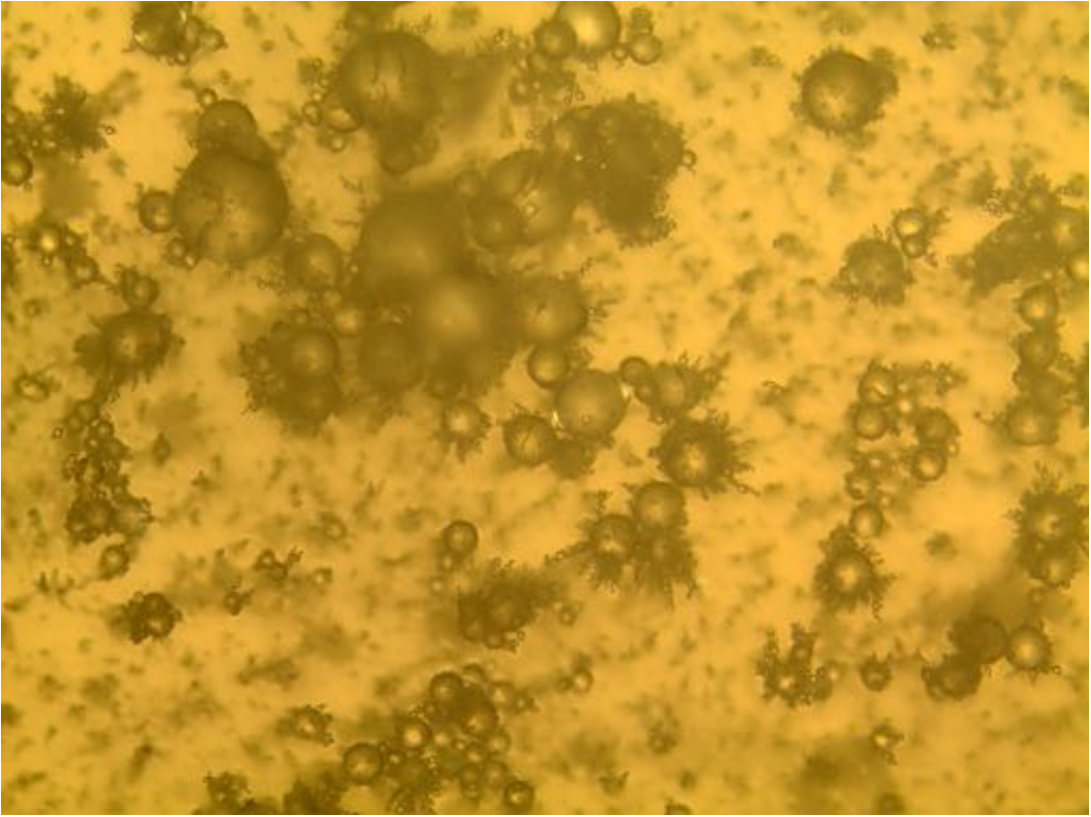

Cold Sample (-80 C) after two weeks       ▬ 50 μm

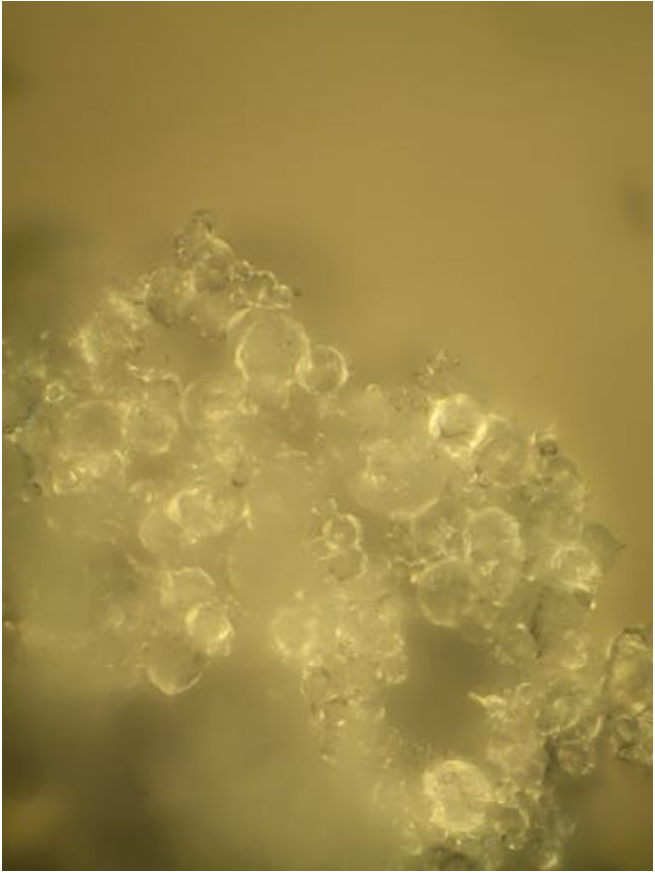
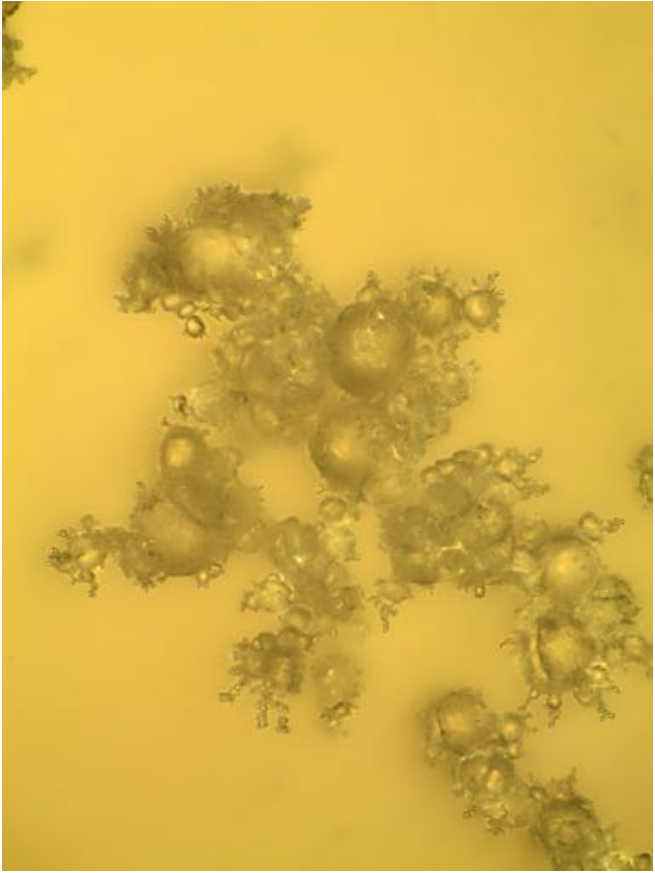
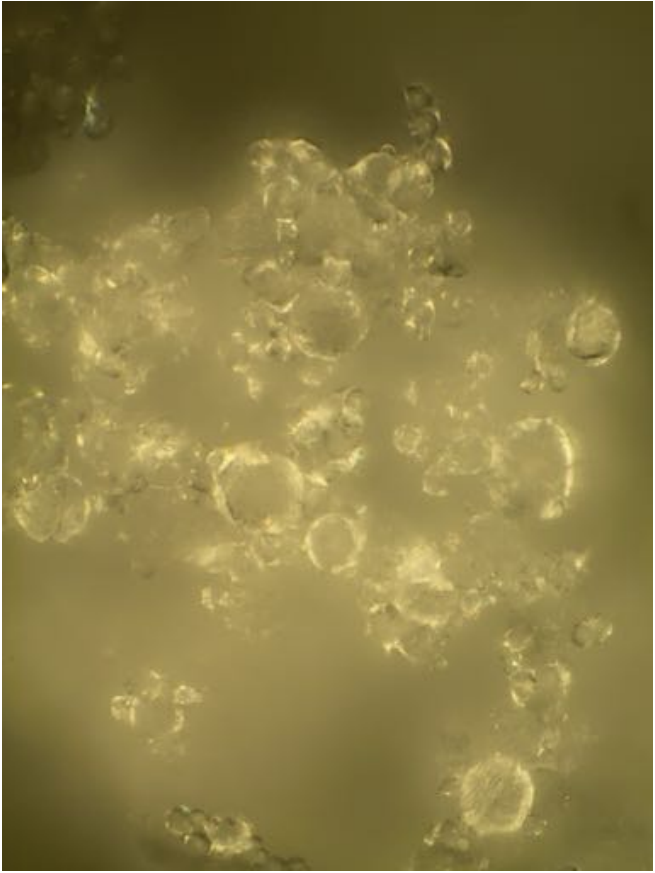
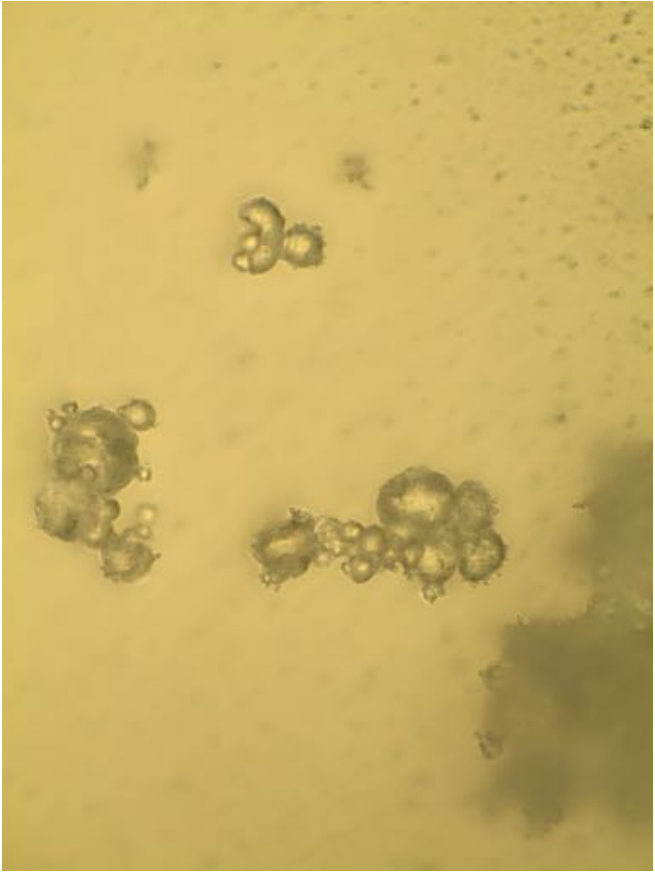

Cold Sample (-80 C) after four weeks — 50 μm

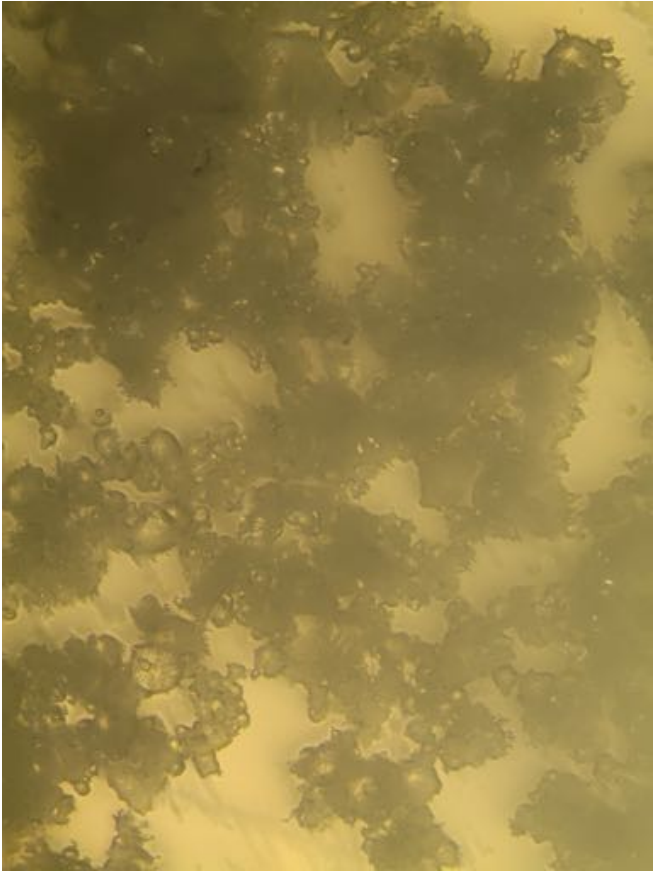
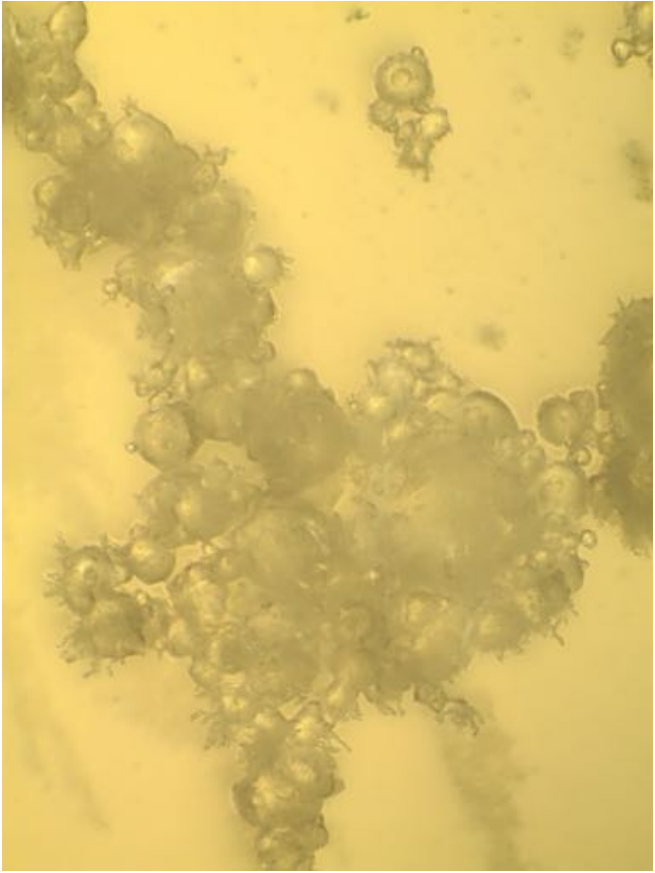
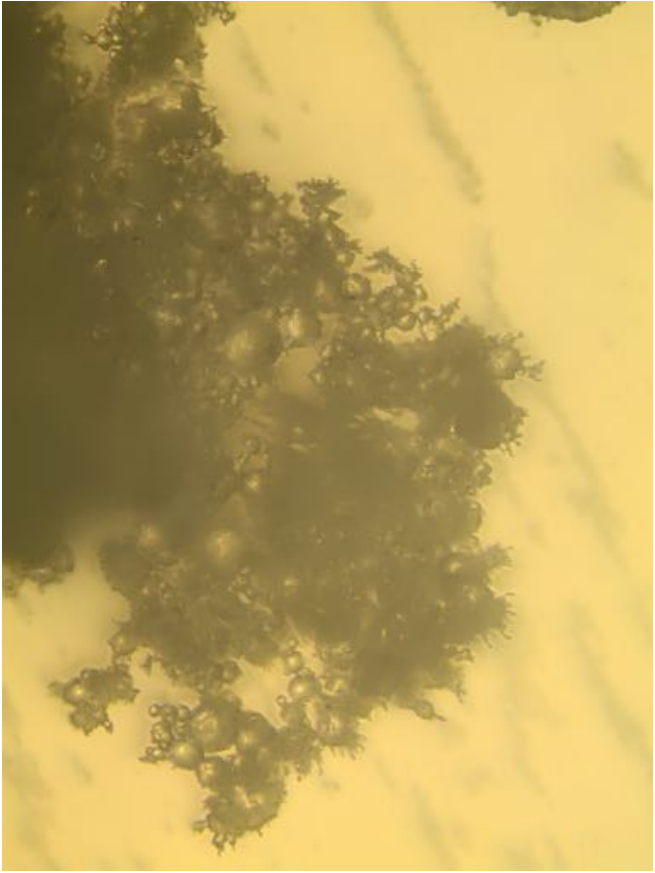
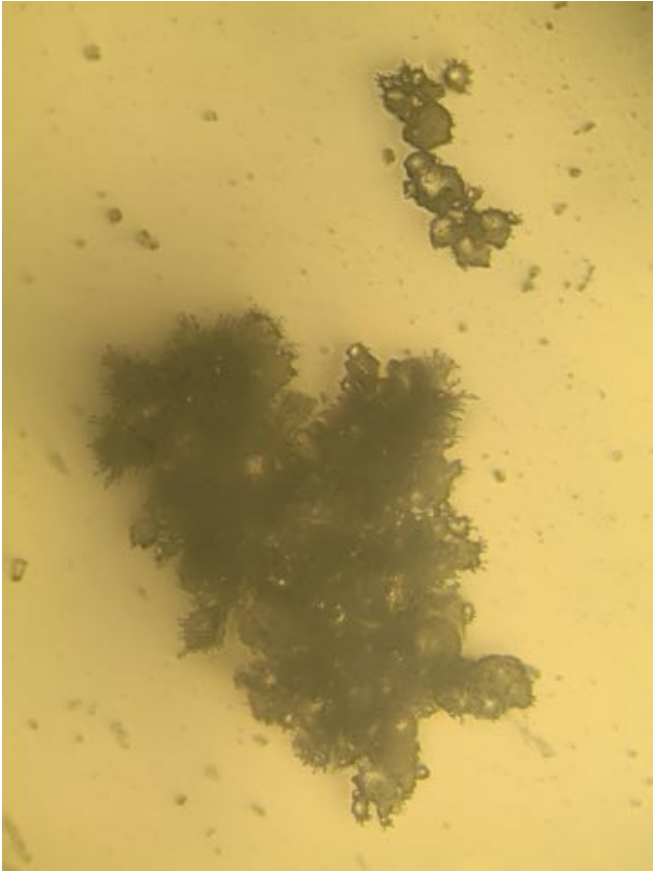

Cold Sample (-80 C) after six weeks ─── 100 μm

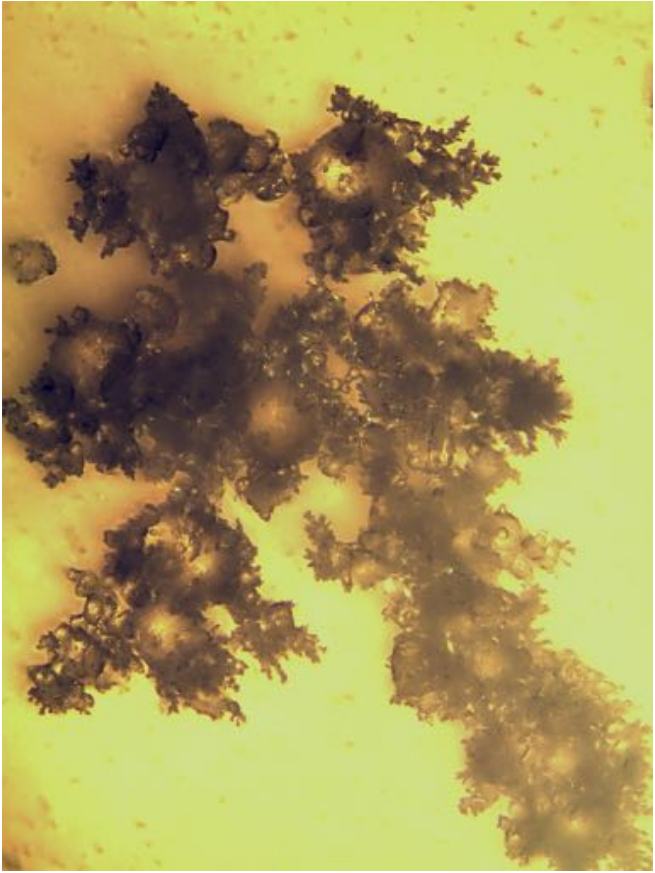
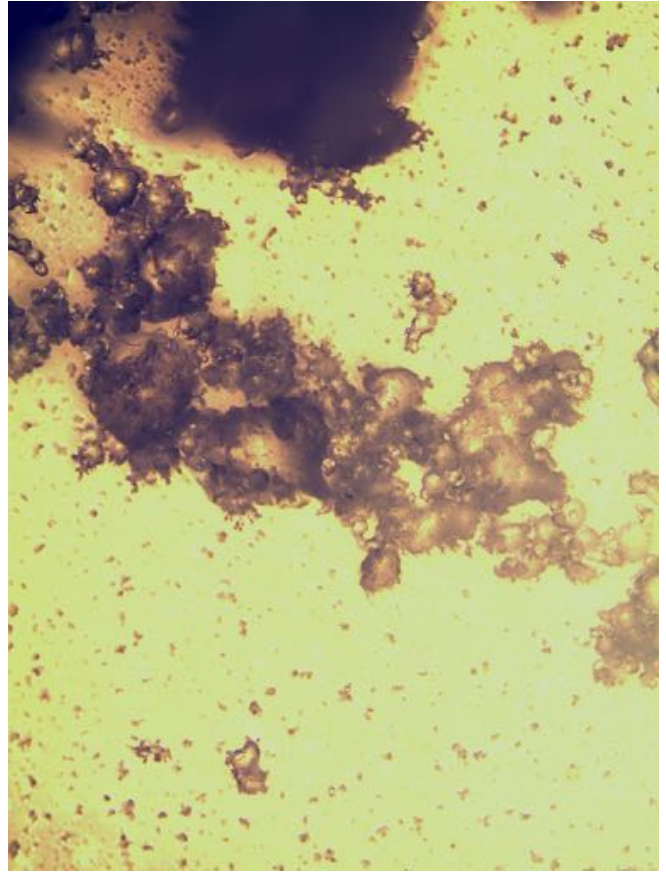
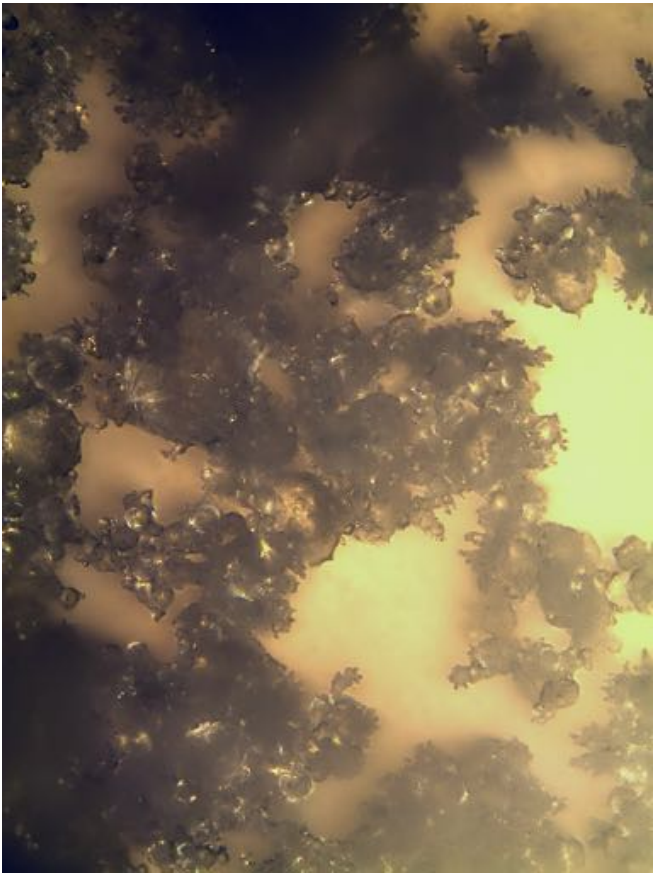
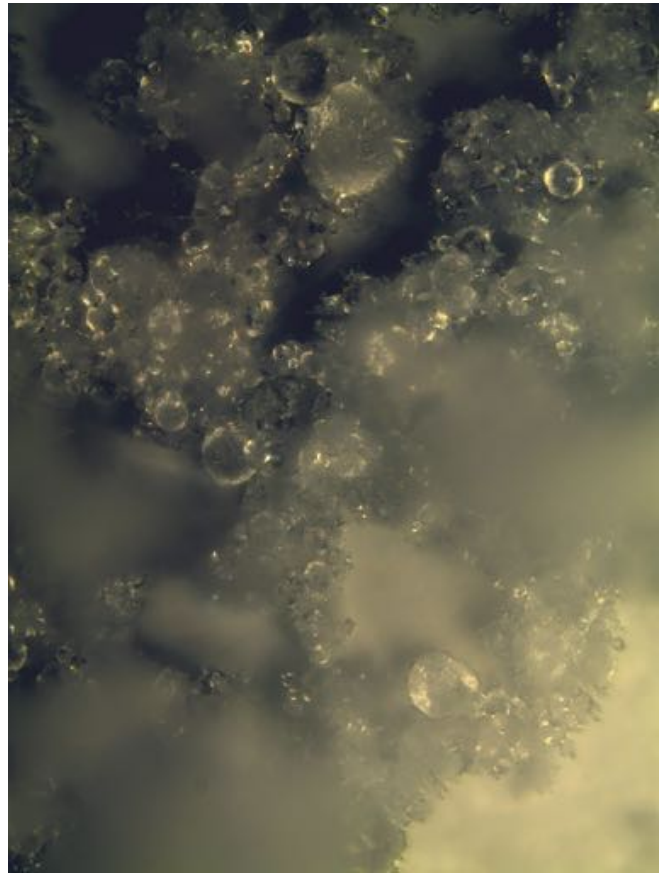

Cold Sample (-80 C) after ten weeks ━ 100 μm

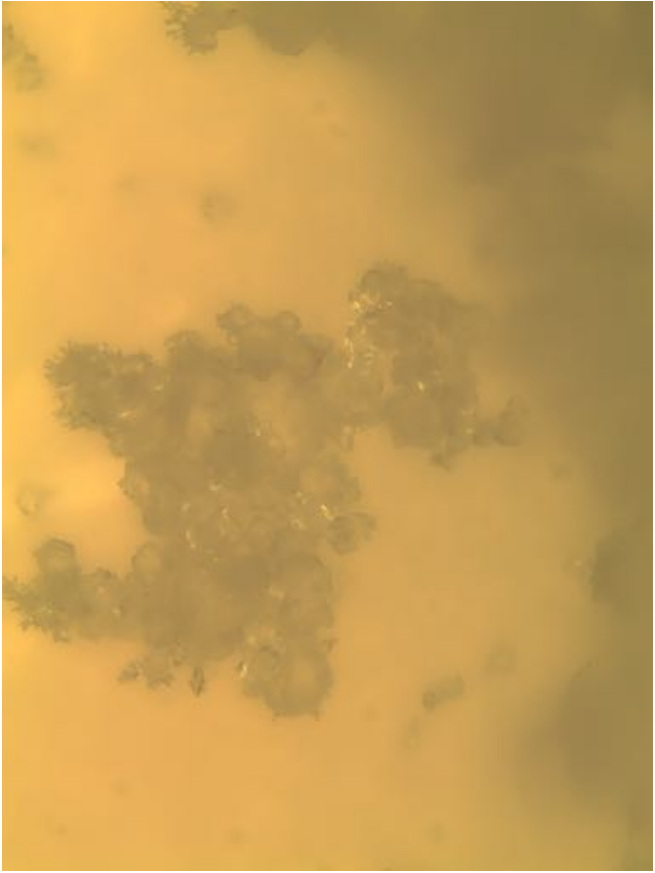
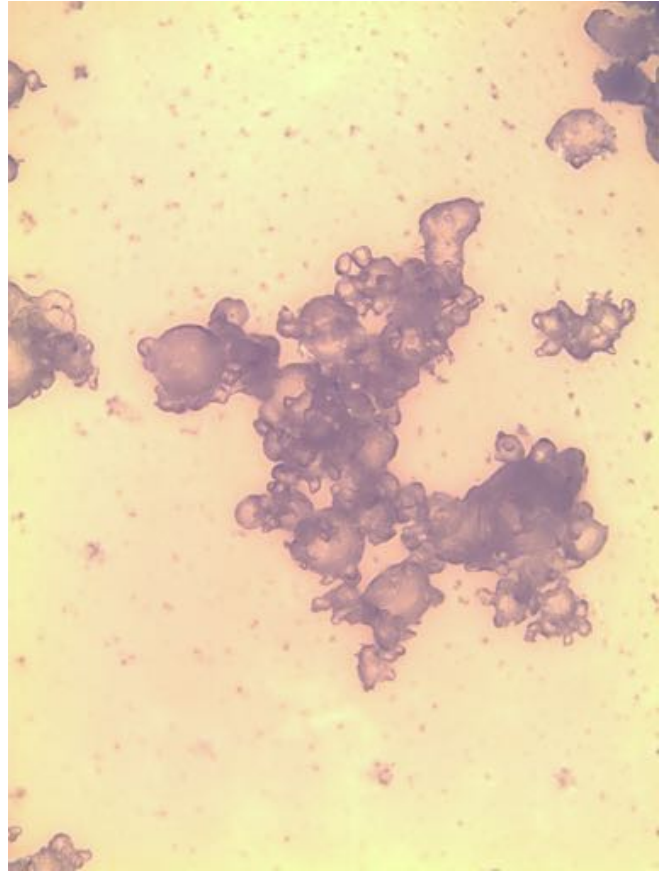
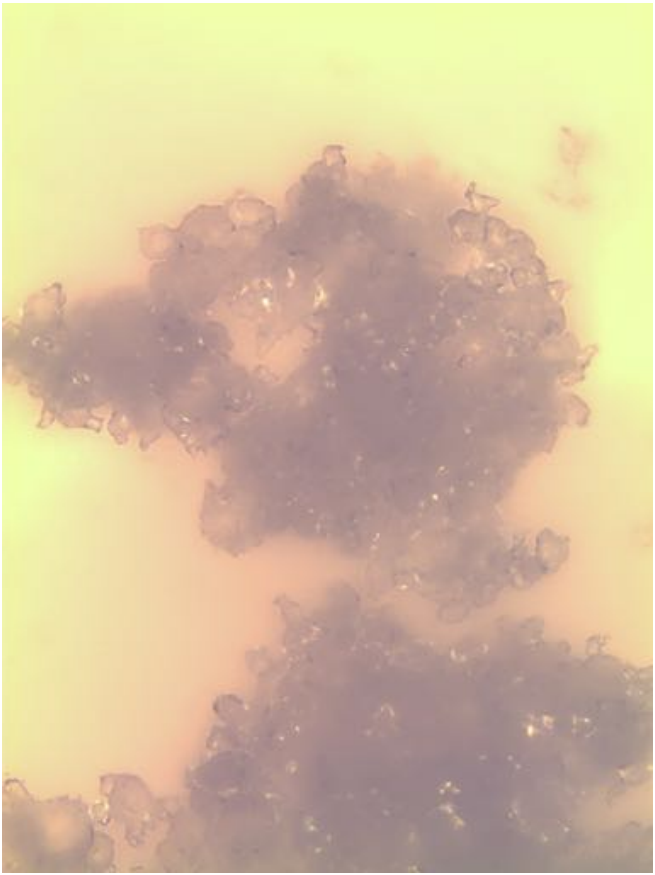
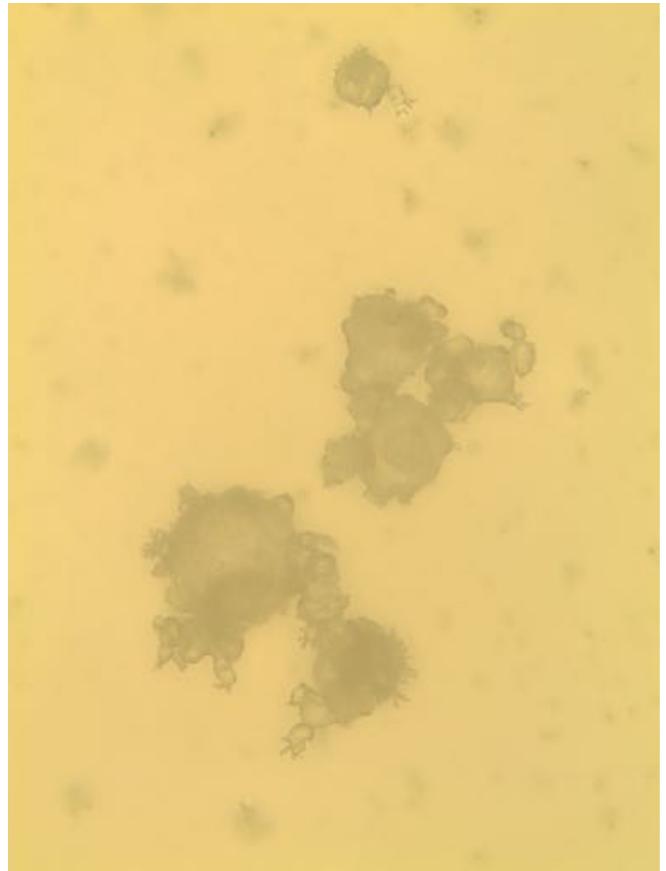

Warm Sample (-20 C) after two weeks      ———100 μm

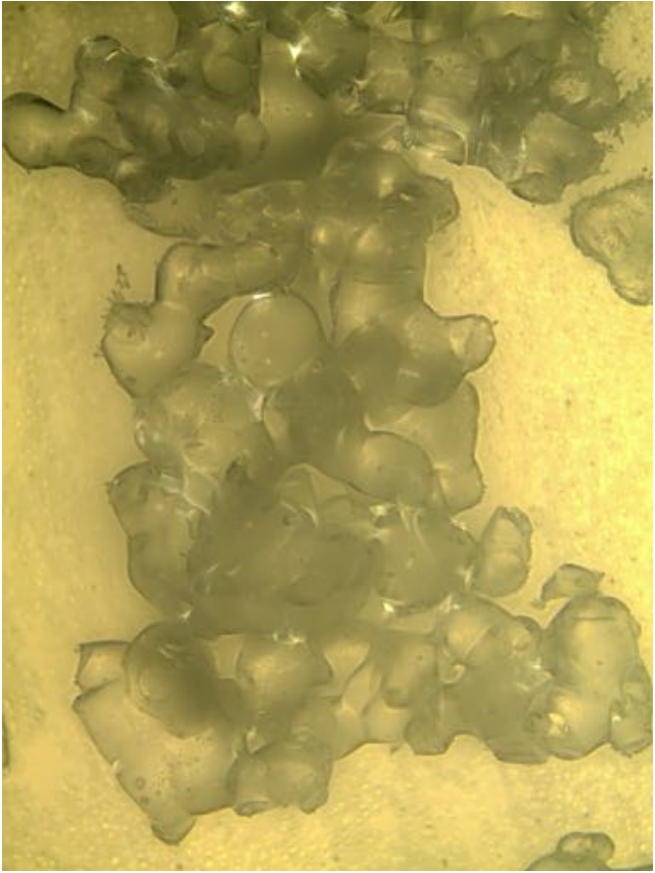
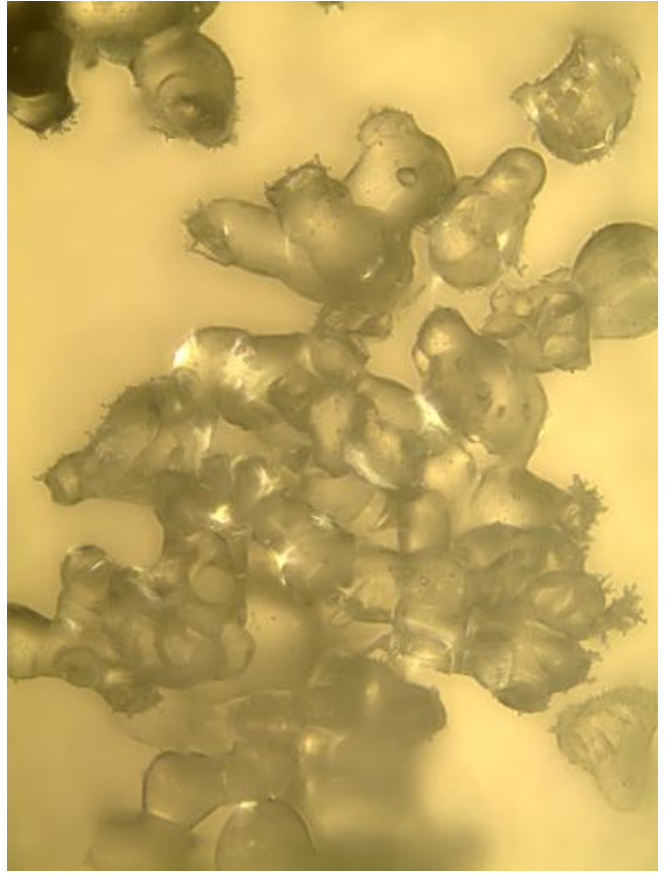
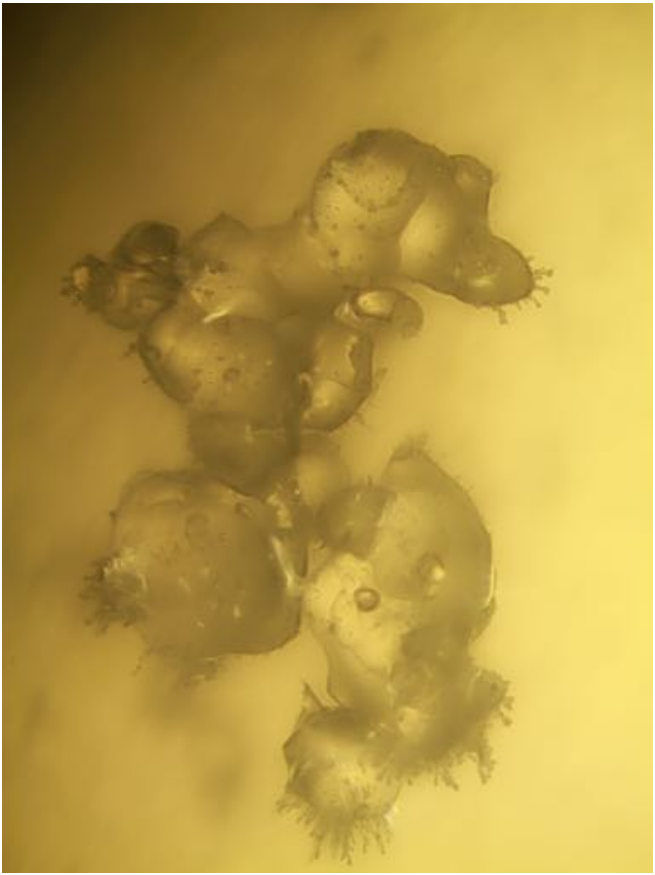
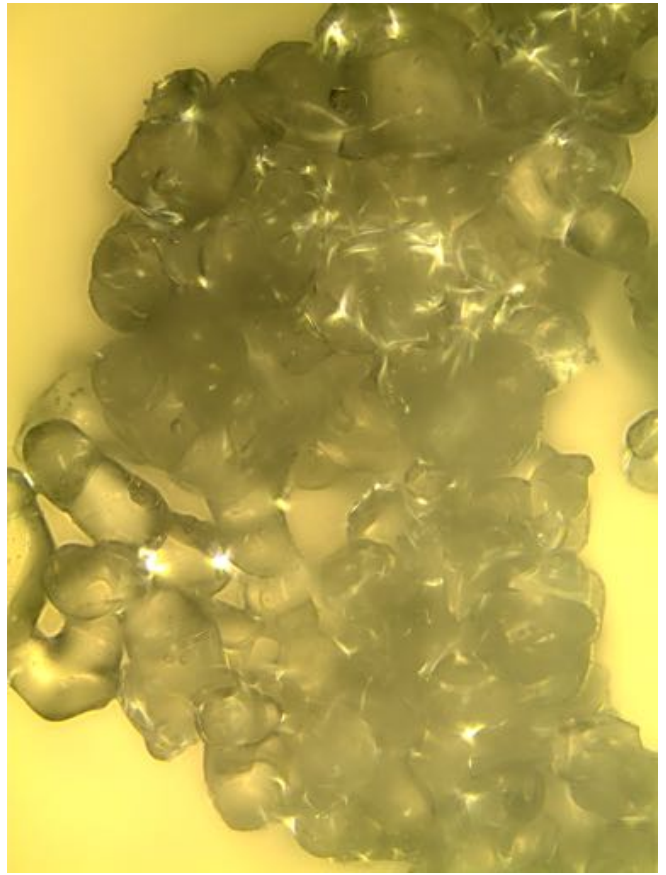

Warm Sample (-20 C) after four weeks

100 μm

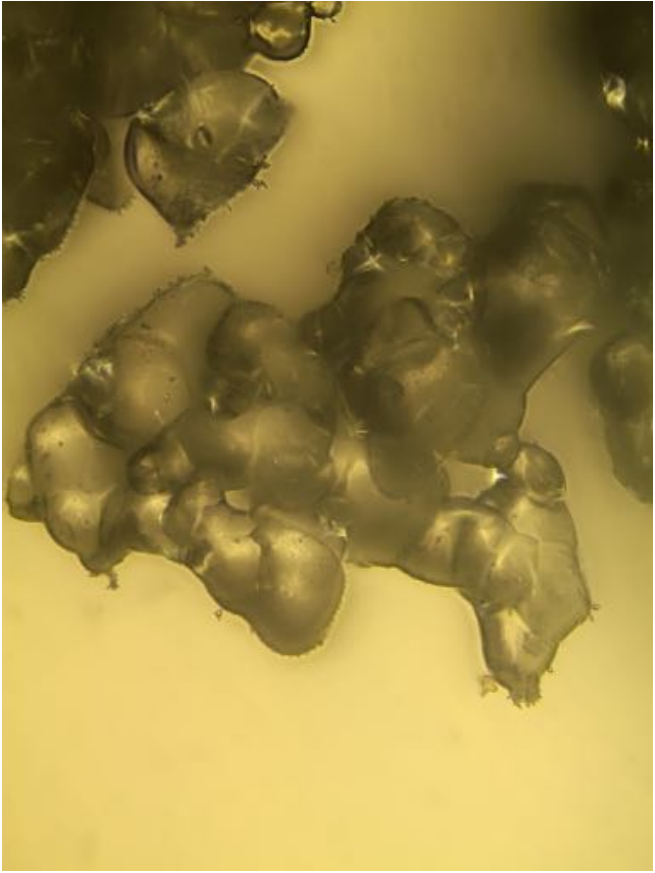
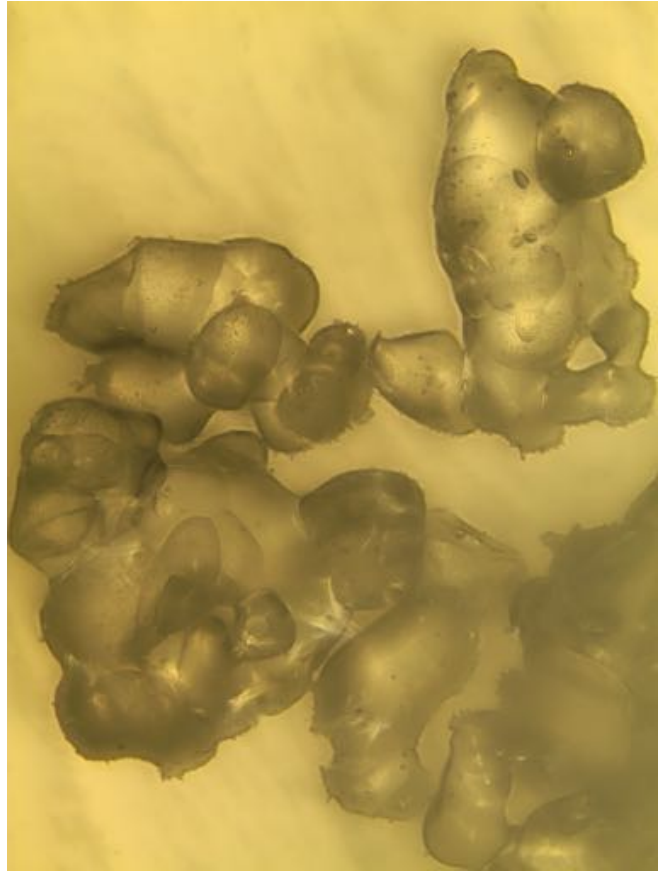
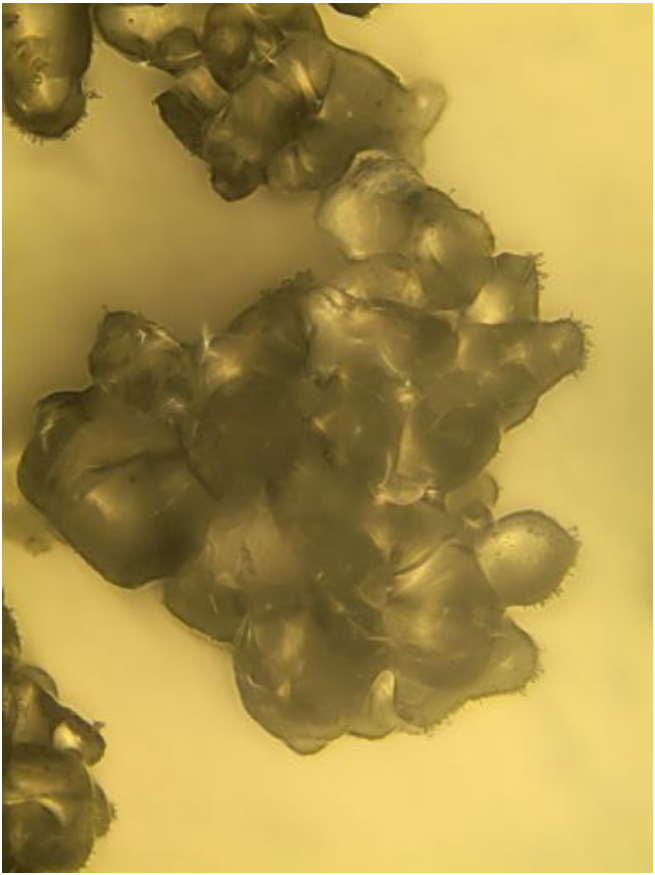
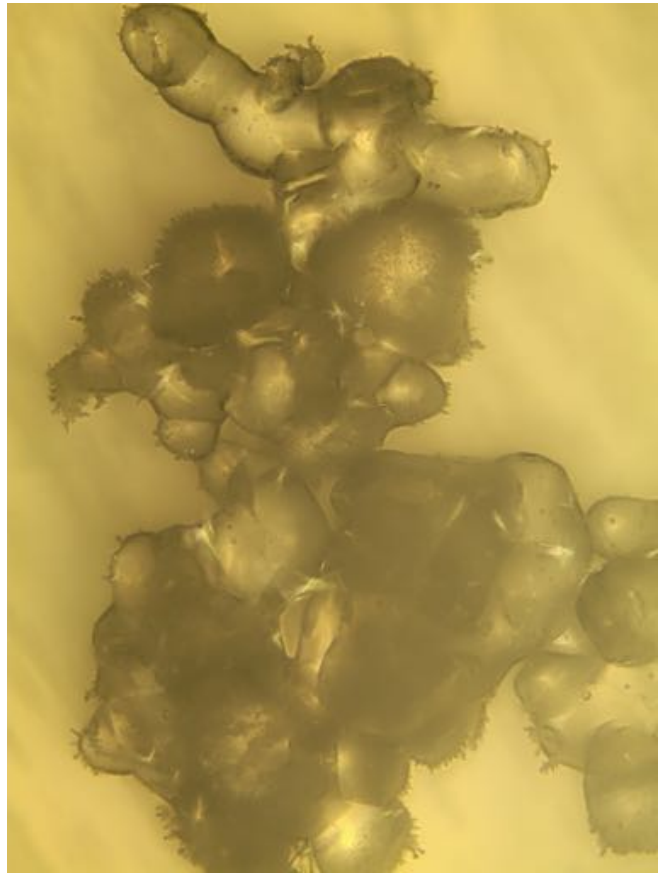

Warm Sample (-20 C) after six weeks     —— 100 μm

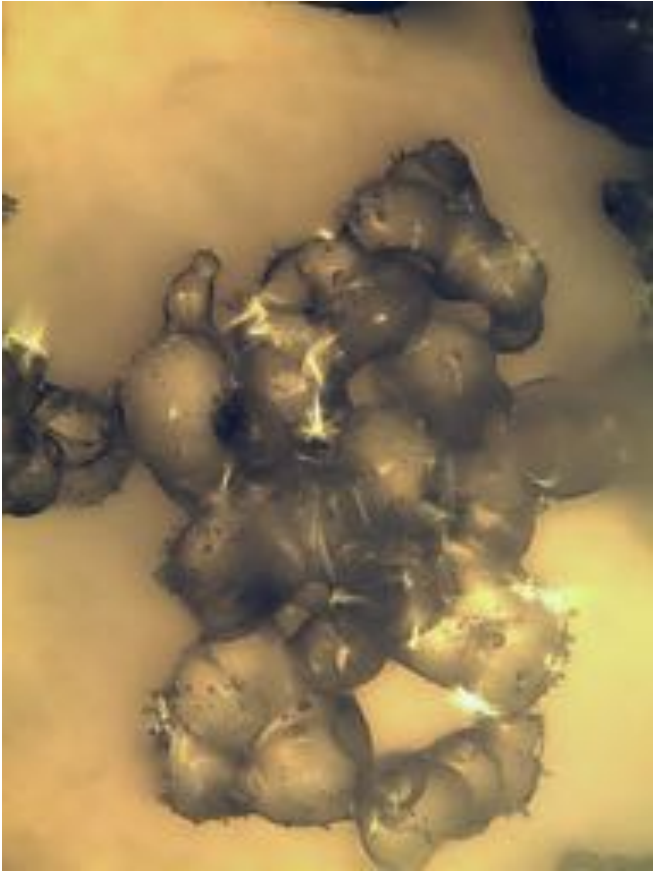
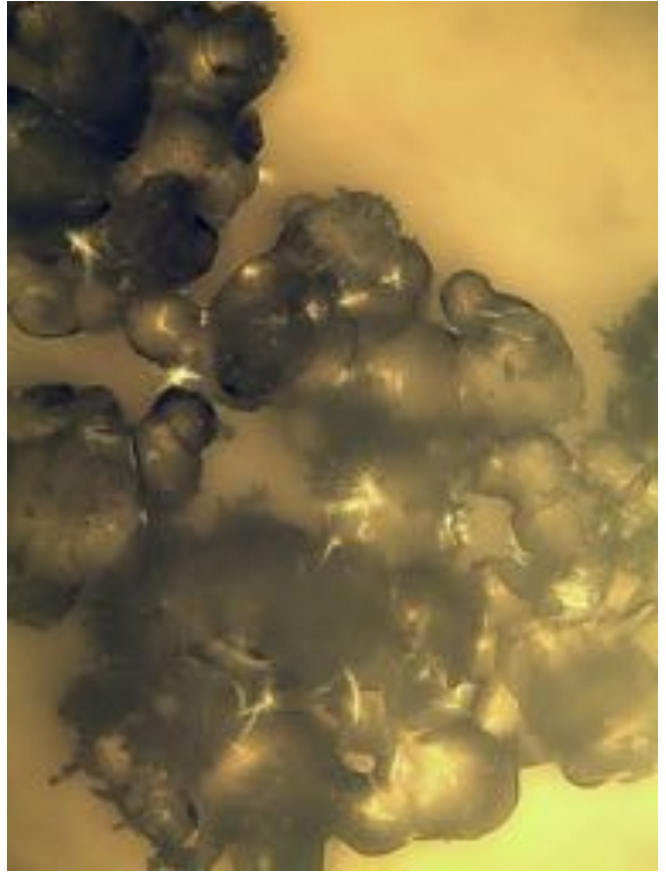
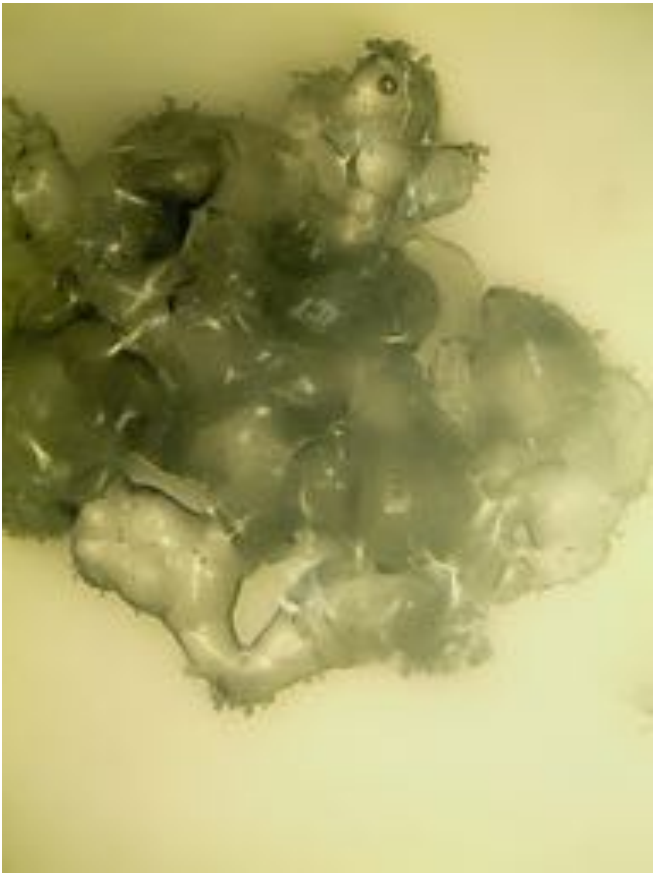
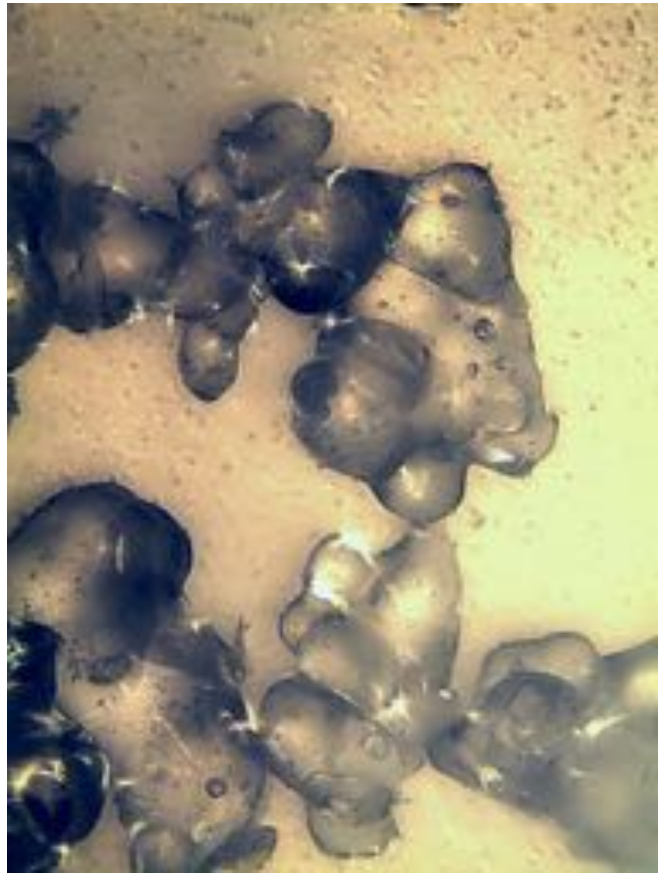

Cold Sample (-80 C)

| Radius Bin (µm) | Starting Ice | | Two Weeks | | Four Weeks | |
|---|---|---|---|---|---|---|
| | Frequency | Cumulative % | Frequency | Cumulative % | Frequency | Cumulative % |
| 0 | 0 | 0% | 0 | 0% | 0 | 0% |
| 1.25 | 42 | 2.71% | 3 | 0.84% | 0 | 0.00% |
| 2.5 | 273 | 20.31% | 13 | 4.47% | 4 | 0.51% |
| 3.75 | 259 | 37.01% | 17 | 9.22% | 13 | 2.16% |
| 5 | 208 | 50.42% | 24 | 15.92% | 40 | 7.23% |
| 6.25 | 166 | 61.12% | 41 | 27.37% | 75 | 16.75% |
| 7.5 | 138 | 70.02% | 38 | 37.99% | 113 | 31.09% |
| 8.75 | 101 | 76.53% | 36 | 48.04% | 77 | 40.86% |
| 10 | 77 | 81.50% | 38 | 58.66% | 77 | 50.63% |
| 11.25 | 61 | 85.43% | 31 | 67.32% | 74 | 60.03% |
| 12.5 | 39 | 87.94% | 22 | 73.46% | 53 | 66.75% |
| 13.75 | 37 | 90.33% | 14 | 77.37% | 41 | 71.95% |
| 15 | 23 | 91.81% | 11 | 80.45% | 32 | 76.02% |
| 16.25 | 30 | 93.75% | 16 | 84.92% | 33 | 80.20% |
| 17.5 | 26 | 95.42% | 11 | 87.99% | 26 | 83.50% |
| 18.75 | 12 | 96.20% | 7 | 89.94% | 19 | 85.91% |
| 20 | 11 | 96.91% | 5 | 91.34% | 26 | 89.21% |
| 21.25 | 8 | 97.42% | 9 | 93.85% | 11 | 90.61% |
| 22.5 | 10 | 98.07% | 2 | 94.41% | 19 | 93.02% |
| 23.75 | 6 | 98.45% | 9 | 96.93% | 6 | 93.78% |
| 25 | 7 | 98.90% | 4 | 98.04% | 10 | 95.05% |
| 26.25 | 4 | 99.16% | 0 | 98.04% | 5 | 95.69% |
| 27.5 | 2 | 99.29% | 1 | 98.32% | 4 | 96.19% |
| 28.75 | 1 | 99.36% | 1 | 98.60% | 7 | 97.08% |
| 30 | 2 | 99.48% | 1 | 98.88% | 4 | 97.59% |
| 31.25 | 1 | 99.55% | 1 | 99.16% | 2 | 97.84% |
| 32.5 | 0 | 99.55% | 1 | 99.44% | 5 | 98.48% |
| 33.75 | 1 | 99.61% | 0 | 99.44% | 4 | 98.98% |
| 35 | 2 | 99.74% | 2 | 100.00% | 1 | 99.11% |
| 36.25 | 0 | 99.74% | 0 | 100.00% | 1 | 99.24% |
| 37.5 | 1 | 99.81% | 0 | 100.00% | 2 | 99.49% |
| 38.75 | 0 | 99.81% | 0 | 100.00% | 1 | 99.62% |
| 40 | 0 | 99.81% | 0 | 100.00% | 0 | 99.62% |
| 41.25 | 0 | 99.81% | 0 | 100.00% | 0 | 99.62% |
| 42.5 | 0 | 99.81% | 0 | 100.00% | 3 | 100.00% |
| 43.75 | 0 | 99.81% | 0 | 100.00% | 0 | 100.00% |
| 45 | 0 | 99.81% | 0 | 100.00% | 0 | 100.00% |
| 46.25 | 0 | 99.81% | 0 | 100.00% | 0 | 100.00% |
| 47.5 | 0 | 99.81% | 0 | 100.00% | 0 | 100.00% |
| 48.75 | 0 | 99.81% | 0 | 100.00% | 0 | 100.00% |
| 50 | 0 | 99.81% | 0 | 100.00% | 0 | 100.00% |
| >50 | 3 | 100.00% | 0 | 100.00% | 0 | 100.00% |
| *Mean Radius (µm)* | 6.57 | | 10.23 | | 11.69 | |
| *Standard Deviation* | 5.57 | | 5.92 | | 6.55 | |

Cold Sample (-80 C)

| Six Weeks | | Ten Weeks | |
|---|---|---|---|
| *Frequency* | *Cumulative %* | *Frequency* | *Cumulative %* |
| 0 | 0% | 0 | 0.00% |
| 2 | 0.29% | 0 | 0.00% |
| 2 | 0.58% | 0 | 0.00% |
| 0 | 0.58% | 0 | 0.00% |
| 5 | 1.30% | 4 | 0.59% |
| 15 | 3.46% | 8 | 1.78% |
| 22 | 6.64% | 6 | 2.67% |
| 42 | 12.70% | 28 | 6.81% |
| 56 | 20.78% | 51 | 14.37% |
| 56 | 28.86% | 51 | 21.93% |
| 58 | 37.23% | 55 | 30.07% |
| 57 | 45.45% | 58 | 38.67% |
| 44 | 51.80% | 46 | 45.48% |
| 51 | 59.16% | 45 | 52.15% |
| 38 | 64.65% | 34 | 57.19% |
| 30 | 68.98% | 40 | 63.11% |
| 34 | 73.88% | 29 | 67.41% |
| 30 | 78.21% | 25 | 71.11% |
| 23 | 81.53% | 26 | 74.96% |
| 19 | 84.27% | 16 | 77.33% |
| 17 | 86.72% | 21 | 80.44% |
| 7 | 87.73% | 17 | 82.96% |
| 13 | 89.61% | 13 | 84.89% |
| 4 | 90.19% | 10 | 86.37% |
| 8 | 91.34% | 16 | 88.74% |
| 7 | 92.35% | 9 | 90.07% |
| 4 | 92.93% | 6 | 90.96% |
| 6 | 93.80% | 7 | 92.00% |
| 6 | 94.66% | 6 | 92.89% |
| 2 | 94.95% | 10 | 94.37% |
| 1 | 95.09% | 6 | 95.26% |
| 1 | 95.24% | 3 | 95.70% |
| 3 | 95.67% | 3 | 96.15% |
| 3 | 96.10% | 5 | 96.89% |
| 2 | 96.39% | 3 | 97.33% |
| 1 | 96.54% | 2 | 97.63% |
| 0 | 96.54% | 1 | 97.78% |
| 1 | 96.68% | 2 | 98.07% |
| 2 | 96.97% | 3 | 98.52% |
| 3 | 97.40% | 1 | 98.67% |
| 0 | 97.40% | 2 | 98.96% |
| 18 | 100.00% | 7 | 100.00% |
| 17.24 | | 18.45 | |
| 10.96 | | 9.57 | |

Cold Sample (-80 C)

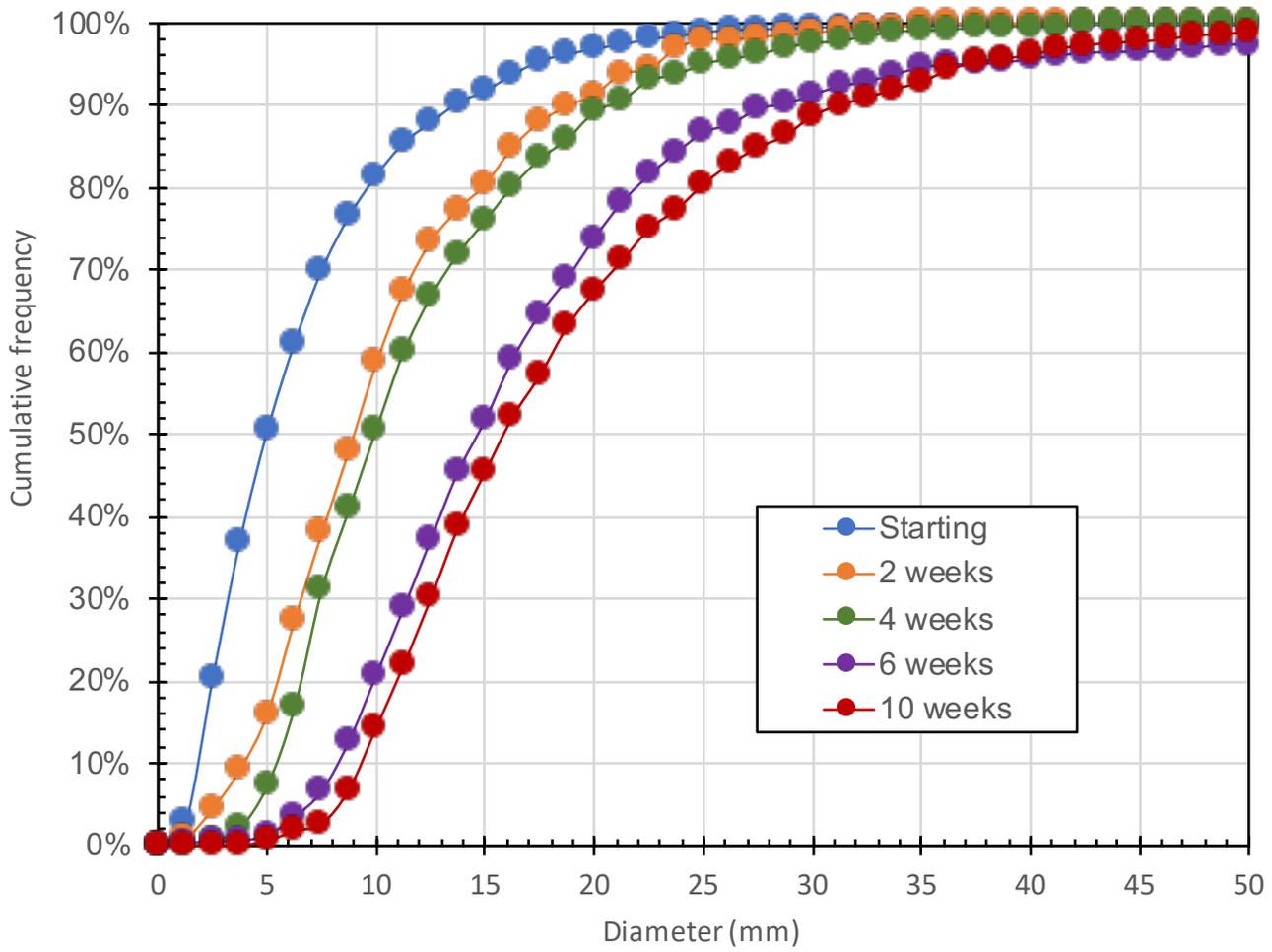

Figure S1. Grain statistics over time for the Cold Sample.

Warm Sample (-20 C)

| Radius Bin (μm) | Two Weeks | | Four Weeks (top) | | Four Weeks (bottom) | |
|---|---|---|---|---|---|---|
| | Frequency | Cumulative % | Frequency | Cumulative % | Frequency | Cumulative % |
| 0 | 0 | 0% | 0 | 0% | 0 | 0% |
| 2.5 | 0 | 0% | 0 | 0% | 0 | 0% |
| 5 | 2 | 1% | 0 | 0% | 0 | 0% |
| 7.5 | 3 | 1% | 0 | 0% | 0 | 0% |
| 10 | 1 | 2% | 0 | 0% | 0 | 0% |
| 12.5 | 3 | 3% | 0 | 0% | 0 | 0% |
| 15 | 4 | 4% | 0 | 0% | 0 | 0% |
| 17.5 | 10 | 7% | 5 | 1% | 0 | 0% |
| 20 | 12 | 10% | 16 | 5% | 1 | 0% |
| 22.5 | 13 | 14% | 9 | 8% | 7 | 4% |
| 25 | 11 | 17% | 14 | 11% | 10 | 9% |
| 27.5 | 28 | 25% | 25 | 18% | 11 | 14% |
| 30 | 30 | 34% | 29 | 26% | 8 | 18% |
| 32.5 | 26 | 42% | 37 | 35% | 10 | 23% |
| 35 | 22 | 48% | 33 | 44% | 16 | 30% |
| 37.5 | 26 | 56% | 33 | 52% | 18 | 39% |
| 40 | 21 | 62% | 26 | 59% | 19 | 48% |
| 42.5 | 22 | 68% | 38 | 69% | 16 | 56% |
| 45 | 24 | 75% | 18 | 74% | 13 | 62% |
| 47.5 | 16 | 80% | 13 | 77% | 9 | 67% |
| 50 | 14 | 84% | 14 | 81% | 10 | 71% |
| 52.5 | 9 | 87% | 20 | 86% | 8 | 75% |
| 55 | 11 | 90% | 10 | 89% | 15 | 83% |
| 57.5 | 6 | 92% | 7 | 91% | 8 | 86% |
| 60 | 5 | 93% | 6 | 92% | 8 | 90% |
| 62.5 | 7 | 95% | 6 | 94% | 5 | 93% |
| 65 | 3 | 96% | 9 | 96% | 4 | 95% |
| 67.5 | 4 | 97% | 3 | 97% | 2 | 96% |
| 70 | 2 | 98% | 1 | 97% | 2 | 97% |
| 72.5 | 4 | 99% | 1 | 97% | 1 | 97% |
| 75 | 1 | 99% | 2 | 98% | 3 | 99% |
| 77.5 | 1 | 99% | 2 | 98% | 1 | 99% |
| 80 | 0 | 99% | 3 | 99% | 0 | 99% |
| 82.5 | 0 | 99% | 1 | 99% | 1 | 100% |
| 85 | 1 | 100% | 0 | 99% | 0 | 100% |
| 87.5 | 1 | 100% | 0 | 99% | 0 | 100% |
| 90 | 0 | 100% | 1 | 100% | 0 | 100% |
| 92.5 | 0 | 100% | 0 | 100% | 0 | 100% |
| 95 | 0 | 100% | 0 | 100% | 0 | 100% |
| 97.5 | 0 | 100% | 1 | 100% | 0 | 100% |
| 100 | 0 | 100% | 0 | 100% | 1 | 100% |
| >100 | 0 | 100% | 0 | 100% | 0 | 100% |
| Mean Radius (μm) | 36.90 | | 42.56 | | 38.85 | |
| Standard Deviation | 14.09 | | 13.52 | | 13.35 | |

| Six Weeks | |
|---|---|
| Frequency | Cumulative % |
| 0 | 0% |
| 0 | 0% |
| 0 | 0% |
| 0 | 0% |
| 0 | 0% |
| 0 | 0% |
| 1 | 0% |
| 7 | 1% |
| 17 | 4% |
| 18 | 7% |
| 22 | 11% |
| 24 | 15% |
| 32 | 20% |
| 31 | 25% |
| 51 | 33% |
| 43 | 40% |
| 28 | 45% |
| 33 | 50% |
| 40 | 57% |
| 29 | 62% |
| 26 | 66% |
| 26 | 70% |
| 23 | 74% |
| 25 | 78% |
| 24 | 82% |
| 12 | 84% |
| 13 | 86% |
| 14 | 88% |
| 6 | 89% |
| 10 | 91% |
| 10 | 93% |
| 13 | 95% |
| 5 | 96% |
| 3 | 96% |
| 4 | 97% |
| 9 | 98% |
| 3 | 99% |
| 0 | 99% |
| 2 | 99% |
| 1 | 99% |
| 1 | 99% |
| 4 | 100% |
| 45.11 | |
| 17.68 | |

Warm Sample (-20 C)

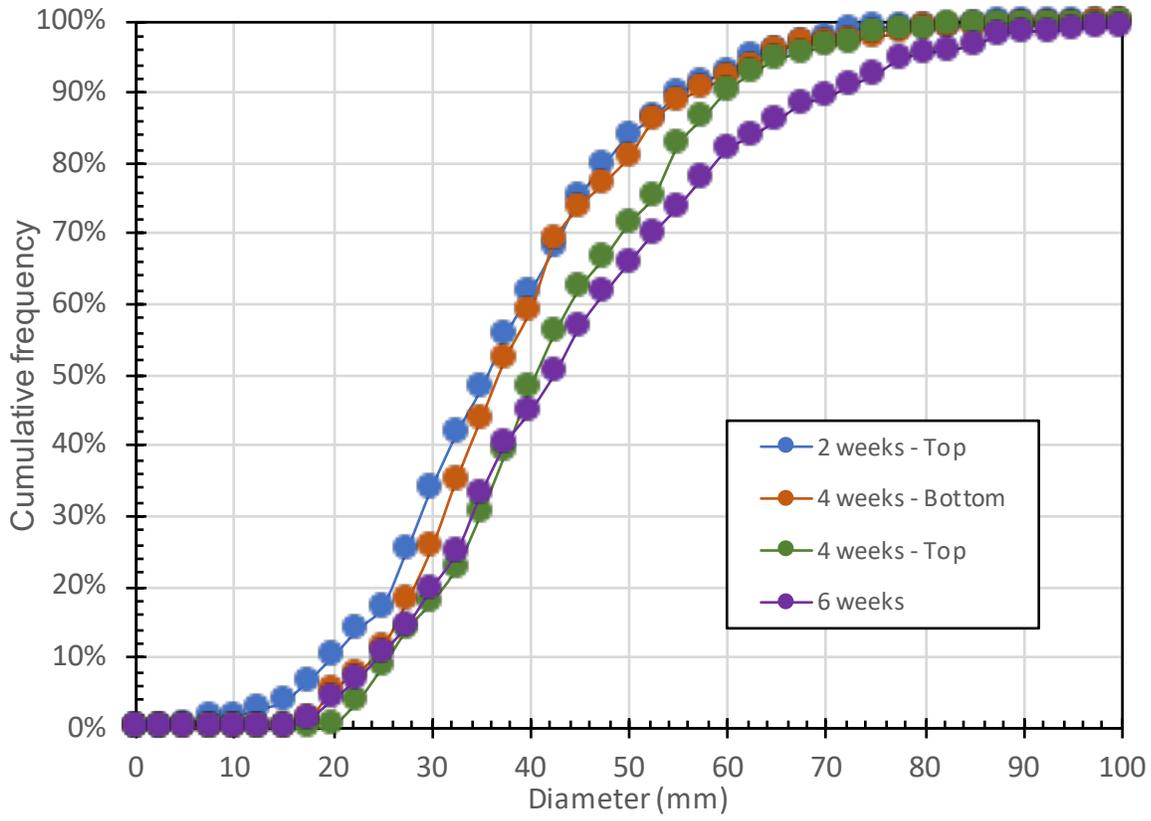

Figure S2. Grain statistics over time for the Warm Sample.

Warm and Cold Samples

|  | # Days | Mean Radius (µm) | Standard Deviation |
|---|---|---|---|
| Starting Ice | 1 | 6.57 | 5.57 |
| -80 C, Two Weeks | 14 | 10.23 | 5.92 |
| -80 C, Four Weeks | 26 | 11.69 | 6.55 |
| -80 C, Six Weeks | 40 | 17.24 | 10.96 |
| -80 C, Ten Weeks | 70 | 18.45 | 9.57 |
| -20 C, Two Weeks | 12 | 36.90 | 14.09 |
| -20 C, Four Weeks (bottom) | 24 | 38.85 | 13.35 |
| -20 C, Four Weeks (top) | 24 | 42.56 | 13.52 |
| -20 C, Six Weeks | 38 | 45.11 | 17.68 |

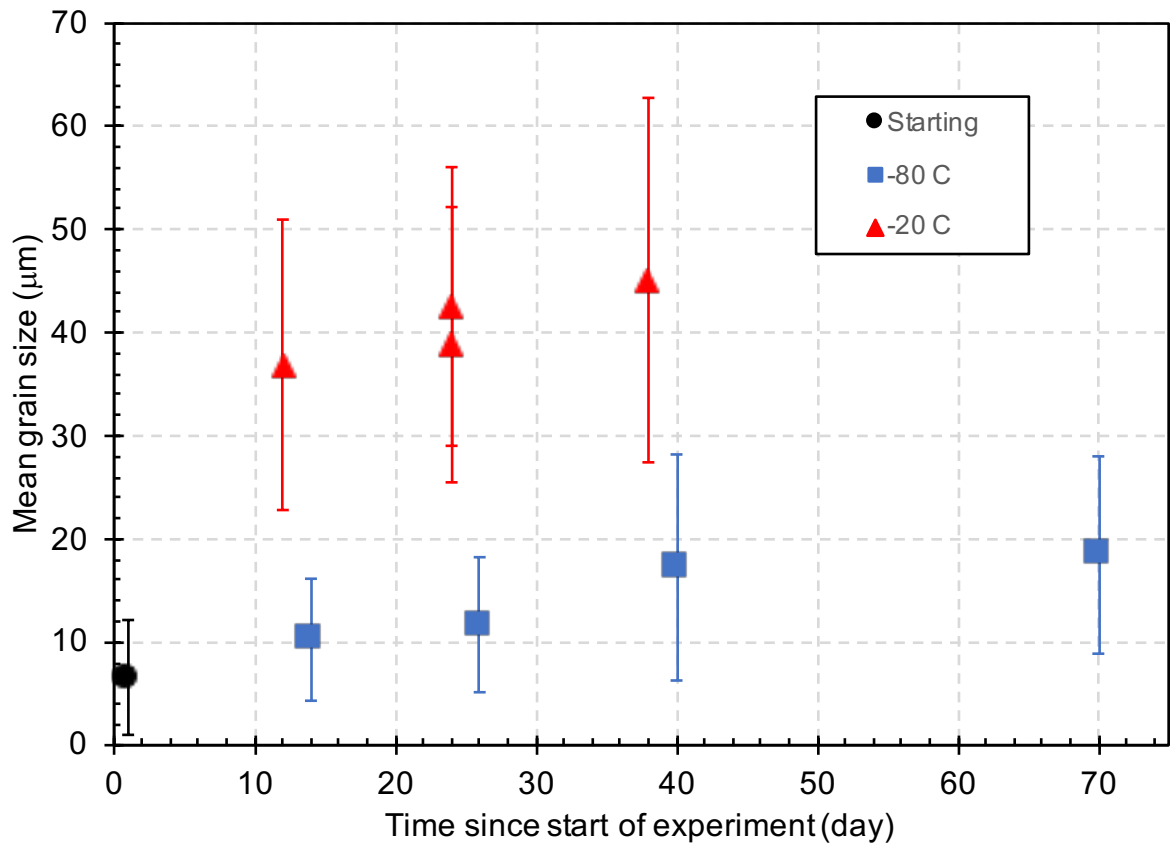

Figure S3. Comparison of grain statistics over time for both Warm and Cold samples.

# DATA FOR FIGURE 6

| Copper | | | Iron | | | Silver | | |
|---|---|---|---|---|---|---|---|---|
| Time (hr) | Relative Neck Size | Relative Density | Time (hr) | Relative Neck Size | Relative Density | Time (hr) | Relative Neck Size | Relative Density |
| 5.58E-10 | 0.0848281 | 0.5001206 | 5.58E-10 | 0.1568942 | 0.5001195 | 5.58E-10 | 0.0185913 | 0.4711248 |
| 2.37E-08 | 0.0848285 | 0.5002318 | 2.37E-08 | 0.1568951 | 0.5002298 | 2.37E-08 | 0.0194866 | 0.4712669 |
| 6.57E-08 | 0.0848292 | 0.5002318 | 6.57E-08 | 0.1568968 | 0.5002298 | 6.574E-08 | 0.0207492 | 0.4712867 |
| 1.42E-07 | 0.0848306 | 0.5002318 | 1.42E-07 | 0.1568998 | 0.5002299 | 1.421E-07 | 0.0224001 | 0.4713146 |
| 2.81E-07 | 0.0848331 | 0.5002319 | 2.81E-07 | 0.1569053 | 0.5002300 | 2.809E-07 | 0.0244240 | 0.4713521 |
| 5.33E-07 | 0.0848376 | 0.5002320 | 5.33E-07 | 0.1569153 | 0.5002302 | 5.33E-07 | 0.0267912 | 0.4714011 |
| 9.91E-07 | 0.0848457 | 0.5002323 | 9.91E-07 | 0.1569334 | 0.5002306 | 9.909E-07 | 0.0294788 | 0.4714637 |
| 1.82E-06 | 0.0848605 | 0.5002327 | 1.82E-06 | 0.1569664 | 0.5002313 | 1.823E-06 | 0.0324787 | 0.4715429 |
| 3.33E-06 | 0.0848874 | 0.5002334 | 3.33E-06 | 0.1570261 | 0.5002325 | 3.334E-06 | 0.0357978 | 0.4716426 |
| 6.08E-06 | 0.0849361 | 0.5002347 | 6.08E-06 | 0.1571342 | 0.5002348 | 6.08E-06 | 0.0394551 | 0.4717676 |
| 1.11E-05 | 0.0850243 | 0.5002372 | 1.11E-05 | 0.1573294 | 0.5002389 | 1.107E-05 | 0.0434786 | 0.4719244 |
| 2.01E-05 | 0.0851831 | 0.5002416 | 2.01E-05 | 0.1576803 | 0.5002464 | 2.013E-05 | 0.0479036 | 0.4721211 |
| 3.66E-05 | 0.0854675 | 0.5002495 | 3.66E-05 | 0.1583057 | 0.5002598 | 3.659E-05 | 0.0527712 | 0.4723676 |
| 6.65E-05 | 0.0859715 | 0.5002636 | 6.65E-05 | 0.1594040 | 0.5002837 | 6.65E-05 | 0.0581282 | 0.4726766 |
| 1.21E-04 | 0.0868479 | 0.5002886 | 0.000121 | 0.1612867 | 0.5003259 | 0.000121 | 0.0640265 | 0.4730640 |
| 2.20E-04 | 0.0883264 | 0.5003317 | 0.000220 | 0.1643934 | 0.5003991 | 0.000220 | 0.0705236 | 0.4735494 |
| 3.99E-04 | 0.0907093 | 0.5004038 | 0.000399 | 0.1692468 | 0.5005224 | 0.000399 | 0.0776826 | 0.4741575 |
| 7.25E-04 | 0.0943192 | 0.5005196 | 0.000725 | 0.1763215 | 0.5007237 | 0.000725 | 0.0855728 | 0.4749191 |
| 1.32E-03 | 0.0994029 | 0.5006967 | 0.001316 | 0.1858861 | 0.5010409 | 0.001316 | 0.0942701 | 0.4758727 |
| 2.39E-03 | 0.1060574 | 0.5009544 | 0.002392 | 0.1979487 | 0.5015257 | 0.002392 | 0.1038574 | 0.4770664 |
| 4.34E-03 | 0.1142436 | 0.5013144 | 0.004345 | 0.2123504 | 0.5022511 | 0.004345 | 0.1144254 | 0.4785607 |
| 7.89E-03 | 0.1238666 | 0.5018026 | 0.007894 | 0.2289073 | 0.5033235 | 0.007894 | 0.1260733 | 0.4804319 |
| 1.43E-02 | 0.1348526 | 0.5024527 | 0.014340 | 0.2475034 | 0.5049010 | 0.014340 | 0.1389090 | 0.4827759 |
| 2.61E-02 | 0.1471852 | 0.5033090 | 0.026052 | 0.2681235 | 0.5072223 | 0.026052 | 0.1530505 | 0.4857152 |
| 4.73E-02 | 0.1609136 | 0.5044309 | 0.047328 | 0.2908513 | 0.5106495 | 0.047328 | 0.1686259 | 0.4894059 |
| 8.60E-02 | 0.1761502 | 0.5058972 | 0.085981 | 0.3158603 | 0.5157365 | 0.085981 | 0.1857749 | 0.4940499 |
| 1.56E-01 | 0.1930681 | 0.5078116 | 0.156201 | 0.3434044 | 0.5233413 | 0.156201 | 0.2046492 | 0.4999104 |
| 2.84E-01 | 0.2119047 | 0.5103100 | 0.283770 | 0.3738038 | 0.5348171 | 0.283770 | 0.2254132 | 0.5073357 |
| 5.16E-01 | 0.2329688 | 0.5135693 | 0.515523 | 0.4074112 | 0.5523604 | 0.515523 | 0.2482449 | 0.5167945 |
| 9.37E-01 | 0.2566507 | 0.5178185 | 0.936549 | 0.4445118 | 0.5797021 | 0.936549 | 0.2733365 | 0.5289308 |
| 1.70E+00 | 0.2834300 | 0.5233544 | 1.701426 | 0.4849950 | 0.6236603 | 1.701426 | 0.3008940 | 0.5446526 |
| 3.09E+00 | 0.3138648 | 0.5305631 | 3.090975 | 0.5269145 | 0.6983491 | 3.090975 | 0.3311357 | 0.5652818 |
| 5.62E+00 | 0.3485340 | 0.5399594 | | | | 5.615364 | 0.3642869 | 0.5928211 |
| 1.02E+01 | 0.3878650 | 0.5522647 | | | | 10.201413 | 0.4005636 | 0.6304619 |
| 1.85E+01 | 0.4317243 | 0.5685808 | | | | 18.532874 | 0.4401184 | 0.6836463 |
| 3.37E+01 | 0.4785403 | 0.5908070 | | | | 33.668610 | 0.4828072 | 0.7626067 |
| 6.12E+01 | 0.5235273 | 0.6227655 | | | | | | |
| 1.11E+02 | 0.5552980 | 0.6739523 | | | | | | |

The materials do not cover exactly the same time intervals, as they finished Stage 1 of the sintering process at different times. The calculated density values for the beginning of Stage 3 (circles in Figure 6) for Copper, Iron, and Silver are 0.9716, 0.9634, and 0.9457, respectively.

# DATA FOR FIGURE 7 (LEFT)

## Diffusive Current (m^3/s)

| Time (hr) | V1 | V2 | V3 | V4 | V5 | V6 |
|---|---|---|---|---|---|---|
| 6.000E-10 | 3.443E-21 | 5.188E-26 | 5.021E-29 | 5.275E-21 | 1.197E-27 | 1.142E-28 |
| 2.630E-08 | 3.461E-21 | 5.240E-26 | 5.078E-29 | 5.275E-21 | 1.197E-27 | 1.167E-28 |
| 7.550E-08 | 3.486E-21 | 5.330E-26 | 5.186E-29 | 5.268E-21 | 1.197E-27 | 1.215E-28 |
| 1.695E-07 | 3.506E-21 | 5.471E-26 | 5.384E-29 | 5.228E-21 | 1.197E-27 | 1.304E-28 |
| 3.488E-07 | 3.469E-21 | 5.645E-26 | 5.732E-29 | 5.089E-21 | 1.196E-27 | 1.465E-28 |
| 6.913E-07 | 3.270E-21 | 5.762E-26 | 6.291E-29 | 4.745E-21 | 1.196E-27 | 1.733E-28 |
| 1.345E-06 | 2.853E-21 | 5.698E-26 | 7.074E-29 | 4.165E-21 | 1.195E-27 | 2.119E-28 |
| 2.594E-06 | 2.319E-21 | 5.432E-26 | 8.032E-29 | 3.470E-21 | 1.194E-27 | 2.580E-28 |
| 4.978E-06 | 1.810E-21 | 5.049E-26 | 9.108E-29 | 2.810E-21 | 1.194E-27 | 3.029E-28 |
| 9.530E-06 | 1.389E-21 | 4.632E-26 | 1.028E-28 | 2.250E-21 | 1.193E-27 | 3.338E-28 |
| 1.822E-05 | 1.062E-21 | 4.222E-26 | 1.154E-28 | 1.795E-21 | 1.192E-27 | 3.306E-28 |
| 3.482E-05 | 8.110E-22 | 3.836E-26 | 1.290E-28 | 1.431E-21 | 1.192E-27 | 2.593E-28 |
| 6.650E-05 | 6.203E-22 | 3.480E-26 | 1.439E-28 | 1.141E-21 | 1.191E-27 | 6.082E-29 |
| 1.270E-04 | 4.751E-22 | 3.155E-26 | 1.601E-28 | 9.091E-22 | 1.190E-27 | 0.000E+00 |
| 2.425E-04 | 3.643E-22 | 2.858E-26 | 1.778E-28 | 7.240E-22 | 1.189E-27 | 0.000E+00 |
| 4.631E-04 | 2.796E-22 | 2.587E-26 | 1.973E-28 | 5.761E-22 | 1.188E-27 | 0.000E+00 |
| 8.842E-04 | 2.148E-22 | 2.341E-26 | 2.186E-28 | 4.580E-22 | 1.187E-27 | 0.000E+00 |
| 1.688E-03 | 1.651E-22 | 2.118E-26 | 2.421E-28 | 3.638E-22 | 1.185E-27 | 0.000E+00 |
| 3.224E-03 | 1.270E-22 | 1.915E-26 | 2.679E-28 | 2.886E-22 | 1.184E-27 | 0.000E+00 |
| 6.155E-03 | 9.767E-23 | 1.730E-26 | 2.963E-28 | 2.287E-22 | 1.182E-27 | 0.000E+00 |
| 1.175E-02 | 7.515E-23 | 1.563E-26 | 3.274E-28 | 1.810E-22 | 1.180E-27 | 0.000E+00 |
| 2.244E-02 | 5.783E-23 | 1.411E-26 | 3.616E-28 | 1.431E-22 | 1.178E-27 | 0.000E+00 |
| 4.285E-02 | 4.450E-23 | 1.273E-26 | 3.990E-28 | 1.131E-22 | 1.176E-27 | 0.000E+00 |
| 8.181E-02 | 3.425E-23 | 1.148E-26 | 4.400E-28 | 8.925E-23 | 1.174E-27 | 0.000E+00 |
| 1.562E-01 | 2.635E-23 | 1.034E-26 | 4.848E-28 | 7.039E-23 | 1.171E-27 | 0.000E+00 |
| 2.982E-01 | 2.026E-23 | 9.317E-27 | 5.338E-28 | 5.548E-23 | 1.168E-27 | 0.000E+00 |
| 5.695E-01 | 1.558E-23 | 8.385E-27 | 5.872E-28 | 4.370E-23 | 1.165E-27 | 0.000E+00 |
| 1.087E+00 | 1.197E-23 | 7.540E-27 | 6.453E-28 | 3.441E-23 | 1.161E-27 | 0.000E+00 |
| 2.076E+00 | 9.187E-24 | 6.774E-27 | 7.086E-28 | 2.709E-23 | 1.157E-27 | 0.000E+00 |
| 3.964E+00 | 7.046E-24 | 6.080E-27 | 7.772E-28 | 2.132E-23 | 1.152E-27 | 0.000E+00 |
| 7.569E+00 | 5.396E-24 | 5.451E-27 | 8.514E-28 | 1.678E-23 | 1.147E-27 | 0.000E+00 |
| 1.445E+01 | 4.127E-24 | 4.881E-27 | 9.315E-28 | 1.321E-23 | 1.142E-27 | 0.000E+00 |
| 2.759E+01 | 3.150E-24 | 4.364E-27 | 1.018E-27 | 1.041E-23 | 1.136E-27 | 0.000E+00 |
| 5.269E+01 | 2.399E-24 | 3.895E-27 | 1.110E-27 | 8.204E-24 | 1.129E-27 | 0.000E+00 |
| 1.006E+02 | 1.821E-24 | 3.470E-27 | 1.208E-27 | 6.474E-24 | 1.121E-27 | 0.000E+00 |
| 1.921E+02 | 1.378E-24 | 3.084E-27 | 1.313E-27 | 5.115E-24 | 1.112E-27 | 0.000E+00 |
| 3.667E+02 | 1.038E-24 | 2.733E-27 | 1.423E-27 | 4.049E-24 | 1.102E-27 | 0.000E+00 |
| 7.002E+02 | 7.784E-25 | 2.413E-27 | 1.537E-27 | 3.211E-24 | 1.091E-27 | 0.000E+00 |
| 1.337E+03 | 5.796E-25 | 2.122E-27 | 1.654E-27 | 2.553E-24 | 1.079E-27 | 0.000E+00 |
| 2.553E+03 | 4.280E-25 | 1.856E-27 | 1.773E-27 | 2.036E-24 | 1.065E-27 | 0.000E+00 |
| 4.874E+03 | 3.125E-25 | 1.612E-27 | 1.887E-27 | 1.629E-24 | 1.050E-27 | 0.000E+00 |
| 9.307E+03 | 2.249E-25 | 1.386E-27 | 1.992E-27 | 1.308E-24 | 1.032E-27 | 0.000E+00 |
| 1.777E+04 | 1.584E-25 | 1.175E-27 | 2.075E-27 | 1.056E-24 | 1.011E-27 | 0.000E+00 |
| 3.393E+04 | 1.080E-25 | 9.713E-28 | 2.113E-27 | 8.559E-25 | 9.884E-28 | 0.000E+00 |
| 6.478E+04 | 6.911E-26 | 7.629E-28 | 2.048E-27 | 6.982E-25 | 9.618E-28 | 0.000E+00 |
| 1.237E+05 | 3.699E-26 | 5.086E-28 | 1.687E-27 | 5.736E-25 | 9.311E-28 | 0.000E+00 |

## DATA FOR FIGURE 7 (CENTER)

### Diffusive Current (m^3/s)

| Time (hr) | V1 | V2 | V3 | V4 | V5 | V6 |
|---|---|---|---|---|---|---|
| 5.583E-10 | 9.979E-16 | 3.099E-20 | 1.429E-22 | 4.093E-18 | 7.150E-22 | 6.822E-23 |
| 2.633E-08 | 2.219E-19 | 1.759E-21 | 2.008E-21 | 1.633E-20 | 7.064E-22 | 0.000E+00 |
| 7.553E-08 | 2.219E-19 | 1.759E-21 | 2.008E-21 | 1.633E-20 | 7.064E-22 | 0.000E+00 |
| 1.695E-07 | 2.219E-19 | 1.759E-21 | 2.008E-21 | 1.633E-20 | 7.064E-22 | 0.000E+00 |
| 3.488E-07 | 2.218E-19 | 1.759E-21 | 2.008E-21 | 1.633E-20 | 7.064E-22 | 0.000E+00 |
| 6.913E-07 | 2.218E-19 | 1.759E-21 | 2.008E-21 | 1.633E-20 | 7.064E-22 | 0.000E+00 |
| 1.345E-06 | 2.217E-19 | 1.759E-21 | 2.008E-21 | 1.633E-20 | 7.064E-22 | 0.000E+00 |
| 2.594E-06 | 2.215E-19 | 1.758E-21 | 2.009E-21 | 1.632E-20 | 7.064E-22 | 0.000E+00 |
| 4.978E-06 | 2.212E-19 | 1.757E-21 | 2.010E-21 | 1.630E-20 | 7.064E-22 | 0.000E+00 |
| 9.530E-06 | 2.207E-19 | 1.756E-21 | 2.011E-21 | 1.628E-20 | 7.064E-22 | 0.000E+00 |
| 1.822E-05 | 2.196E-19 | 1.753E-21 | 2.015E-21 | 1.622E-20 | 7.064E-22 | 0.000E+00 |
| 3.482E-05 | 2.176E-19 | 1.748E-21 | 2.021E-21 | 1.612E-20 | 7.063E-22 | 0.000E+00 |
| 6.650E-05 | 2.139E-19 | 1.738E-21 | 2.032E-21 | 1.594E-20 | 7.062E-22 | 0.000E+00 |
| 1.270E-04 | 2.073E-19 | 1.720E-21 | 2.052E-21 | 1.562E-20 | 7.060E-22 | 0.000E+00 |
| 2.425E-04 | 1.965E-19 | 1.690E-21 | 2.088E-21 | 1.507E-20 | 7.057E-22 | 0.000E+00 |
| 4.631E-04 | 1.800E-19 | 1.641E-21 | 2.146E-21 | 1.423E-20 | 7.052E-22 | 0.000E+00 |
| 8.842E-04 | 1.579E-19 | 1.570E-21 | 2.236E-21 | 1.305E-20 | 7.043E-22 | 0.000E+00 |
| 1.688E-03 | 1.322E-19 | 1.479E-21 | 2.360E-21 | 1.162E-20 | 7.032E-22 | 0.000E+00 |
| 3.224E-03 | 1.063E-19 | 1.372E-21 | 2.519E-21 | 1.009E-20 | 7.016E-22 | 0.000E+00 |
| 6.155E-03 | 8.285E-20 | 1.259E-21 | 2.707E-21 | 8.586E-21 | 6.997E-22 | 0.000E+00 |
| 1.175E-02 | 6.321E-20 | 1.145E-21 | 2.919E-21 | 7.218E-21 | 6.974E-22 | 0.000E+00 |
| 2.244E-02 | 4.751E-20 | 1.035E-21 | 3.150E-21 | 6.020E-21 | 6.947E-22 | 0.000E+00 |
| 4.285E-02 | 3.532E-20 | 9.302E-22 | 3.399E-21 | 4.994E-21 | 6.916E-22 | 0.000E+00 |
| 8.181E-02 | 2.599E-20 | 8.318E-22 | 3.660E-21 | 4.124E-21 | 6.880E-22 | 0.000E+00 |
| 1.562E-01 | 1.891E-20 | 7.394E-22 | 3.932E-21 | 3.391E-21 | 6.839E-22 | 0.000E+00 |
| 2.982E-01 | 1.358E-20 | 6.523E-22 | 4.211E-21 | 2.774E-21 | 6.791E-22 | 0.000E+00 |
| 5.695E-01 | 9.585E-21 | 5.697E-22 | 4.488E-21 | 2.254E-21 | 6.735E-22 | 0.000E+00 |
| 1.087E+00 | 6.606E-21 | 4.908E-22 | 4.752E-21 | 1.815E-21 | 6.671E-22 | 0.000E+00 |
| 2.076E+00 | 4.408E-21 | 4.146E-22 | 4.981E-21 | 1.446E-21 | 6.596E-22 | 0.000E+00 |
| 3.964E+00 | 2.816E-21 | 3.406E-22 | 5.135E-21 | 1.137E-21 | 6.508E-22 | 0.000E+00 |
| 7.569E+00 | 1.697E-21 | 2.685E-22 | 5.142E-21 | 8.804E-22 | 6.406E-22 | 0.000E+00 |
| 1.445E+01 | 9.437E-22 | 1.988E-22 | 4.876E-21 | 6.737E-22 | 6.287E-22 | 0.000E+00 |
| 2.759E+01 | 4.669E-22 | 1.318E-22 | 4.137E-21 | 5.146E-22 | 6.148E-22 | 0.000E+00 |
| 5.269E+01 | 1.846E-22 | 6.840E-23 | 2.669E-21 | 4.029E-22 | 5.988E-22 | 0.000E+00 |
| 1.006E+02 | 2.240E-23 | 9.872E-24 | 4.359E-22 | 3.452E-22 | 5.810E-22 | 0.000E+00 |

# DATA FOR FIGURE 7 (RIGHT)

## Diffusive Current (m^3/s)

| Time (hr) | V1 | V2 | V3 | V4 | V5 | V6 |
|---|---|---|---|---|---|---|
| 5.583E-10 | 1.482E-14 | 9.535E-19 | 2.237E-24 | 1.225E-17 | 4.439E-20 | 1.688E-20 |
| 2.633E-08 | 3.355E-18 | 5.175E-20 | 2.626E-23 | 5.527E-20 | 4.334E-20 | 0.000E+00 |
| 7.553E-08 | 3.354E-18 | 5.175E-20 | 2.626E-23 | 5.527E-20 | 4.334E-20 | 0.000E+00 |
| 1.695E-07 | 3.354E-18 | 5.174E-20 | 2.626E-23 | 5.526E-20 | 4.334E-20 | 0.000E+00 |
| 3.488E-07 | 3.354E-18 | 5.174E-20 | 2.626E-23 | 5.526E-20 | 4.334E-20 | 0.000E+00 |
| 6.913E-07 | 3.353E-18 | 5.174E-20 | 2.626E-23 | 5.525E-20 | 4.334E-20 | 0.000E+00 |
| 1.345E-06 | 3.351E-18 | 5.173E-20 | 2.627E-23 | 5.523E-20 | 4.334E-20 | 0.000E+00 |
| 2.594E-06 | 3.347E-18 | 5.171E-20 | 2.627E-23 | 5.520E-20 | 4.334E-20 | 0.000E+00 |
| 4.978E-06 | 3.341E-18 | 5.167E-20 | 2.629E-23 | 5.513E-20 | 4.334E-20 | 0.000E+00 |
| 9.530E-06 | 3.328E-18 | 5.160E-20 | 2.631E-23 | 5.500E-20 | 4.334E-20 | 0.000E+00 |
| 1.822E-05 | 3.305E-18 | 5.146E-20 | 2.635E-23 | 5.477E-20 | 4.333E-20 | 0.000E+00 |
| 3.482E-05 | 3.261E-18 | 5.121E-20 | 2.643E-23 | 5.433E-20 | 4.333E-20 | 0.000E+00 |
| 6.650E-05 | 3.183E-18 | 5.076E-20 | 2.657E-23 | 5.354E-20 | 4.332E-20 | 0.000E+00 |
| 1.270E-04 | 3.048E-18 | 4.995E-20 | 2.683E-23 | 5.216E-20 | 4.330E-20 | 0.000E+00 |
| 2.425E-04 | 2.834E-18 | 4.862E-20 | 2.727E-23 | 4.992E-20 | 4.326E-20 | 0.000E+00 |
| 4.631E-04 | 2.528E-18 | 4.660E-20 | 2.795E-23 | 4.661E-20 | 4.320E-20 | 0.000E+00 |
| 8.842E-04 | 2.147E-18 | 4.384E-20 | 2.893E-23 | 4.229E-20 | 4.312E-20 | 0.000E+00 |
| 1.688E-03 | 1.739E-18 | 4.050E-20 | 3.018E-23 | 3.733E-20 | 4.299E-20 | 0.000E+00 |
| 3.224E-03 | 1.355E-18 | 3.682E-20 | 3.165E-23 | 3.225E-20 | 4.284E-20 | 0.000E+00 |
| 6.155E-03 | 1.026E-18 | 3.308E-20 | 3.324E-23 | 2.746E-20 | 4.264E-20 | 0.000E+00 |
| 1.175E-02 | 7.618E-19 | 2.945E-20 | 3.486E-23 | 2.318E-20 | 4.239E-20 | 0.000E+00 |
| 2.244E-02 | 5.579E-19 | 2.602E-20 | 3.641E-23 | 1.947E-20 | 4.210E-20 | 0.000E+00 |
| 4.285E-02 | 4.040E-19 | 2.282E-20 | 3.779E-23 | 1.631E-20 | 4.174E-20 | 0.000E+00 |
| 8.181E-02 | 2.893E-19 | 1.986E-20 | 3.889E-23 | 1.365E-20 | 4.130E-20 | 0.000E+00 |
| 1.562E-01 | 2.045E-19 | 1.710E-20 | 3.953E-23 | 1.141E-20 | 4.075E-20 | 0.000E+00 |
| 2.982E-01 | 1.420E-19 | 1.450E-20 | 3.942E-23 | 9.534E-21 | 4.006E-20 | 0.000E+00 |
| 5.695E-01 | 9.558E-20 | 1.194E-20 | 3.803E-23 | 7.966E-21 | 3.920E-20 | 0.000E+00 |
| 1.087E+00 | 6.047E-20 | 9.243E-21 | 3.420E-23 | 6.666E-21 | 3.812E-20 | 0.000E+00 |
| 2.076E+00 | 3.188E-20 | 5.914E-21 | 2.501E-23 | 5.625E-21 | 3.676E-20 | 0.000E+00 |

## DATA FOR FIGURE 8 (MODEL GENERATED)

| (LEFT) | | (CENTER) | | (RIGHT) | |
|---|---|---|---|---|---|
| Time (hr) | Relative Neck Size | Time (hr) | Relative Neck Size | Time (hr) | Relative Neck Size |
| 2.78E-09 | 0.002750 | 2.78E-09 | 0.003792 | 2.78E-09 | 0.003684 |
| 1.40E-07 | 0.002910 | 1.40E-07 | 0.004176 | 1.84E-07 | 0.004066 |
| 3.44E-07 | 0.003099 | 3.44E-07 | 0.004545 | 4.81E-07 | 0.004461 |
| 6.48E-07 | 0.003323 | 6.48E-07 | 0.004923 | 9.70E-07 | 0.004888 |
| 1.10E-06 | 0.003584 | 1.10E-06 | 0.005326 | 1.77E-06 | 0.005365 |
| 1.77E-06 | 0.003890 | 1.77E-06 | 0.005766 | 3.10E-06 | 0.005910 |
| 2.78E-06 | 0.004249 | 2.78E-06 | 0.006256 | 5.27E-06 | 0.006545 |
| 4.27E-06 | 0.004669 | 4.27E-06 | 0.006810 | 8.85E-06 | 0.007296 |
| 6.50E-06 | 0.005162 | 6.50E-06 | 0.007444 | 1.47E-05 | 0.008197 |
| 9.81E-06 | 0.005739 | 9.81E-06 | 0.008174 | 2.44E-05 | 0.009286 |
| 1.47E-05 | 0.006415 | 1.47E-05 | 0.009023 | 4.03E-05 | 0.010606 |
| 2.21E-05 | 0.007202 | 2.21E-05 | 0.010011 | 6.65E-05 | 0.012205 |
| 3.30E-05 | 0.008118 | 3.30E-05 | 0.011164 | 1.10E-04 | 0.014138 |
| 4.93E-05 | 0.009179 | 4.93E-05 | 0.012506 | 1.80E-04 | 0.016460 |
| 7.35E-05 | 0.010404 | 7.35E-05 | 0.014067 | 2.97E-04 | 0.019238 |
| 0.000110 | 0.011814 | 0.000110 | 0.015876 | 0.000488 | 0.022546 |
| 0.000163 | 0.013433 | 0.000163 | 0.017966 | 0.000803 | 0.026470 |
| 0.000243 | 0.015289 | 0.000243 | 0.020373 | 0.001322 | 0.031110 |
| 0.000362 | 0.017412 | 0.000362 | 0.023137 | 0.002174 | 0.036584 |
| 0.000540 | 0.019839 | 0.000540 | 0.026306 | 0.003575 | 0.043029 |
| 0.000803 | 0.022610 | 0.000803 | 0.029930 | 0.005880 | 0.050604 |
| 0.001196 | 0.025770 | 0.001196 | 0.034068 | 0.009670 | 0.059492 |
| 0.001781 | 0.029373 | 0.001781 | 0.038787 | 0.015905 | 0.069905 |
| 0.002652 | 0.033476 | 0.002652 | 0.044163 | 0.026157 | 0.082081 |
| 0.003949 | 0.038148 | 0.003949 | 0.050278 | 0.043019 | 0.096290 |
| 0.005880 | 0.043462 | 0.005880 | 0.057230 | 0.070751 | 0.112829 |
| 0.008755 | 0.049503 | 0.008755 | 0.065122 | 0.116360 | 0.132025 |
| 0.013034 | 0.056366 | 0.013034 | 0.074073 | 0.191370 | 0.154222 |
| 0.019407 | 0.064155 | 0.019407 | 0.084211 | 0.314733 | 0.179774 |
| 0.028894 | 0.072987 | 0.028894 | 0.095679 | 0.517619 | 0.209021 |
| 0.043019 | 0.082990 | 0.043019 | 0.108630 | 0.851294 | 0.242252 |
| 0.064050 | 0.094305 | 0.064050 | 0.123230 | 1.400065 | 0.279648 |
| 0.095362 | 0.107085 | 0.095362 | 0.139651 | 2.302591 | 0.321192 |
| 0.141981 | 0.121493 | 0.141981 | 0.158075 | 3.786915 | 0.366523 |
| 0.211391 | 0.137702 | 0.211391 | 0.178682 | 6.228081 | 0.414721 |
| 0.314733 | 0.155894 | 0.314733 | 0.201647 | 10.242901 | 0.463992 |
| 0.468594 | 0.176248 | 0.468594 | 0.227123 | 16.845801 | 0.511310 |
| 0.697673 | 0.198940 | 0.697673 | 0.255228 | 27.705141 | 0.552271 |
| 1.038740 | 0.224128 | 1.038740 | 0.286019 | 45.564757 | 0.581927 |
| 1.546543 | 0.251934 | | | 74.937249 | 0.597611 |
| 2.302591 | 0.282423 | | | 123.244185 | 0.602393 |
| 3.428245 | 0.315565 | | | 202.691309 | 0.602967 |
| 5.104190 | 0.351188 | | | | |
| 7.599443 | 0.388905 | | | | |

# DATA FOR FIGURE 8 (FROM LITERATURE)

| Hobbs and Mason (1963) | | Kingery (1960) | | Thomas et al. (1994) | |
|---|---|---|---|---|---|
| Time (hr) | Relative Neck Size | Time (hr) | Relative Neck Size | Time (hr) | Size |
| 0.05 | 0.1063 | 0.03 | 0.2493 | 2.49 | 0.0962 |
| 0.67 | 0.1224 | 0.06 | 0.2969 | 4.85 | 0.1489 |
| 2.55 | 0.1466 | 0.09 | 0.3135 | 19.86 | 0.1782 |
| 5.33 | 0.1723 | 0.14 | 0.3365 | 26.20 | 0.2105 |
| | | | | 44.02 | 0.2697 |
| | | | | 53.72 | 0.2734 |
| | | | | 68.40 | 0.3101 |
| | | | | 77.20 | 0.3200 |
| | | | | 97.58 | 0.3764 |
| | | | | 163.99 | 0.4446 |
| | | | | 190.44 | 0.4406 |

## DATA FOR FIGURES 9 & 10

| Time (hr) | Relative Neck Size | Relative Density | Diffusive Current (m^3/s) | | | | | |
|---|---|---|---|---|---|---|---|---|
| | | | V1 | V2 | V3 | V4 | V5 | V6 |
| 2.33E-10 | 0.0062428 | 0.5001164 | 5.19E-20 | 3.89E-24 | 6.29E-22 | 1.72E-23 | 1.02E-25 | 1.26E-26 |
| 9.24E-09 | 0.0074407 | 0.5002237 | 3.04E-20 | 3.25E-24 | 7.50E-22 | 1.21E-23 | 1.02E-25 | 1.69E-26 |
| 2.51E-08 | 0.0083584 | 0.5002237 | 2.13E-20 | 2.89E-24 | 8.42E-22 | 9.55E-24 | 1.02E-25 | 2.00E-26 |
| 5.32E-08 | 0.0092252 | 0.5002238 | 1.58E-20 | 2.61E-24 | 9.28E-22 | 7.83E-24 | 1.02E-25 | 2.24E-26 |
| 1.03E-07 | 0.0101082 | 0.5002238 | 1.20E-20 | 2.38E-24 | 1.02E-21 | 6.52E-24 | 1.02E-25 | 2.41E-26 |
| 1.90E-07 | 0.0110414 | 0.5002238 | 9.19E-21 | 2.18E-24 | 1.11E-21 | 5.46E-24 | 1.02E-25 | 2.45E-26 |
| 3.43E-07 | 0.0120491 | 0.5002238 | 7.06E-21 | 1.99E-24 | 1.21E-21 | 4.58E-24 | 1.02E-25 | 2.26E-26 |
| 6.14E-07 | 0.0131536 | 0.5002239 | 5.41E-21 | 1.82E-24 | 1.32E-21 | 3.84E-24 | 1.02E-25 | 1.68E-26 |
| 1.09E-06 | 0.0143796 | 0.5002239 | 4.13E-21 | 1.66E-24 | 1.44E-21 | 3.22E-24 | 1.02E-25 | 3.99E-27 |
| 1.93E-06 | 0.0157575 | 0.5002240 | 3.13E-21 | 1.52E-24 | 1.57E-21 | 2.68E-24 | 1.02E-25 | 0.00E+00 |
| 3.42E-06 | 0.0173278 | 0.5002240 | 2.34E-21 | 1.38E-24 | 1.73E-21 | 2.21E-24 | 1.02E-25 | 0.00E+00 |
| 6.04E-06 | 0.0191453 | 0.5002241 | 1.73E-21 | 1.24E-24 | 1.90E-21 | 1.81E-24 | 1.02E-25 | 0.00E+00 |
| 1.07E-05 | 0.0212849 | 0.5002242 | 1.25E-21 | 1.11E-24 | 2.11E-21 | 1.46E-24 | 1.02E-25 | 0.00E+00 |
| 1.88E-05 | 0.0238483 | 0.5002243 | 8.85E-22 | 9.89E-25 | 2.36E-21 | 1.16E-24 | 1.02E-25 | 0.00E+00 |
| 3.31E-05 | 0.0269698 | 0.5002245 | 6.07E-22 | 8.71E-25 | 2.66E-21 | 9.09E-25 | 1.02E-25 | 0.00E+00 |
| 5.85E-05 | 0.0308187 | 0.5002246 | 4.03E-22 | 7.58E-25 | 3.02E-21 | 6.95E-25 | 1.02E-25 | 0.00E+00 |
| 0.000103 | 0.0355985 | 0.5002247 | 2.58E-22 | 6.51E-25 | 3.47E-21 | 5.20E-25 | 1.02E-25 | 0.00E+00 |
| 0.000182 | 0.0415399 | 0.5002249 | 1.60E-22 | 5.53E-25 | 4.02E-21 | 3.81E-25 | 1.02E-25 | 0.00E+00 |
| 0.000321 | 0.0488958 | 0.5002250 | 9.63E-23 | 4.64E-25 | 4.69E-21 | 2.74E-25 | 1.02E-25 | 0.00E+00 |
| 0.000565 | 0.0579395 | 0.5002252 | 5.65E-23 | 3.86E-25 | 5.49E-21 | 1.94E-25 | 1.01E-25 | 0.00E+00 |
| 0.000997 | 0.0689706 | 0.5002253 | 3.25E-23 | 3.18E-25 | 6.44E-21 | 1.36E-25 | 1.01E-25 | 0.00E+00 |
| 0.001757 | 0.0823241 | 0.5002255 | 1.84E-23 | 2.60E-25 | 7.55E-21 | 9.49E-26 | 1.01E-25 | 0.00E+00 |
| 0.003098 | 0.0983788 | 0.5002256 | 1.03E-23 | 2.12E-25 | 8.83E-21 | 6.58E-26 | 1.01E-25 | 0.00E+00 |
| 0.005463 | 0.1175610 | 0.5002258 | 5.70E-24 | 1.71E-25 | 1.03E-20 | 4.56E-26 | 1.00E-25 | 0.00E+00 |
| 0.009633 | 0.1403416 | 0.5002259 | 3.12E-24 | 1.37E-25 | 1.18E-20 | 3.16E-26 | 1.00E-25 | 0.00E+00 |
| 0.016986 | 0.1672210 | 0.5002261 | 1.69E-24 | 1.09E-25 | 1.35E-20 | 2.19E-26 | 9.95E-26 | 0.00E+00 |
| 0.029951 | 0.1986983 | 0.5002264 | 9.05E-25 | 8.57E-26 | 1.52E-20 | 1.52E-26 | 9.89E-26 | 0.00E+00 |
| 0.052811 | 0.2352136 | 0.5002266 | 4.76E-25 | 6.64E-26 | 1.68E-20 | 1.06E-26 | 9.82E-26 | 0.00E+00 |
| 0.093120 | 0.2770467 | 0.5002270 | 2.46E-25 | 5.05E-26 | 1.80E-20 | 7.43E-27 | 9.73E-26 | 0.00E+00 |
| 0.164195 | 0.3241447 | 0.5002274 | 1.23E-25 | 3.73E-26 | 1.86E-20 | 5.25E-27 | 9.62E-26 | 0.00E+00 |
| 0.289520 | 0.3758317 | 0.5002279 | 5.98E-26 | 2.65E-26 | 1.83E-20 | 3.76E-27 | 9.50E-26 | 0.00E+00 |
| 0.510501 | 0.4303436 | 0.5002285 | 2.75E-26 | 1.77E-26 | 1.64E-20 | 2.75E-27 | 9.36E-26 | 0.00E+00 |
| 0.900150 | 0.4841715 | 0.5002294 | 1.17E-26 | 1.06E-26 | 1.29E-20 | 2.07E-27 | 9.21E-26 | 0.00E+00 |
| 1.587205 | 0.5315204 | 0.5002306 | 4.34E-27 | 5.35E-27 | 8.01E-21 | 1.65E-27 | 9.06E-26 | 0.00E+00 |
| 2.798668 | 0.5651957 | 0.5002325 | 1.27E-27 | 1.93E-27 | 3.34E-21 | 1.41E-27 | 8.95E-26 | 0.00E+00 |
| 4.934802 | 0.5812460 | 0.5002354 | 2.21E-28 | 3.74E-28 | 6.90E-22 | 1.31E-27 | 8.90E-26 | 0.00E+00 |
| 8.701377 | 0.5848784 | 0.5002403 | 1.26E-29 | 2.19E-29 | 4.10E-23 | 1.29E-27 | 8.88E-26 | 0.00E+00 |
| 15.342860 | 0.5850963 | 0.5002490 | 3.73E-32 | 6.47E-32 | 1.21E-25 | 1.29E-27 | 8.88E-26 | 0.00E+00 |

## DATA FOR FIGURE 12 (MODEL GENERATED)

| | (Left) | | (Right) | |
|---|---|---|---|---|
| Time (min) | Relative Neck Size (dashed) | Relative Neck Size (solid) | Relative Neck Size (dashed) | Relative Neck Size (solid) |
| 3.35E-07 | 0.0039579 | 0.0042882 | 0.0039124 | 0.0043775 |
| 1.13E-05 | 0.0048576 | 0.0054845 | 0.0044038 | 0.0051623 |
| 2.93E-05 | 0.0056141 | 0.0064045 | 0.0048767 | 0.0058132 |
| 5.9E-05 | 0.0063987 | 0.0073406 | 0.0053720 | 0.0064601 |
| 0.00011 | 0.0072797 | 0.0083848 | 0.0059160 | 0.0071548 |
| 0.00019 | 0.0083078 | 0.0095995 | 0.0065330 | 0.0079354 |
| 0.00032 | 0.0095308 | 0.0110412 | 0.0072495 | 0.0088387 |
| 0.00054 | 0.0109970 | 0.0127663 | 0.0080966 | 0.0099062 |
| 0.00089 | 0.0127574 | 0.0148333 | 0.0091116 | 0.0111852 |
| 0.00148 | 0.0148667 | 0.0173055 | 0.0103374 | 0.0127294 |
| 0.00244 | 0.0173854 | 0.0202524 | 0.0118229 | 0.0145986 |
| 0.00403 | 0.0203818 | 0.0237533 | 0.0136216 | 0.0168579 |
| 0.00664 | 0.0239352 | 0.0278998 | 0.0157920 | 0.0195779 |
| 0.01093 | 0.0281375 | 0.0327983 | 0.0183982 | 0.0228370 |
| 0.01799 | 0.0330965 | 0.0385731 | 0.0215125 | 0.0267229 |
| 0.02960 | 0.0389380 | 0.0453687 | 0.0252173 | 0.0313369 |
| 0.04869 | 0.0458082 | 0.0533529 | 0.0296083 | 0.0367961 |
| 0.08009 | 0.0538767 | 0.0627188 | 0.0347974 | 0.0432371 |
| 0.13173 | 0.0633386 | 0.0736879 | 0.0409151 | 0.0508187 |
| 0.21665 | 0.0744177 | 0.0865122 | 0.0481133 | 0.0597247 |
| 0.35632 | 0.0873677 | 0.1014752 | 0.0565681 | 0.0701666 |
| 0.58603 | 0.1024748 | 0.1188932 | 0.0664823 | 0.0823858 |
| 0.96382 | 0.1200570 | 0.1391131 | 0.0780878 | 0.0966553 |
| 1.58514 | 0.1404634 | 0.1625088 | 0.0916475 | 0.1132810 |
| 2.60698 | 0.1640692 | 0.1894726 | 0.1074568 | 0.1326003 |
| 4.28753 | 0.1912676 | 0.2204015 | 0.1258431 | 0.1549791 |
| 7.05142 | 0.2224553 | 0.2556752 | 0.1471635 | 0.1808051 |
| 11.59700 | 0.2580100 | 0.2956233 | 0.1717999 | 0.2104755 |
| 19.07280 | 0.2982568 | 0.3404788 | 0.2001480 | 0.2443781 |
| 31.36774 | 0.3434208 | 0.3903103 | 0.2326017 | 0.2828620 |
| 51.58840 | 0.3935577 | 0.4449252 | 0.2695269 | 0.3261960 |
| 84.84393 | 0.4484545 | 0.5037221 | 0.3112248 | 0.3745076 |
| 139.53704 | 0.5074779 | 0.5654574 | 0.3578780 | 0.4276970 |

## Log of Sintering Timescale (yr)

| Temperature (K) | 1 μm Radius | 10 μm Radius | 100 μm Radius |
|---|---|---|---|
| 69  | 12.5989 | 15.8874 | 17.9661 |
| 74  | 11.5191 | 14.6946 | 16.7656 |
| 79  | 10.5099 | 13.5798 | 15.6437 |
| 84  | 9.5627  | 12.5334 | 14.5907 |
| 89  | 8.6703  | 11.5476 | 13.5986 |
| 94  | 7.8266  | 10.6156 | 12.6607 |
| 99  | 7.0267  | 9.7320  | 11.7714 |
| 104 | 6.2662  | 8.8919  | 10.9260 |
| 109 | 5.5414  | 8.0913  | 10.1203 |
| 114 | 4.8492  | 7.3265  | 9.3507  |
| 119 | 4.1866  | 6.5947  | 8.6141  |
| 124 | 3.5514  | 5.8929  | 7.9079  |
| 129 | 2.9412  | 5.2189  | 7.2296  |
| 134 | 2.3543  | 4.5705  | 6.5771  |
| 139 | 1.7888  | 3.9459  | 5.9485  |
| 144 | 1.2434  | 3.3433  | 5.3421  |
| 149 | 0.7165  | 2.7613  | 4.7564  |
| 154 | 0.2071  | 2.1986  | 4.1900  |
| 159 | -0.2861 | 1.6538  | 3.6418  |
| 164 | -0.7640 | 1.1259  | 3.1105  |
| 169 | -1.2275 | 0.6138  | 2.5952  |
| 174 | -1.6775 | 0.1167  | 2.0949  |
| 179 | -2.1148 | -0.3664 | 1.6088  |
| 184 | -2.5400 | -0.8361 | 1.1360  |
| 189 | -2.9539 | -1.2933 | 0.6760  |
| 194 | -3.3569 | -1.7385 | 0.2279  |
| 199 | -3.7497 | -2.1723 | -0.2087 |

# DATA FOR FIGURE 16

## Log of Sintering Timescale (yr)

| Grain Radius ($\mu$m) | 100% Sintered | 98% Sintered | 50% Sintered |
|---|---|---|---|
| 1 | 3.29E+03 | 6.94E+02 | 6.94E+09 |
| 31 | 4.02E+06 | 6.60E+05 | 6.60E+12 |
| 61 | 1.63E+07 | 2.55E+06 | 2.55E+13 |
| 91 | 3.74E+07 | 5.67E+06 | 5.67E+13 |
| 121 | 6.74E+07 | 1.00E+07 | 1.00E+14 |
| 151 | 1.07E+08 | 1.56E+07 | 1.56E+14 |
| 181 | 1.55E+08 | 2.24E+07 | 2.24E+14 |
| 211 | 2.13E+08 | 3.04E+07 | 3.04E+14 |
| 241 | 2.81E+08 | 3.96E+07 | 3.96E+14 |
| 271 | 3.58E+08 | 5.01E+07 | 5.01E+14 |
| 301 | 4.44E+08 | 6.18E+07 | 6.18E+14 |
| 331 | 5.41E+08 | 7.47E+07 | 7.47E+14 |
| 361 | 6.47E+08 | 8.88E+07 | 8.88E+14 |
| 391 | 7.64E+08 | 1.04E+08 | 1.04E+15 |
| 421 | 8.90E+08 | 1.21E+08 | 1.21E+15 |
| 451 | 1.03E+09 | 1.39E+08 | 1.39E+15 |
| 481 | 1.17E+09 | 1.58E+08 | 1.58E+15 |
| 511 | 1.33E+09 | 1.78E+08 | 1.78E+15 |
| 541 | 1.50E+09 | 1.99E+08 | 1.99E+15 |
| 571 | 1.67E+09 | 2.22E+08 | 2.22E+15 |
| 601 | 1.86E+09 | 2.46E+08 | 2.46E+15 |
| 631 | 2.06E+09 | 2.71E+08 | 2.71E+15 |
| 661 | 2.26E+09 | 2.97E+08 | 2.97E+15 |
| 691 | 2.48E+09 | 3.25E+08 | 3.25E+15 |
| 721 | 2.71E+09 | 3.54E+08 | 3.54E+15 |
| 751 | 2.95E+09 | 3.84E+08 | 3.84E+15 |
| 781 | 3.20E+09 | 4.15E+08 | 4.15E+15 |
| 811 | 3.46E+09 | 4.47E+08 | 4.47E+15 |
| 841 | 3.73E+09 | 4.81E+08 | 4.81E+15 |
| 871 | 4.01E+09 | 5.16E+08 | 5.16E+15 |
| 901 | 4.30E+09 | 5.52E+08 | 5.52E+15 |
| 931 | 4.60E+09 | 5.89E+08 | 5.89E+15 |
| 961 | 4.91E+09 | 6.28E+08 | 6.28E+15 |
| 991 | 5.24E+09 | 6.67E+08 | 6.67E+15 |

| Normalized Sintering Timescale | Depth (skin depths) |
|---|---|
| 0.12 | 0.00 |
| 0.13 | 0.06 |
| 0.14 | 0.12 |
| 0.16 | 0.19 |
| 0.18 | 0.25 |
| 0.20 | 0.31 |
| 0.22 | 0.37 |
| 0.23 | 0.43 |
| 0.24 | 0.50 |
| 0.26 | 0.56 |
| 0.28 | 0.62 |
| 0.29 | 0.68 |
| 0.30 | 0.75 |
| 0.32 | 0.82 |
| 0.34 | 0.89 |
| 0.36 | 0.96 |
| 0.37 | 1.04 |
| 0.39 | 1.13 |
| 0.41 | 1.21 |
| 0.43 | 1.31 |
| 0.45 | 1.40 |
| 0.47 | 1.50 |
| 0.49 | 1.61 |
| 0.52 | 1.72 |
| 0.54 | 1.84 |
| 0.57 | 1.96 |
| 0.60 | 2.09 |
| 0.63 | 2.23 |
| 0.66 | 2.37 |
| 0.69 | 2.52 |
| 0.72 | 2.68 |
| 0.76 | 2.84 |
| 0.79 | 3.01 |
| 0.83 | 3.20 |
| 0.86 | 3.39 |
| 0.89 | 3.59 |
| 0.92 | 3.80 |
| 0.95 | 4.02 |
| 0.97 | 4.25 |
| 0.99 | 4.49 |
| 1.00 | 4.75 |
| 1.00 | 5.02 |